\documentclass[twocolumn]{aastex63}
\usepackage{amsmath}

\newcommand{\delx}{\Delta x}
\newcommand{\grad}{\mathbf{\nabla}}

\newcommand{\nhat}{\mathbf{\hat{n}}}
\newcommand{\rhat}{\mathbf{\hat{r}}}
\newcommand{\shat}{\mathbf{\hat{s}}}
\newcommand{\vout}{ v_{\rm out}}
\newcommand{\voutavg}{\left\langle v_{\rm out} \right\rangle}
\newcommand{\vcond}{v_{\rm cond}}
\newcommand{\vturb}{v_{\rm turb}}

\newcommand{\foldedness}{\left\langle \mathbf{\hat{r}}\cdot  \mathbf{\hat{n}}\right\rangle}

\newcommand{\Tpk}{T_{\rm pk}}
\newcommand{\Ptpk}{P(\Tpk)}
\newcommand{\fdip}{f_{\rm dip}}
\newcommand{\Ltpk}{\Lambda(\Tpk)}
\newcommand{\edotpk}{\dot{E}_{\rm cool, pk}}
\newcommand{\Vpk}{V_{\rm pk}}

\newcommand{\vel}{\mathbf{v}}
\newcommand{\magfield}{\mathbf{B}}
\newcommand{\rhobar}{\overline{\rho}}
\newcommand{\mdotw}{\dot{M}_w}
\newcommand{\pdotw}{\dot{p}_w}
\newcommand{\Lwind}{L_w}
\newcommand{\Vwind}{V_w}
\newcommand{\fwind}{f_{\rm wind}}

\newcommand{\Rbub}{R_{\rm bub}}
\newcommand{\Wbub}{W_{\rm bub}}
\newcommand{\Wbubavg}{\left\langle W_{\rm bub}\right\rangle}
\newcommand{\Rfree}{R_{\rm f}}
\newcommand{\Rdotbub}{\dot{R}_{\rm bub}}
\newcommand{\Vbub}{V_{\rm bub}}

\newcommand{\Abub}{A_{\rm bub}}

\newcommand{\Phot}{P_{\rm hot}}
\newcommand{\rhohot}{\rho_{\rm hot}}

\newcommand{\dsteady}{\varphi_{\rm bub}}

\newcommand{\rhops}{\rho_{\rm ps}}
\newcommand{\vps}{v_{\rm ps}}
\newcommand{\Pps}{P_{\rm ps}}

\newcommand{\lfield}{\lambda_F}
\newcommand{\lcool}{\ell_{\rm cool}}
\newcommand{\tcool}{t_{\rm cool}}
\newcommand{\teddy}{t_{\rm eddy}}
\newcommand{\wcool}{w_{\rm cool}}
\newcommand{\tschot}{t_{\rm sc, hot}}

\newcommand{\texp}{t_{\rm exp}}
\newcommand{\tns}{t_{\rm ns}}

\newcommand{\cshot}{c_{s,{\rm hot}}}

\newcommand{\Rcl}{R_{\rm cl}}
\newcommand{\Mcl}{M_{\rm cl}}
\newcommand{\Nres}{N_{\rm res}}
\newcommand{\Lbox}{L_{\rm box}}

\newcommand{\Mstar}{M_*}

\newcommand{\kms}{\,{\rm km}\,{\rm s}^{-1}}
\newcommand{\cc}{\,{\rm cm}^{-3}}

\newlength{\subcolumnwidth}

\newcommand{\nextsubcolumn}[1][]{%
  \cr\noalign{\hfill}
  \if\relax\detokenize{#1}\relax\else\hsize=#1\setlength{\subcolumnwidth}{\hsize}\fi
}

\received{XXX}
\revised{XXX}
\accepted{XXX}

\submitjournal{ApJ}

\shorttitle{Geometry, Dissipation, and Cooling in WBBs}
\shortauthors{Lancaster et al.}
\graphicspath{{./}{figures/}}

\begin{document}

\title{Geometry, Dissipation, Cooling, and the Dynamical Evolution of Wind-Blown Bubbles}

\correspondingauthor{Lachlan Lancaster}
\email{ltl2125@columbia.edu}

\author[0000-0002-0041-4356]{Lachlan Lancaster}
\thanks{Simons Fellow}
\affiliation{Department of Astronomy, Columbia University,  550 W 120th St, New York, NY 10025, USA}
\affiliation{Center for Computational Astrophysics, Flatiron Institute, 162 5th Avenue, New York, NY 10010, USA}
\affiliation{Department of Astrophysical Sciences, Princeton University, 4 Ivy Lane, Princeton, NJ 08544, USA}

\author[0000-0002-0509-9113]{Eve C. Ostriker}
\affiliation{Department of Astrophysical Sciences, Princeton University, 4 Ivy Lane, Princeton, NJ 08544, USA}
\affiliation{Institute for Advanced Study, 1 Einstein Drive, Princeton, NJ 08540, USA}

\author[0000-0003-2896-3725]{Chang-Goo Kim}
\affiliation{Department of Astrophysical Sciences, Princeton University, 4 Ivy Lane, Princeton, NJ 08544, USA}

\author[0000-0001-6228-8634]{Jeong-Gyu Kim}
\affiliation{Division of Science, National Astronomical Observatory of Japan, Mitaka, Tokyo 181-0015, Japan}
\affiliation{Korea Astronomy and Space Science Institute, Daejeon 34055, Republic Of Korea}
\affiliation{Department of Astrophysical Sciences, Princeton University, 4 Ivy Lane, Princeton, NJ 08544, USA}

\author[0000-0003-2630-9228]{Greg L. Bryan}
\affiliation{Department of Astronomy, Columbia University,  550 W 120th St, New York, NY 10025, USA}
\affiliation{Center for Computational Astrophysics, Flatiron Institute, 162 5th Avenue, New York, NY 10010, USA}

\begin{abstract}
Bubbles driven by energy and mass injection from small scales are ubiquitous in astrophysical fluid systems and essential to feedback across multiple scales. In particular, O stars in young clusters produce high velocity winds that create hot bubbles in the surrounding gas. We demonstrate that the dynamical evolution of these bubbles is critically dependent upon the geometry of their interfaces with their surroundings and the nature of heat transport across these interfaces. These factors together determine the amount of energy that can be lost from the interior through cooling at the interface, which in turn determines the ability of the bubble to do work on its surroundings. We further demonstrate that the scales relevant to physical dissipation across this interface are extremely difficult to resolve in global numerical simulations of bubbles for parameter values of interest. This means the dissipation driving evolution of these bubbles in numerical simulations is often of a numerical nature. We describe the physical and numerical principles that determine the level of dissipation in these simulations; we use this, along with a fractal model for the geometry of the interfaces, to explain differences in convergence behavior between hydrodynamical and magneto-hydrodynamical simulations presented here. We additionally derive an expression for momentum as a function of bubble radius expected when the relevant dissipative scales are resolved and show that it still results in efficiently-cooled solutions as postulated in previous work.
\end{abstract}

\keywords{ISM, Stellar Winds, Star forming regions}

\section{Introduction}
\label{sec:intro}

In the modern theory of galaxy evolution and star formation, energetic feedback from stars \citep{McKeeOstriker07,PRFM22,WalchSILCC15,KrumholzSFR18,Krumholz14,Sun23,Hopkins14,Kruijssen19,Chevance23} and central black holes or active galactic nuclei \citep[AGN,][]{FGQ12,ZG14,Andrew15} is essential in regulating the flow of mass and energy, and hence the growth of a galaxy and its stellar population \citep{SomervilleDave15,NaabOstriker17}.

Feedback plays multiple roles on multiple scales. AGN and `super-bubble' feedback from repeated supernovae (SNe) drive outflows from galaxies overall \citep{TomisakaIkeuchi86,maclow88,McCrayKafatos87,KooMcKee92a,FGQ12,Sharma14,KrauseDiehl14,Cicone14,Andrew15,Yadav17,KimOstrikerRaileanu17,ElBadry19,Orr22,Gentry17,Gentry19}, set the energy and mass- and metal- loading of galactic winds \citep{Schneider2020,FieldingBryan22,arkenstone1,CGK_TIGRESS2,CGK2020,CGK20_SMAUG,Pandya20_SMAUG}, and regulate the energy balance of the galaxy's circum-galactic medium \citep[CGM, ][]{Fielding17,FGO23CGM,Pandya_CGMSAM}. Expanding remnants from individual and clustered SNe are also believed to bear primary responsibility for driving turbulence on large scales in the ISM \citep{McKeeOstriker77,Elmegreen2004,Agertz2009, OstrikerShetty2011,kok11,kok13,hennebelle14,Girichidis16,PRFM22}. Radiation feedback from stars locally acts to disperse the molecular gas around star-forming regions \citep{Dale05,Dale12,Walch12,Haid18,Grudic21Starforge,StarforgeResults22,JGK16,JGK18,JGK21,Geen15b,Geen16,Geen17,he19} but also provides Far Ultraviolet (FUV) radiation which is essential to the equilibrium thermal balance of the surrounding ISM \citep{Ostriker10,JGK_NCR23,PRFM22,TIGRESS_NCR}. Kinetic feedback from stellar winds \citep{Lancaster21a,Lancaster21b,Fierlinger16,Krause13,HCM09,Rosen14,Lopez11,RogersPittard13,Wareing17,Dale14} and proto-stellar jets \citep{Cunningham11,Offner16,OffnerChaban17,MGC18} can act to regulate the accretion of mass onto individual stars \citep{Rosen20,Guszejnov21}, as well as help to disrupt the parent giant molecular clouds (GMCs) in which these stars form \citep{Lancaster21c,Dale15Review}.

In many of the above situations, high velocity flows shock to create a `bubble' of very hot, very diffuse gas that is `blown' into the surrounding medium, so-called Wind-Blown Bubbles (WBBs). While in this work we will generally explore the parameter space of mechanical feedback relevant to WBBs from stellar winds, much of the material presented here is relevant to feedback from AGNs, super-bubbles at late stages, or proto-stellar jets with appropriate changes to the relevant cooling physics and pressures to be considered. It should be noted that the case of an individual SN remnant (as well as most of the evolution of a super-bubble) is quite distinct from WBB dynamics, in that the majority of the work imparted by feedback is done by individual shocks propagating into surrounding gas, rather than by application of a continuous pressure force by the hot bubble interior. Because SN feedback is impulsive, the strict resolution requirements we discuss in this work are not necessary to accurately model the momentum injection by SNe in the ISM.

It is clear from recent work that the effectiveness of feedback from WBBs is strongly dependent upon the exchange of energy between the hot gas in a bubble's interior and the cold gas in its radiative shell\footnote{We have implicitly specified to a given region of the parameter space relevant to the expansion of WBBs; we will explain further in \autoref{sec:theory} that this is generally the regime of greatest interest.} \citep{Gentry17,Gentry19,ElBadry19,Lancaster21a,Lancaster21b}. In particular, such energy exchange, mediated by either micro-(conduction) or macro-(turbulent) dissipation, can potentially lead to much of the feedback energy being radiated away. Additionally, energy exchange across this interface can have drastic consequences for the mass and density of the hot phase, which in turn has strong implications for observational signatures of these interfaces \citep{ChuMacLow90,ParkinPittard10,RogersPittard14,Krause14,ToalaArthur18}. Understanding the exchange of energy across such interfaces is therefore a fundamental problem to the theory of star formation and galaxy evolution.

As we review in \autoref{sec:theory}, correctly resolving this energy exchange should entail resolving the scales appropriate to the dissipative physics that mediate this exchange. In some situations, including shear-driven turbulent boundary layers, numerical simulations indicate that it may not be necessary to fully resolve the relevant dissipative scales in order for the cooling rate to be converged \citep{FieldingFractal20,TOG21}.

However, in the present work we will show through detailed analysis of three-dimensional hydrodynamical (HD) and magneto-hydrodynamical (MHD) simulations that this property does not hold for the case of simulated WBBs. In particular, in \autoref{sec:theory} we will expand upon the fluid structure analysis of \citet{Lancaster21a} relevant for such bubbles to show exactly how the geometry of and dissipation at the WBB's interface determines its dynamical evolution. We combine this with an extensive discussion of the physics relevant to setting the geometry and dissipative properties of this interface, and how they should set the transfer of energy when they are properly resolved in simulations. We quantify the numerical requirements for the relevant physical scales to be resolved, and demonstrate that this is difficult to achieve even for very high numerical resolution. Finally, we present a consistency condition on cooling in our simulations which can act to determine the degree of dissipation in the `unresolved' case.

In \autoref{sec:simulations} we describe the numerical methods used in our simulations and present the details of the main suite of simulations that we perform. In \autoref{sec:results} we use these simulations to extensively validate the physical picture described in \autoref{sec:theory} and demonstrate how the resolution dependence of the bulk properties of the WBBs can be explained through the resolution dependence of the fractal geometry of the bubbles and dissipation at their interfaces governed through the consistency condition laid out in \autoref{subsec:vout_cooling}. Finally, in \autoref{sec:discussion} we discuss this work in the context of past results both in regards to WBBs and the mixing layer literature as well as avenues for future work. We summarize our main conclusions in \autoref{sec:conclusions}.

\section{Background and Theory}
\label{sec:theory}

In this section we review the theory of WBBs relevant to the present work and present several novel interpretations and results. In \autoref{subsec:problem_def} we provide a formal definition of the problem under consideration. In \autoref{subsec:structure_highlights} we review the results presented in \autoref{app:structure}, in particular how the fluid structure in the shocked wind determines the dynamical evolution of the bubble. In \autoref{subsec:contact} and \autoref{subsec:diss_scales} we discuss the processes and physical scales relevant for setting the boundary conditions on the fluid flow interior to the bubble. In \autoref{subsec:vout_cooling} we consider how a consistency condition that the boundary layer and bubble interior must mutually satisfy, and how this is affected by the limited numerical resolution of the simulations we present in \autoref{sec:results}. Finally, in \autoref{subsec:momentum_resolved} we derive the momentum evolution expected for a WBB where the boundary conditions on its surface are determined by physically resolved turbulent dissipation.

\subsection{Problem Definition}
\label{subsec:problem_def}

We consider the action of a constant mass-loss rate, $\mdotw$, and mechanical wind luminosity, $\Lwind$, source term into a general background medium.  Specifically, the equations of motion describing this problem are:
\begin{equation}
    \label{eq:mass_cons}
    \frac{\partial \rho}{\partial t} 
    + \grad \cdot \left( \rho \vel \right) = \mdotw K(|\mathbf{x}|) \, ,
\end{equation}
\begin{multline}
    \label{eq:momentum_cons}
    \frac{\partial \left(\rho \vel\right)}{\partial t}
    + \grad \cdot \left[ \left(P + \frac{B^2}{8\pi} \right)\mathbb{I}
    + \rho \vel \otimes \vel  - \frac{\magfield \otimes \magfield}{4\pi} \right] \\ = \dot{p}_w(|\mathbf{x}|)\rhat K(|\mathbf{x}|) \, ,
\end{multline}
\begin{multline}
    \label{eq:energy_cons}
    \frac{\partial E}{\partial t}
    + \grad \cdot \left[ \vel \left(E + P + \frac{B^2}{8\pi} \right)  - \frac{\magfield\left(\magfield \cdot \vel\right)}{4\pi}\right] = \\(\mathcal{G} - \mathcal{L}) + \Lwind K(|\mathbf{x}|) \, .
\end{multline}
Equations \ref{eq:mass_cons}, \ref{eq:momentum_cons}, and \ref{eq:energy_cons} are the equations for the conservation of mass, momentum, and energy respectively. Here, $\rho$ is the mass-density of the fluid, $\vel$ is the fluid velocity, $P$ is its thermal pressure, $K(|\mathbf{x}|)$ is some spherically symmetric kernel function with compact support\footnote{$K(|\mathbf{x}|) \neq 0$ only in some small region and $\int K(|\mathbf{x}|) d\mathbf{x} = 1$.} centered on the origin,  $\magfield$ is the magnetic field in units with a magnetic permeability of unity, $\mathbb{I}$ is the identity matrix, $\otimes$ denotes the tensor product, and $E$ is the total energy density of the fluid given by
\begin{equation}
    \label{eq:energy_def}
    E = e + \frac{1}{2}\rho v^2 + \frac{B^2}{8\pi} \, ,
\end{equation}
where $e$ is the internal energy density, which we will assume to be given by an ideal gas law $e = P/(\gamma -1)$; we assume $\gamma = 5/3$. The source terms $\mdotw K(\mathbf{x})$, $\dot{\mathbf{p}}_w(\mathbf{x}) K(\mathbf{x})$, and $\Lwind K(\mathbf{x})$ in \autoref{eq:mass_cons}, \autoref{eq:momentum_cons}, and \autoref{eq:energy_cons}, respectively, denote the input of mass, momentum, and energy from the central source, which we assume to be constant in time. $\mdotw$ and $\Lwind$ together define the wind velocity
\begin{equation}
    \label{eq:vwind_def}
    \Vwind = \sqrt{\frac{2 \Lwind}{\mdotw}} \, .
\end{equation}
and wind momentum input rate
\begin{equation}
    \pdotw = \mdotw \Vwind=2 \Lwind/\Vwind = \sqrt{2\Lwind\mdotw}
\end{equation}
We assume that these source terms are ``turned on'' at some time $t=0$, before which they are zero and after which they are constant.

Finally, $\mathcal{G}$ and $\mathcal{L}$ in \autoref{eq:energy_cons} denote thermal heating and cooling of the gas, respectively, which depend on the local (global) distribution of the gas temperature and density for optically thin (thick) radiative cooling and heating. Generally we will assume optically-thin cooling and heating \citep{JGK_NCR23}. The above equations are supplemented and made complete by the ideal induction equation and divergence-free constraint on the magnetic field.

Under the action of the above source terms a hot `bubble' naturally develops around the source resulting from the injected mass and energy; this is commonly referred to as a Wind-Blown Bubble (WBB). At the edge of the bubble will be an interface with cooler gas.  For stellar wind bubbles (and potentially other applications as well), the bubble will be surrounded by gas which is photo-ionized. Depending on the regime in which feedback is taking place this photo-ionized region may or may not be ``trapped" by the shell of the WBB \citep{Peters10,GeendeKoter22,Geen23,Geen24}, in which case the WBB shell would consist of both photo-ionized `warm' gas and cool, neutral gas. In the present work, we shall not explicitly include the effects of photo-ionized gas, but the analysis here may be generalized to allow for it.

Depending on the exact values of $\Lwind$, $\mdotw$, and properties of the stresses (pressure, density, velocity, magnetic field) in the surrounding medium, different regimes explaining the expansion of this bubble will be relevant. Here we focus on the intermediate stage of evolution, which applies to most high-energy stellar wind sources.  This stage has previously been referred to as the `pressure-driven snowplow' (PDS) stage \citep{Weaver77}.   However, as argued in \citet{Lancaster21a,Lancaster21b} \citep[see also][]{ElBadry19}, in fact the dynamics differ from the original PDS conception due to cooling at the interface between the hot gas and the shell.  In contrast, \citet{Weaver77}  assumed that the only energy loss from the bubble's hot interior is due to $PdV$ work done in expanding the shell.

What we specifically mean by the intermediate stage is that we assume the following:
\begin{itemize}
    \item[1.] We have left the `free-expansion' stage where the bubble expands ballistically at $\Vwind$. This occurs very early for all parameter values of interest.

    \item[2.] We are dealing with a `fast-wind' in the language of \citet{KooMcKee92a}, where cooling in the wind bubble interior is negligible.

    \item[3.] The exterior stresses outside the bubble and shell are negligible compared to the pressure in the bubble interior. This implies that the bubble expands super-sonically (and super-Alfv\'{e}nically) into the background medium.

    \item[4.] The outer shell of the WBB, mostly consisting of gas that was initially shock-heated when swept up by the bubble, has had enough time to cool back to ambient temperature.
\end{itemize}

\begin{figure*}
    \centering
    \includegraphics[width=0.8\textwidth]{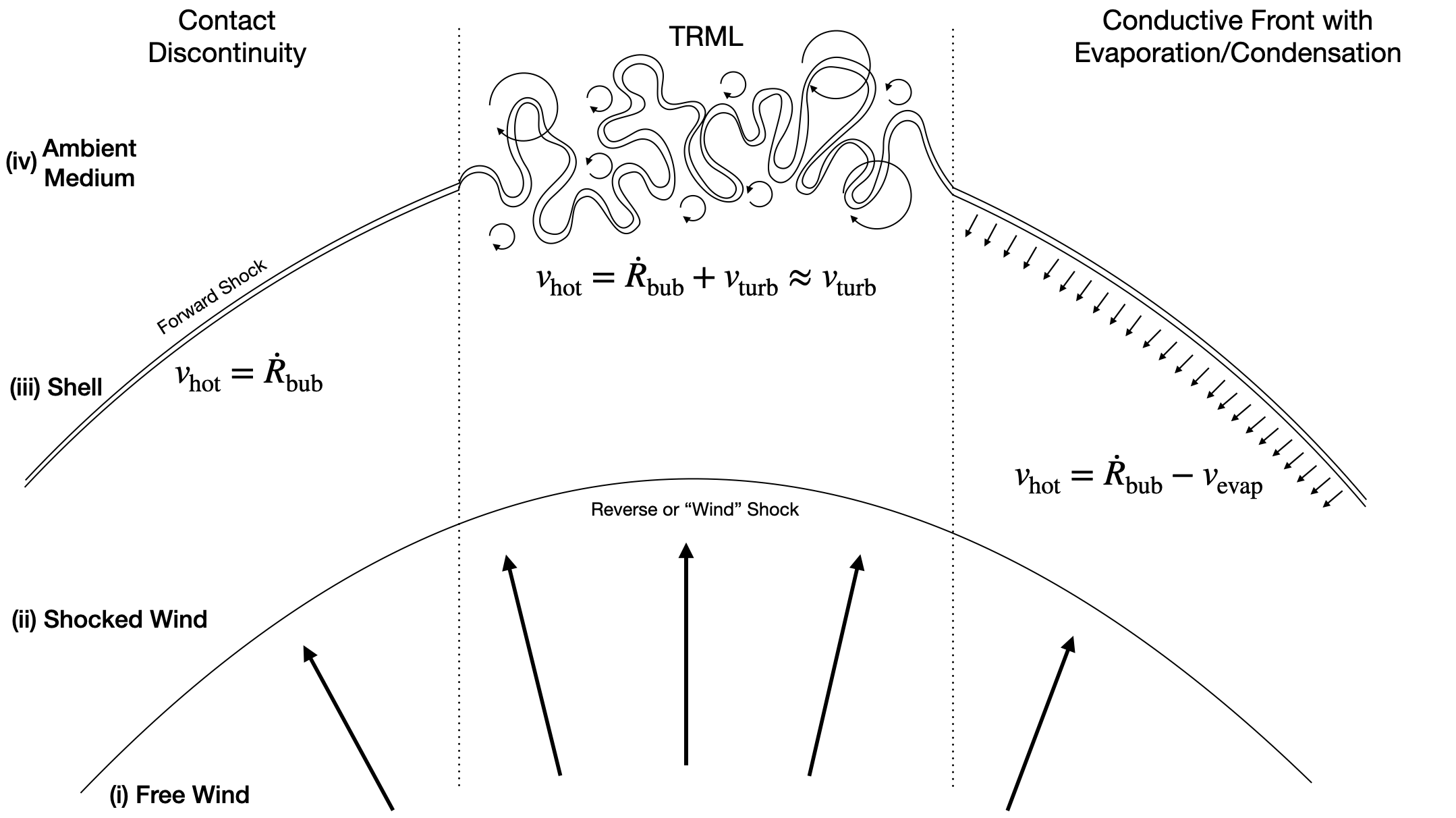}
    \caption{A schematic representation of the interface of a wind-blown bubble with the shell consisting of the ambient material it has swept up, in the thin shell limit. Three aspects of this interface are emphasized.  \textit{Left}: A contact discontinuity between the ambient material and the wind material, with no movement of mass or energy across this surface. The hot gas velocity then must match the velocity of the surface $v_{\rm hot} = \dot{R}_{\rm bub}$. \textit{Right}: The inclusion of conductive heat transport across this interface causes mass-loading of the bubble interior. The flow is usually strongly subsonic resulting in the hot gas velocity still matching the surface velocity. \textit{Middle}: Turbulence in the hot gas, seeded by hydrodynamical instabilities and inhomogeneities in the background, modifies the geometry of the bubble's surface and causes mixing between the hot wind and the surrounding cold gas, changing the nature of energy transport. This turbulent mixing and subsequent cooling can be drastic enough to draw in gas much faster than the bubble itself is expanding, acting as a new boundary condition on the velocity of the hot gas. The relative volume between the free and shocked wind (the placement of the reverse shock) can change depending on the boundary conditions.}
    \label{fig:schematic}
\end{figure*}

With these assumptions the bubble naturally has 4 distinct regions, as described in \citet{Weaver77}: (i) the Free Wind, (ii) the Shocked Wind, (iii) the shell of the WBB, and (iv) the background medium. The presence of the shocked wind arises from our first assumption above: the bubble is expanding slower than $\Vwind$ so the wind must shock. As we will discuss below, the exact relative volume of (ii) as opposed to (i) will be determined by the properties of the flow in (ii) and boundary conditions. The structure of such a bubble, discussed further below, is laid out schematically in \autoref{fig:schematic}.

We will define the volume of the free and shocked winds as $\Vbub$, and define its volume-equivalent radius as
\begin{equation}
    \label{eq:Rbub_def}
    \Rbub \equiv \left( \frac{3\Vbub}{4\pi}\right)^{1/3} \, .
\end{equation}
By equating the $d\Vbub/dt$ derived from a time derivative of the above with the $d\Vbub/dt$ derived from considering the motion of the bubble's surface it is straightforward to show that
\begin{equation}
    \label{eq:dr_dt}
    \frac{d \Rbub}{dt} = 
    \Wbubavg
    \frac{\Abub}{4\pi \Rbub^2} \, .
\end{equation}
where $\Wbubavg = (d\Vbub/dt)/\Abub$ is the expansion velocity of the bubble boundary averaged over the whole surface (this average is denoted by the brackets $\left\langle\cdot \right\rangle$) and $\Abub$ is the area of the surface of the WBB. This states that the rate of change of the bubble's linear dimension is its average outward velocity at its surface `modulated' by the extent to which the bubble is aspherical, denoted here by the ratio of the bubble's surface area to its surface area if it were a perfect sphere. It was shown in \citet{Lancaster21b} that, for fractal bubbles, $4\pi \Rbub^2/\Abub \approx \foldedness$, the average value of the dot product of the radial and normal vectors over the surface. This would imply that $\frac{d\Rbub}{dt} \approx \Wbubavg/\foldedness$, the factor in the denominator corrects for the reduced surface velocity when $\Abub \gg 4\pi \Rbub^2$.

With the assumptions stated above we have now restricted ourselves to the case of a bubble with an adiabatic interior (a ``fast wind") and a radiative exterior (a shell that has cooled). We will see in the next section that the evolution of such a bubble depends crucially on the physics at the interface between these two regions.

\subsection{Bubble Structure Highlights}
\label{subsec:structure_highlights}

In \autoref{app:structure} we expand upon the Appendix of \citet{Lancaster21a} to derive the fluid structure in the wind bubble's interior without assuming a purely radial flow. In particular we assume only that the structure of the shocked wind is determined by the properties of a steady, adiabatic, isobaric flow and discuss the justification for these assumptions and the time-scales over which they apply.

In the context of a general, aspherical wind bubble and taking into account the `dilation' of streamlines through a deformed, messy, shocked wind region, we show that the only relevant stress communicated from the wind bubble interior to the shell is the thermal pressure force, locally $\Phot d\Abub$, where $\Phot$ is the (isobaric) pressure of the shocked wind and $d\Abub$ is an area element of the bubble surface.  Locally, this force is in the direction normal to the bubble (or shell) surface, $\nhat$. We show that the radial pressure force exerted by the bubble is 
\begin{eqnarray}
    \label{eq:bubble_radial_force}
    F_r &\approx&  4\pi \Rbub^2 \Phot \\
    &=& \frac{3}{4}\pdotw\left( \frac{\Rbub}{\Rfree}\right)^2 \, 
    \label{eq:bubble_force2}
\end{eqnarray}
(see \autoref{eq:Fr_eff} and \autoref{eq:app_pr3}, respectively). The second expression shows that a larger force corresponds to a larger  radius of the bubble, $\Rbub$, relative to that of the free wind shock, $\Rfree$.  As we discuss in \autoref{app:structure}, this equation is missing a term due to the Reynolds stress which is only relevant in the extreme cooling limit where $F_r = \pdotw$ and $\Rbub/\Rfree \approx 1$. We note that the derivation  fully accounts for the `foldedness' of the wind bubble \citep[see Figure 16 of][and also \autoref{eq:fold_scaling} here]{Lancaster21b}.  Even though the bubble surface may be highly folded ($\foldedness\ll 1$ and $4 \pi \Rbub^2 \ll \Abub$),  the net radial force in \autoref{eq:bubble_radial_force} may still be written only in terms of the volume equivalent bubble radius $\Rbub$. 

The interior pressure, as we see from \autoref{eq:bubble_radial_force},  determines the bulk dynamics of the bubble. As we discuss further in \autoref{app:structure}, equating the enthalpy flux through the interior shock, $15\Lwind/16$, with the total enthalpy flux through the bubble's surface, $5 \Phot\Abub\voutavg/2$,  allows us to write (in the context of a steady bubble)
\begin{equation}
    \label{eq:Phot_steady}
    \Phot = \frac{3}{8} \frac{\Lwind}{\voutavg \Abub} \, ,
\end{equation}
where
\begin{equation}
    \label{eq:voutavg_def}
    \voutavg \equiv \Abub^{-1} \int_{\rm \Abub} \left(\mathbf{v} - \mathbf{W}\right) \cdot \nhat dA \, . 
\end{equation}
Here, $\mathbf{v}$ is the fluid velocity\footnote{When $\Rdotbub \ll |\mathbf{v}|$ near the surface, the bubble's exterior surface is moving very slowly compared to the velocity at which gas is being advected into the surface and we may consider $\mathbf{v}- \mathbf{W} \approx \mathbf{v}$ in \autoref{eq:voutavg_def}}, and it is dotted into the normal to the surface $\nhat$. The above assumes a clear definition of the bubble's surface, which requires additional care in the case that the surface has a fractal structure. In all physical cases, this fractal structure will be regulated or smoothed-out at some small scale determined by dissipative physics. However, as we discuss in \autoref{app:frac}, one can give a meaningful definition to both $\Abub$ and $\voutavg$ measured on scales larger than this dissipative scale, provided that both are measured using the same `ruler.'\footnote{We shall analyze measurement scale dependencies of $\Abub$ from our simulations in \autoref{subsec:Abub_scaling}. Our measurements of $\voutavg$ will be at the smallest scale available in the simulation, $\Delta x$.} In \autoref{app:structure} we discuss how, if \autoref{eq:Phot_steady} is not satisfied at a given time, the energy dynamics of the bubble interior will act to restore it by losing or gaining energy, causing a change in pressure.

Using \autoref{eq:bubble_radial_force} and \autoref{eq:Phot_steady} we can write a momentum equation for the evolution of the shell surrounding the bubble as
\begin{equation}
    \label{eq:pdot_main}
    \dot{p}_r = \frac{3\Lwind}{8} \frac{4\pi\Rbub^2}{\voutavg\Abub}\, .
\end{equation}
The momentum equation of \citet{Lancaster21a} for the shell surrounding a bubble (Equation 18 of that paper with $\alpha_R = 1$) is 
\begin{equation}
    \label{eq:evolution}
    \frac{dp_r}{dt}=\frac{d}{dt} \left(\frac{4\pi}{3} \bar{\rho} \Rbub^3 \frac{d\Rbub}{dt}  \right) = \alpha_p \pdotw,    
\end{equation}
where $\bar{\rho}$ is the mean density in the environment\footnote{\autoref{eq:evolution} has solution $\Rbub = \left[3\alpha_p \pdotw t^2/(2\pi \bar{\rho})  \right]^{1/2}$, and the limit of maximally efficient cooling (`EC') takes $\alpha_p\rightarrow 1$.}. Comparing these and using $\pdotw=2\Lwind/\Vwind$, we can identify the momentum enhancement factor in terms of the bubble parameters as
\begin{equation}
    \label{eq:alphap_derive}
    \alpha_p = \frac{3}{4} \frac{\Vwind/4}{\voutavg} \frac{4\pi \Rbub^2}{\Abub}\, .
\end{equation}
We can see from the above that whether or not a bubble appears to be ``momentum-driven'' (i.e. $\alpha_p \approx {\rm const.}$) depends solely on how $4\pi\Rbub^2/\Abub$ and $\voutavg/V_w$ scale in time with respect to one another; these factors are driven by geometry\footnote{The $4\pi \Rbub^2/\Abub$ dependence in \autoref{eq:alphap_derive} can be thought of straightforwardly as the result of enthalpy flux being dependent on the total area (increase in $\Abub$ results in further losses) whereas the radial momentum is dependent only on the radially outward component.} and dissipative processes respectively. We will next discuss the physics that govern each of these quantities.

\subsection{The Bubble-Shell Interface}
\label{subsec:contact}

In \autoref{fig:schematic} we have laid out three different ways of thinking about the boundary between the bubble's interior and its shell: a contact discontinuity (ideal fluid dynamics), a turbulent, radiative mixing layer (TRML) which simply emphasizes the importance of hydrodynamical instabilities in modifying the geometry of the interface and enhancing micro-physical dissipation, or a conductive evaporation or condensation front (non-ideal fluid dynamics, with cooling). In reality, the boundary between the wind-blown bubble and its shell may combine all of these features. Ignoring for a moment the diffusion of mass, there is a `contact surface' that separates the injected mass from the surrounding mass. This surface is distorted and amplified in area by dynamical instabilities as well as the larger scale inhomogeneities in the background medium. Across this distorted surface, heat dissipation acts to carry energy from the hot wind to the cold shell. As we can see from \autoref{eq:alphap_derive}, the details of the physics of this interface, through its geometry (given by $4\pi \Rbub^2/\Abub$) and the rate at which energy is diffused across the surface (which determines $\voutavg$) have a profound impact on the dynamics of the bubble.

We will show in \autoref{subsec:Abub_scaling} that the geometry of these interfaces is well described by a fractal model. In the next section we discuss in depth the various diffusive physics that can set $\voutavg$.

\begin{figure*}
    \centering
    \includegraphics[width=\textwidth]{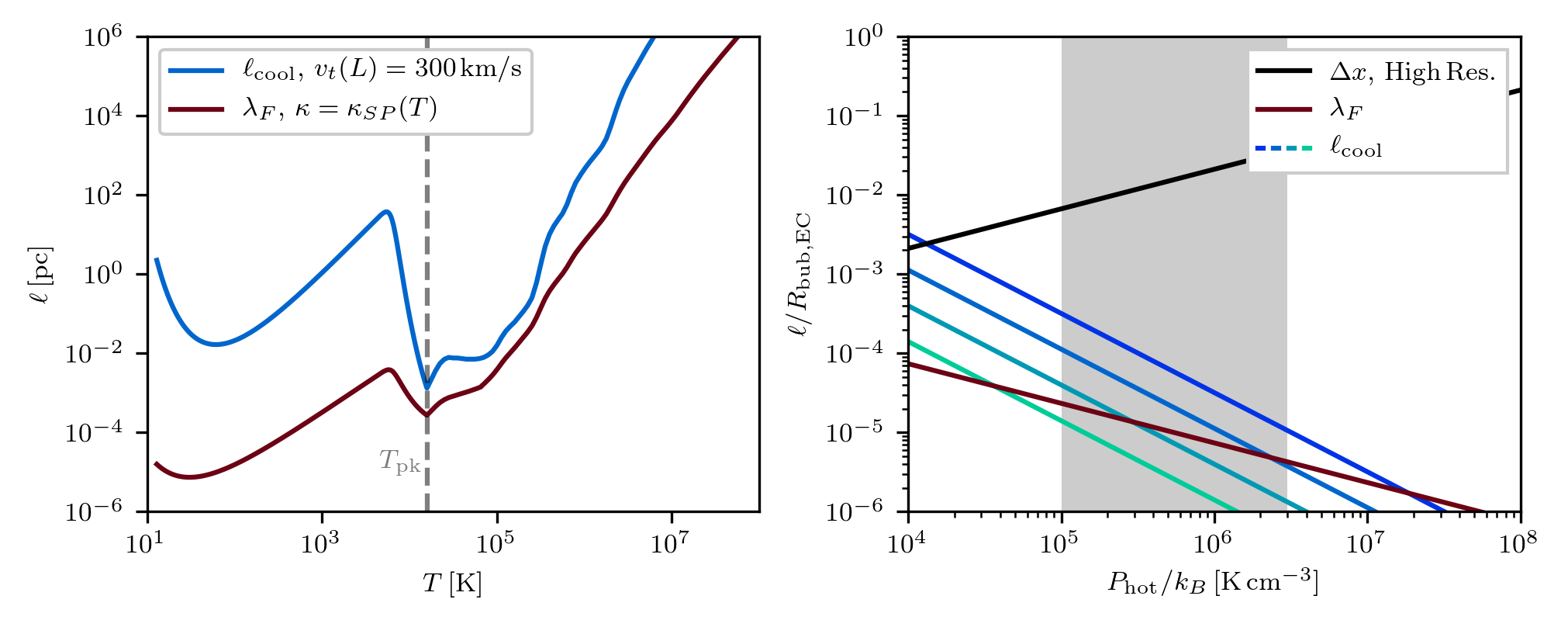}
    \caption{Dissipative scales as a function of temperature (left) at $P/k_B = 10^5\, {\rm K}\, {\rm cm}^{-3}$ and pressure (right) using the values at $\Tpk$. \textit{Left panel}: $\lcool$ from turbulence-mediated mixing  as defined in \autoref{eq:ellcool} with $L = 4\, {\rm pc}$ and $v_t(L) = 300\, {\rm km/s}$; $\lfield$ as defined in \autoref{eq:lambdaF} with temperature dependent $\kappa$ (see \autoref{eq:kappa_SP}). The vertical gray dashed lin indicates the temperature which yields the shortest cooling time $\approx 1.6 \times 10^4\, {\rm K}$. \textit{Right panel}: The diffusive scales at the temperatures responsible for mixing as a function of pressure.  Each length scale is displayed relative to the bubble radius for the `efficiently cooled' solution (inverse of \autoref{eq:Pps} with $\Rfree\to R_{\rm bub, EC}$ and $\pdotw$ corresponding to our fiducial simulations). The turbulent diffusive scale $\lcool$ is shown with $v_t(L = 4\, {\rm pc}) = 75,\, 150,\, 300,\, \& \ 600\, {\rm km/s}$ which correspond to the lighter to darker blue lines, respectively. 
    $\lfield(\Tpk)$ is shown in brown (see \autoref{eq:lambdaF}) and the resolution of our highest resolution simulations is shown in black. The rough range in pressures spanned by our simulations are shown as a grey band.}
    \label{fig:scales}
\end{figure*}

\subsection{Mechanisms and Physical Scales for Dissipation}
\label{subsec:diss_scales}

The setting of $\voutavg$ on the smallest scales has to be regulated by balancing the transport of heat through dissipative processes and the rate at which energy can be cooled away. The latter is straightforwardly given by a characteristic cooling time
\begin{equation}
    \label{eq:tcooldef}
    \tcool(T; P) \equiv 
    \frac{P}{n^2 \Lambda(T)}
    =\frac{(k_B T)^2}{P\Lambda(T)} \, ,
\end{equation}
where $\Lambda(T)$ is the cooling function\footnote{In the simulations presented here, and in reality, $\Lambda$ is strictly speaking not only a function of $T$ but also of, primarily, the electron fraction $x_e$. However, we ignore that dependence here since the phases we care most about are fully ionized.},  $n$ is the number density of the gas, $k_B$ is Boltzmann's constant, and $T$ is the temperature. Generally we will consider isobaric layers (though this does not hold strictly in our simulations) so $\tcool(T;P)$ may be regarded solely as a function of temperature (at given pressure $P$). The minimum cooling time as a function of pressure determines the temperature of gas the cools most rapdily, which we will refer to as $\Tpk$
\begin{equation}
    \label{eq:Tpk_def}
    \Tpk \equiv {{\rm arg}\,{\rm min}}_{T} \tcool(T)\, .
\end{equation}
Since the dependence of $\tcool$ on pressure does not change the temperature dependence, $\Tpk$ is independent of pressure and is equal to $1.59 \times 10^4\, {\rm K}$ for our adopted cooling function (which assumes solar metallicity).

We next discuss two forms of physical dissipation and their relevant spatial and velocity scales. Where appropriate, we discuss how magnetic fields alter these means of dissipation. The dissipative scales defined below are functions of the pressure of the medium, $P$, and the temperature, $T$, which may vary throughout the medium, although the bubble interior and boundary layer are roughly isobaric. In \autoref{fig:scales} we show these scales as functions of both temperature (left panel) and pressure (right panel) for given choices of the parameters of the dissipative physics, defined and explained below.

\subsubsection{Thermal Conduction}
\label{subsubsec:vout_cond_theory}

The first dissipation mechanism is thermal conduction mediated by individual particle collisions over a mean free path \citep{SpitzerHarm53,Spitzer62}, parameterized by the (temperature-dependent) thermal conductivity $\kappa(T)$, which produces a heat flux $-\kappa \nabla T$. In balancing heat transport through conduction against cooling in the energy equation, one can derive a characteristic length scale $\lfield$, which we will refer to as the `Field Length' \footnote{Note that this differs by order unity from the length at which thermal instability is suppressed by thermal conduction \citep{Field65}.}\citep{Field65,ZDPN69,KimKim13,TOG21}:
\begin{equation}
    \label{eq:lambdaF}
    \lfield(T) = \sqrt{\frac{\kappa(T)T}{ n^2\Lambda(T)}} = \sqrt{\frac{\kappa(T)T\tcool(T)}{P}} \, .
\end{equation}
Below we take $\kappa$ as given by either
\begin{equation}
    \label{eq:kappa_SP}
    \frac{\kappa_{SP}(T)}{{\rm erg}/{\rm s}/{\rm cm}/{\rm K}} = 
    \begin{cases} 
      2.5\times 10^3 \left(\frac{T}{\rm K}\right)^{1/2} & T <  T_{\rm eq} \\
      6\times 10^{-7} \left(\frac{T}{\rm K}\right)^{5/2} & T\geq T_{\rm eq}
   \end{cases}
\end{equation}
which is a combination of Spitzer (high temperatures) and Parker \citep[low temperatures, ][]{Parker53} conductivity, where the point of switching between the two $T_{\rm eq} = 6.45\times 10^{4}\,{\rm K}$ is the point at which they are equal. 

In the left panel of \autoref{fig:scales}, we show $\lfield$ as a function of temperature $T$ at $P/k_B = 10^5{\rm K\ cm^{-3}}$. The cooling function used here, $\Lambda(T)$, is that from \citet{CGK_TIGRESS1}. Since $\tcool \propto P^{-1}$, it is clear that $\lfield\propto P^{-1}$ if one considers a fixed temperature. In the right panel of \autoref{fig:scales}, we show $\lfield$ relative to the bubble radius $\Rbub$, selecting $\lfield(\Tpk)$. The bubble radius is calculated as a function of $\Phot$ assuming the efficiently cooled solution (\autoref{eq:Pps} with $\Rfree\to R_{\rm bub, EC}$) with $\pdotw = 4.8\times 10^4 M_{\odot}/{\rm km}/{\rm s}/{\rm Myr}$. Really, we should choose the value at which $\lfield$ is minimized in the range of temperatures probed by the conductive interface, roughly $5000\, {\rm K} < T < 10^8\, {\rm K}$, which is slightly different from $\Tpk$ due to the temperature dependence of $\kappa(T)T$ in the numerator of \autoref{eq:lambdaF}. However, the dependence of $\lfield$ on $T$ is so strongly influenced by $\Lambda(T)$ that this temperature is essentially $\Tpk$, so we will use the latter for simplicity.

Note that, in the presence of a magnetic field, $\magfield$, thermal conduction across field lines is expected to be suppressed, due to suppression of electron transport across field lines. In the context of WBBs expanding into a magnetized medium, the process of `magnetic draping' of the field lines around the bubble  will virtually guarantee that the magnetic field is mostly parallel to the bubble's surface, suppressing conduction. Furthermore, recent plasma physics simulations and experiments have demonstrated that thermal conduction can be strongly suppressed due to plasma instabilities in high-$\beta$, weakly collisional and turbulent plasmas, as may be the case in some regimes of these mixing layers \citep{RobergClark16,Komarov18,Drake21,Meinecke22}. Additionally, as derived in \citet{CowieMcKee77}, there is a maximum heat flux that can be carried by electrons, $q_{\rm max} \sim (3/2)\rho c_{s,{\rm iso}}^3$ (where $c_{s,{\rm iso}}$ is the isothermal sound speed), which limits the flux when the temperature gradient scale length, $T/|\nabla T|$, is comparable to the mean free path of electrons.  All of these effects can decrease $\kappa$ by an order of magnitude, making $\lfield$ a factor of 3 or more smaller.

We can derive a rough estimate of the value of the velocity that would be induced by resolved thermal conduction by assuming that all cooling occurs in a layer of thickness $\lfield (\Tpk)$. We assume the cooling proceeds at a rate per unit area of $\lfield n^2\Ltpk$ and balance this against advection of enthalpy from the hot gas into the layer at a rate $5P\vcond/2$. This gives us
\begin{equation}
    \label{eq:vout_conduction}
    \vcond \approx \frac{2}{5} \sqrt{\frac{\kappa(\Tpk)\Ltpk}{k_B^2 \Tpk}} \approx \frac{\lfield}{\tcool} \, ,
\end{equation}
which is independent of pressure.

We should note that there is another scale associated with the action of thermal conduction, which comes from the balancing of the advection term with the conductive heat transport term in a one-dimensional energy equation of a conductive cooling layer \citep[e.g.][]{TOG21,TanOh21}. This `evaporation length' scale is given by
\begin{equation}
    \lambda_{\rm evap} = \frac{2\kappa T}{5 P v}
\end{equation}
where $v$ is the gas velocity. Though the value of this length scale depends on the details of the evaporative mass loss \citep[which is subject to the energy lost in the cooling layer -- see section 5.3 of][]{ElBadry19}, this scale is generally much larger than $\lfield$. It is therefore generally possible to resolve an evaporative mass flow from the surface even if the details of cooling in the surface are unresolved in a given simulation.

\subsubsection{Turbulent Dissipation}
\label{subsubsec:vout_turb_theory}

The second dissipation mechanism is turbulent diffusivity, which has been explored extensively in recent literature \citep{FieldingFractal20,TOG21,Lancaster21a}. In particular, we imagine a turbulent cascade in the hot gas with energy containing scale $L$ and velocity normalization $v_t(L)$, following a spectrum 
\begin{equation}
    \label{eq:vt_structure}
    v_t(\ell) = v_t(L) \left(\frac{\ell}{L}\right)^{p}\, ,
\end{equation}
with $p = 1/3$ for subsonic, Kolmogorov turbulence. We define the eddy turn-over time at scale $\ell$ as
\begin{equation}
    \label{eq:teddy}
    \teddy(\ell) \equiv \frac{\ell}{v_t(\ell)} \, .
\end{equation}

The mixing that leads to cooling is expected to take place on a scale $\lcool$ such that $\teddy(\lcool) = \tcool$, where \autoref{eq:tcooldef} gives $\tcool$. Combining \autoref{eq:vt_structure} and \autoref{eq:teddy} with this definition we obtain the cooling length 
\begin{equation}
    \label{eq:ellcool}
    \lcool \equiv 
    L \left( \frac{\tcool}{\teddy(L)}\right)^{\frac{1}{1-p}}\, .
\end{equation}

In this picture, eddies on scales larger than $\lcool$ mix too slowly to contribute significantly to energy losses. At scales $\lesssim\lcool$, the surface area of the interface is rapidly enhanced on a time-scale shorter than $\tcool$ (since $\teddy(\ell) < \tcool$ for all $\ell < \lcool$). In particular, this enhancement in area is so rapid that it does not matter what the micro-physical means of energy transport across the surface is; the area will always become large enough to carry the heat being mixed by the turbulence. In the language of the combustion literature \citep{TOG21}, we can think of mixing on the scale $\lcool$ as occurring in the `well-stirred reactor' limit, at which point an effective turbulent diffusivity may be applied. Either way, this results in the $\lcool$-scale eddy being `mixed' to the intermediate temperatures where cooling is rapid.

The velocity scale associated with turbulent dissipation is built in:
\begin{equation}
    \label{eq:vout_turb}
    \vturb \approx v_t(\lcool)  = v_t(L) \left( \frac{\tcool}{\teddy(L)}\right)^{\frac{p}{1-p}}\, .
\end{equation}
We can see that this is very similar to \autoref{eq:vout_conduction} if we remember that $v_t(\lcool) = \lcool/\tcool$.

In the presence of a magnetic field, turbulent heat diffusion is also altered due to the modification of instabilities which drive the turbulence \citep{DasGronke23,ZhaoBai23}. Roughly speaking, instabilities will be suppressed if the velocity scale associated with the instability (by which we mean the length scale associated with the instability, divided by its linear theory growth time) is less than the Alfv\`{e}n speed in the medium \citep{ChandrasekharBook}. This likely has drastic consequences for the WBB as, due to compression of the magnetic field, the shell of the WBB will be strongly magnetically dominated. This is likely the cause of the large differences in fractal structure (\autoref{subsec:Abub_scaling}) and turbulent properties (\autoref{subsec:vout_turb}) between our HD and MHD simulations.

In this section and the previous one we have defined two mechanisms of physical heat dissipation and the associated length scales relevant to resolving them. These length scales are functions of temperature, as shown in the left panel of \autoref{fig:scales}. Considering one dissipation mechanism alone, it is relevant to resolve the minimum such length scale in the relevant temperature range for the problem. This is essentially determined by where the cooling rate is maximized, due to the strong temperature dependence of $\Lambda(T)$. However, considering a realistic case where both dissipation mechanisms are acting, it is only important to resolve the larger of the two length scales, as the associated dissipation mechanism will dominate the heat transport. For example, in the right panel of \autoref{fig:scales} we can see that, for $v_t(L) = 300\, {\rm km/s}$, the turbulent dissipation scale, $\lcool$, is always greater than length scale associated with thermal conduction, $\lfield$. In this case, it is only necessary to resolve $\lcool$, which is can be orders of magnitude larger than $\lfield$ at lower pressures (later in the bubble evolution). If instead we were simulating a bubble where $\lcool < \lfield$, as is the case for the $v_t(L) = 75\, {\rm km/s}$ case in \autoref{fig:scales}, it would then be necessary to resolve $\lfield$, though we will see that this is often not the case.

\subsubsection{Numerical Dissipation}
\label{subsubsec:numerical_dissipation}

When the appropriate dissipative scales are resolved (the maximum of $\lcool$ or $\lfield$ for a given condition of the gas) physical dissipation controls the pace at which energy can be transported across the interface and therefore sets $\voutavg$. As we can see from comparing $\Delta x \approx 0.1\, {\rm pc}$ at our highest resolution to the relevant dissipative scales in the right panel of \autoref{fig:scales}, we are only just near resolving the turbulent dissipation at the latest times in our simulations for relevant choices of the turbulent dissipative parameters (see \autoref{subsec:vout_turb}). The question then arises: how is $\voutavg$ set when the relevant physical dissipative scales are not properly resolved?

In keeping with the previous two subsections, let us first suppose that $\voutavg$ is set by dissipation, this time of a numerical nature. In numerical dissipation the relevant dissipative velocity scale associated with dissipation across contact discontinuities (as is relevant here) is roughly the local gas velocity relative to the grid. For example, consider the advection equation in one-dimension, $\partial_t q + a \partial_x q =0$ with $q$ some scalar field and $a$ the advection speed. If we solve this equation on a fixed grid with spatial discretization $\delx$, using a first-order upwind scheme one can analytically show that the actual equation being solved (to second order accuracy) is $\partial_t q + a \partial_x q =C_{\rm num} \partial_x^2 q$ with numerical diffusivity $C_{\rm num} = \delx a(1 -|c|)/2$ with $c$ the Courant number (see e.g. Chapter 5 of \citet{ToroRiemann} or Chapter 20 of \citet{NumericalRecipes}). From $C_{\rm num}$ we can read off the relevant spatial and velocity scale for the dissipation: $\delx$ and $a$, the resolution and the advection velocity. While it is much more difficult to derive an exact numerical diffusivity for a higher-order method, let alone a system as complicated as the MHD equations, the same basic principle still holds. That is, for diffusion across contact discontinuities, the numerical diffusive velocity scale, $v_{\rm num}$, is comparable to the local velocity of the gas relative to the grid:
\begin{equation}
    \label{eq:vnum_def}
    v_{\rm num} \approx v_{\rm gas}\, ,
\end{equation}
in analogy to \autoref{eq:vout_conduction} and \autoref{eq:vout_turb}.

In the previous sections we posited that the local gas velocity is set by the relevant diffusive velocity at the interface $v_{\rm gas} \approx v_{\rm diff}$ with $v_{\rm diff} = \vcond,\, \vturb$. However, here, this provides no constraint, as the relevant diffusive velocity is the local gas velocity. We will see in the next section that a consistency condition on cooling at the wind bubble's interface will lead to a  meaningful estimate of the numerical diffusive velocity as
\begin{equation}
    \label{eq:vnum_def2}
    v_{\rm num} \approx \frac{\delx}{\tcool}\, .
\end{equation}

\subsection{Relation Between $\voutavg$ and Cooling}
\label{subsec:vout_cooling}

As we discuss in \autoref{app:structure}, the assumption of an entirely `steady' flow in the bubble interior necessitates that most of the mass and energy in the bubble is advected out of it. As discussed in \autoref{subsec:diss_scales}, this transfer can be accomplished by the diffusive flux associated with a turbulent boundary layer in the physical case; numerical diffusion can alternatively play this role if the turbulence is not sufficiently resolved.  Whether physically or numerically driven, mixing of the energy from the interior with mass from the shell can produce temperatures where cooling is very strong, so the energy mixed into the shell can be largely radiated away. This loss of energy is essentially a consistency condition on the assumption of a `steady' interior: if energy were not lost, there would be no movement of energy out of the hot gas bubble, as is required by the steady flow assumption.

If we assume for simplicity that this cooling is dominated by gas at $\Tpk$, given in \autoref{eq:Tpk_def}, and that the volume of gas at this temperature is $\Vpk$ then the cooling rate of this gas is
\begin{equation}
    \label{eq:edot_cool}
    \edotpk = \left(\frac{\Ptpk}{k_B \Tpk} \right)^2 \Ltpk \Vpk 
    = \frac{\Ptpk \Vpk}{\tcool(\Tpk)}
\end{equation}
where $\Lambda(T)$ is the cooling function (we have ignored heating here), $\Ptpk$ is the pressure of the gas in $\Vpk$, and $\tcool$ is the cooling time (\autoref{eq:tcooldef}). If we assume that the cooling layer is in pressure equilibrium with the hot gas, as seems reasonable from resolved simulations \citep{ElBadry19,FieldingFractal20}, then we have $\Ptpk = \Phot$.

We can define the thickness of the cooling region as
\begin{equation}
    \label{eq:wcool_defe}
    \wcool \equiv \frac{\Vpk}{\Abub} \, .
\end{equation}
Then, with  $\edotpk \approx \Lwind$ from the assumption of strong cooling, and using \autoref{eq:Phot_steady} and \autoref{eq:edot_cool}, we have
\begin{equation}
    \label{eq:vout_cool}
    \voutavg \approx \frac{3}{8} \frac{\sqrt{\Ltpk\Vpk\Lwind}}{\Abub k_B\Tpk} \propto \left(\frac{\wcool}{\Abub}\right)^{1/2} \, .
\end{equation}
The last proportionality has ignored factors having to do with the exact properties of the cooling (as well as the source luminosity), assuming them to be roughly constant.

It is worth discussing what the above relation is telling us about both the bubble and boundary layer dynamics and thermodynamics. Suppose that, at a given pressure, $\voutavg$ were too high for \autoref{eq:vout_cool} to hold. This incongruence can be seen as a failure from either the assumption of cooling matching the energy flux being provided ($\edotpk \approx \Lwind$), or in the context of a steady state bubble (\autoref{eq:Phot_steady}). From the perspective of the bubble, $\voutavg$ being too large means that the bubble is losing energy at its outer surface faster than it gains energy through the wind shock, which  will lead to pressure loss. If we take $\voutavg$ as given, steady-state can be re-established when de-pressurization reaches the point at which \autoref{eq:Phot_steady} is again satisfied (which would also involve an increase in $\Rfree$ from \autoref{eq:Pps}). In our picture, this decrease in pressure also affects the rate of cooling in the interface layer, with the new lower pressure driving down cooling from \autoref{eq:edot_cool}. From the perspective of the cooling layer, $\voutavg$ being too large means that it is being provided energy faster than it can cool it away.
If we enforce, in this thought experiment, fixed $\voutavg$ and isobaric conditions, the build up of energy resulting from insufficient cooling would cause an expansion of the cooling layer (an increase in $\wcool$). This would continue until the cooling layer was once again in equilibrium according to \autoref{eq:vout_cool}.


We emphasize that in the resolved case, \autoref{eq:vout_cool} is not a \textit{prediction} of the value of $\voutavg$ but rather a \textit{consistency condition} on the relationship between $\voutavg$, the properties of the cooling, and the geometry of the cooling layer ($\wcool$ \& $\Abub$). In fact, if we replace $\Ptpk$ in \autoref{eq:edot_cool} with \autoref{eq:Phot_steady} and assume $\edotpk = \Lwind$ as above we have
\begin{equation}
    \label{eq:vout_consistency_cond}
    \voutavg = \frac{3}{8} \frac{\wcool}{\tcool} \sim \frac{\wcool}{\tcool}\, .
\end{equation}
Comparing this with \autoref{eq:vout_conduction} and \autoref{eq:vout_turb} we see that, with the appropriate choices of $\wcool = \lfield$ and $\lcool$ respectively, this condition approximately holds.

In the unresolved case, \autoref{eq:vout_cool} must still hold, and therefore acts as a condition on $\voutavg$ that effectively contributes to setting its value if $\wcool$ is numerically rather than physically controlled. In fact, if we assume $\wcool \approx \delx$ in \autoref{eq:vout_consistency_cond}, we get a similar relation to the resolved cases: $\voutavg \sim \delx/\tcool$, as foreshadowed by \autoref{eq:vnum_def2}.

In the present simulations, the enhanced cooling in the shell can also lead to a pressure dip if $\wcool$ is not fully resolved. We will return to the measured scaling behavior of $\voutavg$ in the context of our simulations in \autoref{subsec:voutavg_explained}.

\begin{figure*}
    \includegraphics{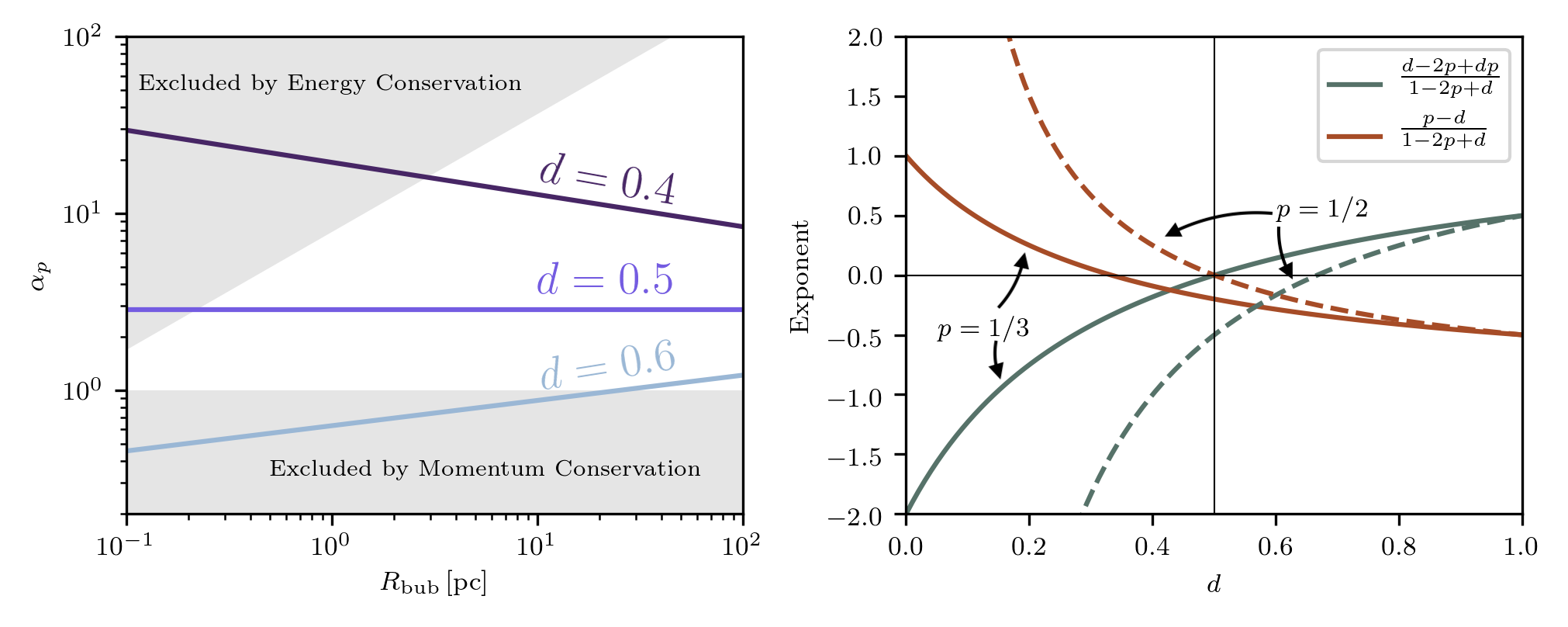}
    \caption{Scaling relations relevant to the theoretical value for $\alpha_p$ given in \autoref{eq:alphap_theory}. \textit{Left}: Evolution of $\alpha_p$ with $\Rbub$ for $v_t(L) = 10 \, \kms$ and three values of the excess fractal dimension, $d$ (other relevant parameters given in the text). Areas that are dis-allowed by energy and momentum conservation are marked as shaded regions; the upper limit is specific to the parameter choices made for this plot. \textit{Right}: Exponents that appear in \autoref{eq:alphap_theory} as a function of $d$ for $p=1/3$ and $p=1/2$ in solid and dashed lines respectively.}
    \label{fig:alphap_theory}
\end{figure*}

\subsection{Momentum Evolution with Turbulent Dissipation}
\label{subsec:momentum_resolved}

We now wish to give a model for how the bubble should evolve in the situation that dissipative physics is controlled by turbulent mixing at small scales, as in \autoref{subsubsec:vout_turb_theory}, rather than numerics. We specify here to the case of turbulent mixing since, as demonstrated by the right panel of \autoref{fig:scales}, we expect turbulent heat transport to dominate over conduction in most real WBBs. To this end, we can assume $\voutavg$ is equal to a resolved, turbulent dissipation velocity $v_t(\lcool)$ (see \autoref{eq:vout_turb}). We can also replace $\Abub$ with a fractal model for its evolution, given by
\begin{equation}
    \label{eq:Abub_frac}
    \Abub(\ell) = 4\pi \Rbub^2 \left(\frac{\Rbub}{\ell}\right)^d \, ,
\end{equation}
where $d$ is the so-called `excess fractal dimension.' This is discussed in depth in \citet{FieldingFractal20,Lancaster21a} and we return to it in \autoref{subsec:Abub_scaling}.

Substituting for $\voutavg$ and using $\Abub(\lcool)$ in \autoref{eq:Phot_steady}, we arrive at
\begin{equation}
    \Phot = \frac{3\Lwind}{32\pi \Rbub^2 v_t(L)} \left(\frac{\Rbub}{L} \right)^{-d} \left(\frac{\tcool}{\teddy(L)} \right)^{\frac{d-p}{1-p}} \, .
\end{equation}
Since $\tcool$ depends on $\Phot$ (assuming the cooling layer is isobaric; we will return to this below), we can rearrange the above to solve for the dependence of $\Phot$ on the other parameters. Doing this, and setting $\alpha_p = 4\pi \Rbub^2 \Phot/\pdotw$ we can derive
\begin{multline}
    \label{eq:alphap_theory}
    \alpha_p = \frac{3\Vwind}{16 v_t(L)} \left[ \frac{3\Lwind \Ltpk}{32\pi L v_t^2(L)(k_B\Tpk)^2}\right]^{\frac{p-d}{1-2p+d}}\\ \times \left(\frac{\Rbub}{L}\right)^{\frac{d-2p+dp}{1-2p+d}} \, .
\end{multline}
It should be noted that the above relation puts no restrictions on physically allowable values of $\alpha_p$ or the pressure. In reality, $\alpha_p \geq 1$ and has an upper limit imposed by energy conservation from the \citet{Weaver77} solution.

If we choose $p=1/3$ for Kolmogorov turbulence and $d=1/2$ as seems reasonable in our hydrodynamical simulations (see \autoref{subsec:Abub_scaling}), we have
\begin{equation}
    \label{eq:alphap_specific}
    \alpha_p = \frac{3}{16}
    \frac{\Vwind}{v_t(L)} 
    \left[ \frac{3\Lwind \Ltpk}{32\pi L v_t^2(L)(k_B\Tpk)^2}\right]^{-1/5} \, ;
\end{equation}
i.e. $\alpha_p$ is independent of the bubble size $\Rbub$, and also (based on the $-1/5$ power) insensitive to the details of the cooling function. As we can see in the right panel of \autoref{fig:alphap_theory}, the physically motivated choice of $p$ and $d$ that results in a $\alpha_p$ which is constant in time is somewhat serendipitous: large changes from $d=1/2$ result in large changes in the exponent of $\alpha_p$ with $\Rbub$. Fortunately, for reasonable values for the HD case, near $d=1/2$, this exponent does not change drastically. 

We can re-write \autoref{eq:alphap_specific} in a more physically motivated way as
\begin{equation}
    \alpha_p = \frac{3}{16}
    \frac{\Vwind}{v_t(L)} 
    \left[\frac{32\pi}{3} \frac{\Phot L^3 (\tcool/\teddy(L))}{\Lwind \teddy(L)} \right]^{1/5} \, .
\end{equation}
Consider the last fraction in the square bracket term. The numerator is a measure of the amount of thermal energy that is able to be stored in the shocked gas: $\Phot L^3$ is the thermal energy in a turbulent eddy and $\tcool/\teddy(L)$ is a measure of the fraction of energy that can be retained before being lost to dissipation and cooling. The denominator is a measure of how much energy is being provided to the bubble. If the energy that the bubble can retain (numerator) is large compared to the rate at which energy is provided to the shocked wind (denominator), then the momentum enhancement is higher.

If we take quantities relevant to the high-resolution HD simulations that we present below ($L = 4\, {\rm pc}$, $v_t(L) = 600\, {\rm km/s}$, $\Tpk = 1.59\times 10^4\, {\rm K}$, $\Ltpk = 2\times 10^{-22}\, {\rm erg/s}\, {\rm cm}^3$, $\Lwind = 4.875\times 10^{37}\, {\rm erg/s}$, $\Vwind = 3230\, {\rm km/s}$, $\overline{n}_{\rm H} = 100\, {\rm cm}^{-3}$) the quantity in square brackets above is roughly $1/5$ and the prefactor in front is roughly unity, implying $\alpha_p \approx 0.2$ and constant in time. The fact that $\alpha_p < 1$ is simply because we have not explicitly enforced momentum conservation in the derivation given above. Physically, however, $\alpha_p \geq 1$ and is bounded from above by the corresponding energy-driven solution from \citet{Weaver77}. These bounds are shown as grey regions in the left-panel of \autoref{fig:alphap_theory} for the $\Lwind$ and $\Vwind$ as quoted above and an assumed mean background density that is identical to our simulations.

The fact that the derived $\alpha_p < 1$ for the parameters quoted above simply means that turbulent mixing over a fractal surface is likely more than capable of dispersing enough energy for the bubble's evolution to approach the momentum-conserving solution ($\alpha_p=1$) as posited in \citet{Lancaster21a}. Motivated by the fact that turbulence is likely reduced in the, more realistic, MHD case we instead plot the $\alpha_p$ evolution with $\Rbub$ according to \autoref{eq:alphap_theory} for an assumed $v_t(L) = 10\, \kms$, slightly above the observed Alfv\'{e}n speeds in our MHD simulations. Though the turbulence in the hot gas in our MHD simulations lies far above this value (see \autoref{fig:vt_vs_vout}), this choice is likely more representative of turbulence in resolved MHD mixing layers. This choice results in moderate $\alpha_p$ enhancement above unity. Interestingly, in this formalism, fixing other parameters and reducing the fractal dimension of the bubble results in a higher $\alpha_p$, but one that decreases with time ($\Rbub$).

\begin{figure*}
    \includegraphics[width=\textwidth]{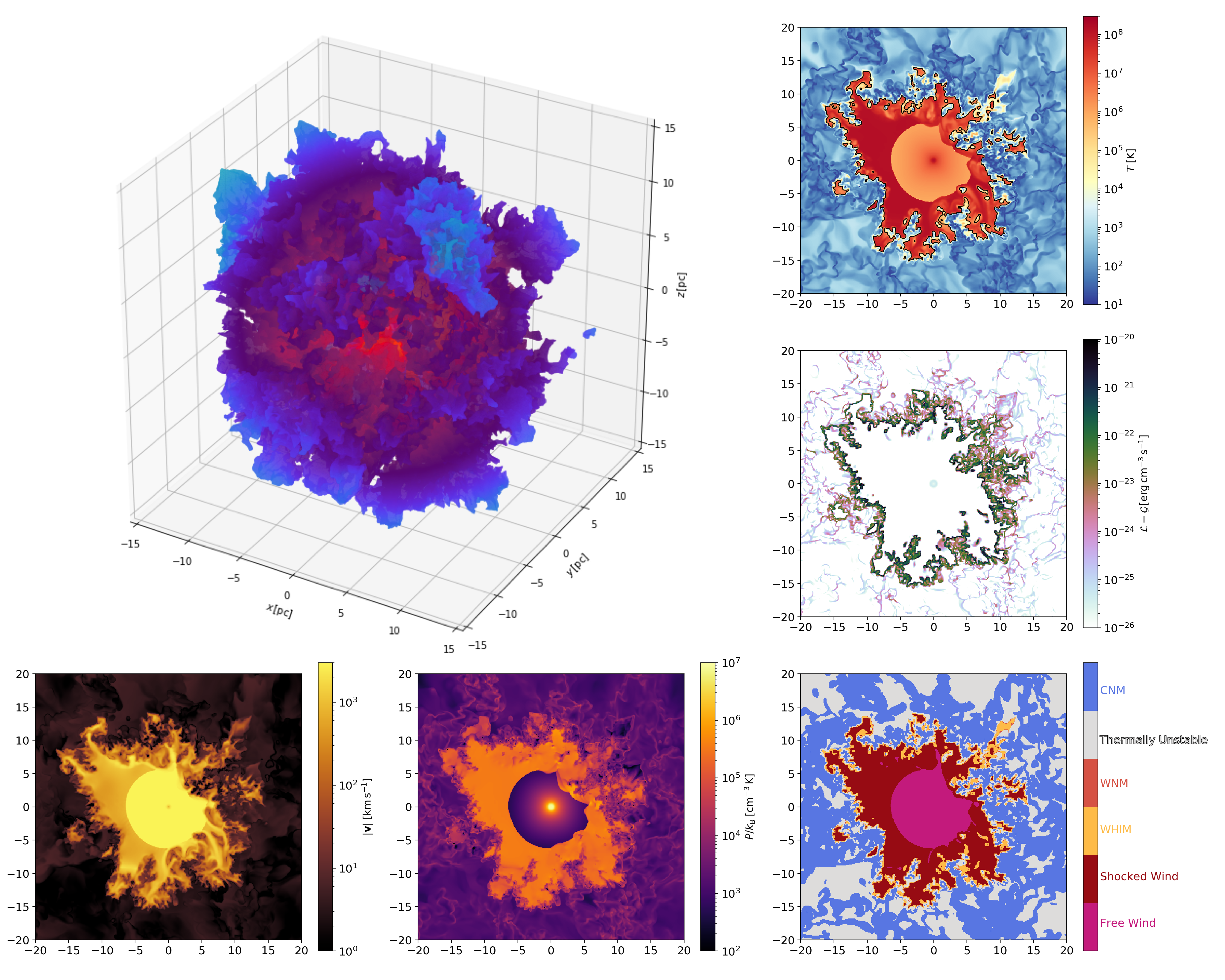}
    \caption{Visualization from our fiducial simulation at a resolution of $\Nres = 512$ approximately $t=0.80\, {\rm Myr}$ after the wind has been turned on. All smaller panels depict slices through the $z=0$ midplane of the simulation (clockwise from top-right): the gas temperature, the black line is a $T=10^6\, {\rm K}$ iso-contour; the net cooling rate in the gas; loci in the slice-plot of the thermal phases we use here; the thermal pressure of the gas; the magnitude of the velocity field. The large panel in the top left shows a three-dimensional rendering of the $T=10^6\, {\rm K}$ iso-temperature surface within the simulation. The color of the surface corresponds to the radius of that part of the surface and is included for clarity of viewing the geometry of the surface. Light blue corresponds to surface elements far from the wind source and bright red corresponds to surface elements close to the wind source. }
    \label{fig:simplot}
\end{figure*}

\begin{figure*}
    \includegraphics[width=\textwidth]{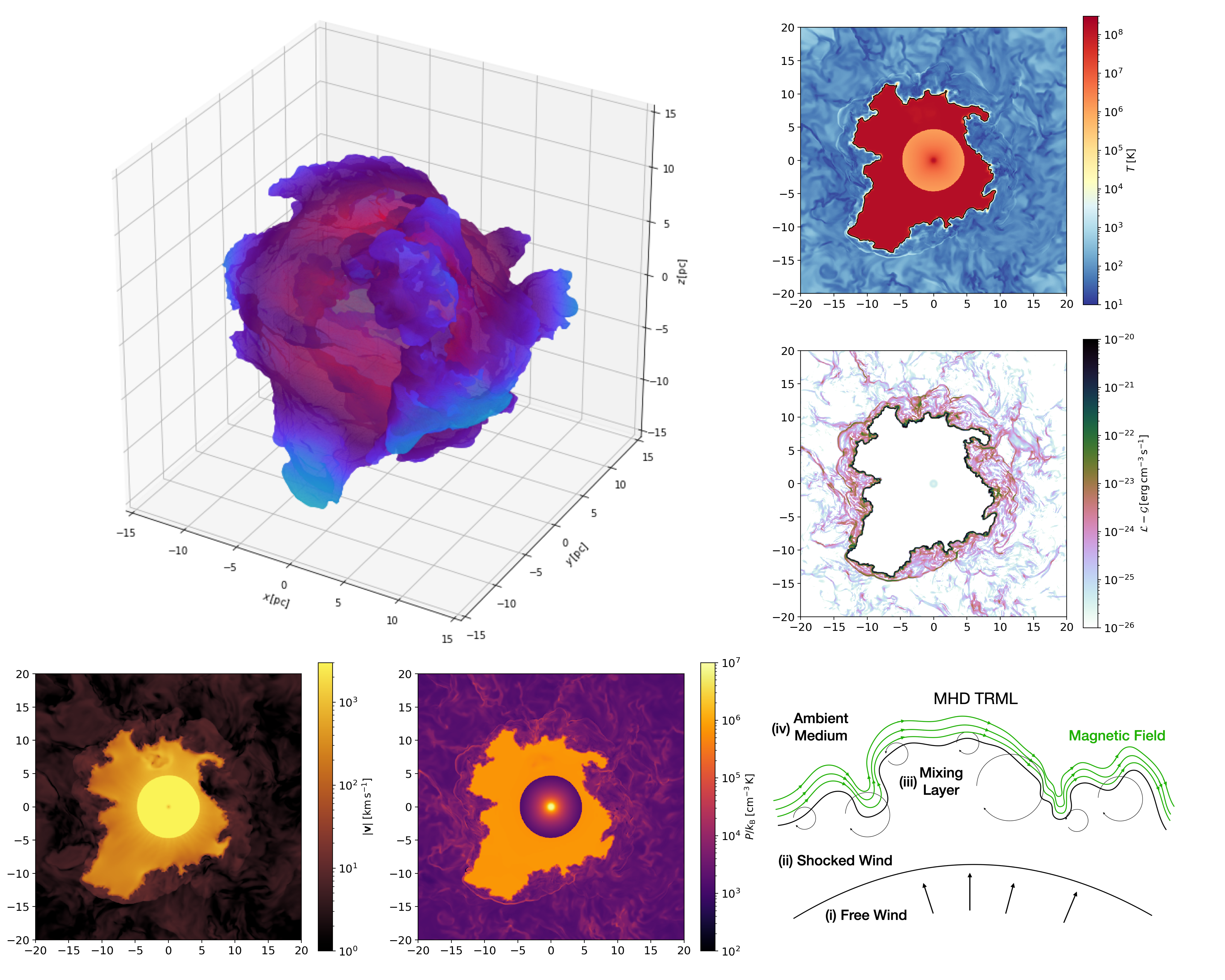}
    \caption{Visualization exactly analogous to \autoref{fig:simplot} but now for the \texttt{N512\_mhd} simulation. We have additionally replaced the phase splitting slice in the bottom right panel with a schematic diagram that is similar to the middle panel of \autoref{fig:schematic} but for the MHD case. We can see that structure in the interface is suppressed by the presence of the magnetic field.}
    \label{fig:simplot_mhd}
\end{figure*}

\section{Methods}
\label{sec:simulations}

To provide a test of the theory of WBB evolution discussed above, we run three-dimensional (magneto)hydrodynamic simulations, detailed below.

\subsection{Numerical Methods}
\label{subsec:num_methods}

We use the \textit{Athena} code \citep{Stone08_Athena}. The numerical methods employed in these simulations are nearly identical to those used in \citet{Lancaster21b}; in particular we use the same `hybrid/subcell method' for injecting wind mass (tracked as a passive scalar field $\rho_{\rm wind}$) and energy onto the grid and the same basic setup, which we will detail further below. However, unlike \citet{Lancaster21b}, we employ the chemistry and radiative cooling module of \citet{JGK_NCR23}, which tracks the non-equilibrium photochemistry of molecular, atomic, and ionized hydrogen as well as the equilibrium chemistry of carbon- and oxygen-bearing species (${\rm C}$, ${\rm C}^+$, ${\rm CO}$, ${\rm O}$, and ${\rm O}^+$), and their effects on the thermal heating and cooling of the medium. For the heating and ionization of neutral gas, we adopt uniform FUV background radiation of \citet{Draine78} and cosmic ionization rate per hydrogen $\xi_{\rm cr} = 2\times 10^{-16}\,{\rm s}^{-1}$ \citep{Indriolo15}. For cooling by helium and metals at high temperature ($T > 2\times 10^{4}\,{\rm K}$), we use the tabulated collisional ionization equilibrium (CIE) cooling function from \citet{Gnat12}. Our hydrogen cooling is fully dynamical, dependent on the local radiation field and ionization state of the gas and therefore not assumed to be CIE.

Our simulations are all run using the linearized Roe Riemann solver for flux calculations, the piecewise linear method (PLM) for second-order spatial reconstruction, and the unsplit second-order van Leer integrator for temporal evolution \citep[][and references therein]{Stone08_Athena}. We also include the `H-correction' of \citet{SandersHcorr98} to prevent carbuncle instability when the difference in signal speeds in different grid-aligned directions is greater than $20~\kms$. We do not include gravity in these simulations, and we use open/outflow boundary conditions.

We run simulations of both hydrodynamics (HD) and magneto-hydrodynamics (MHD). In our MHD simulations, we employ Athena's default constrained transport (CT) method for evolving the magnetic field, which we will refer to as the \texttt{CT-Contact} scheme \citep{GardinerStone05}. However, we found that this method is subject to numerical instabilities in regions of our simulation with strong source terms such as the bubble's edge (where cooling is strong) and the wind injection region. In order to suppress these numerical instabilities we implement the upwind constrained transport (UCT) HLL scheme of \citet{LondrilloDelZanna04} as described in \citet{MignoneDelZanna21}. As the \texttt{UCT-HLL} method is more diffusive\footnote{This diffusion only applies to the magnetic field, not to the other fluid variables.} than the \texttt{CT-Contact} scheme, we only employ it in the regions where these instabilities were observed: in the wind-feedback region and at the wind bubble interface, which we identify by regions with a strong gradient in the magnetic field $\delta B/B > 10$. We leave investigation as to the origin of these numerical instabilities to future work.

\begin{deluxetable}{ccccc}
    \tablecaption{Simulation List.\label{tab:simulations}}
    \tablewidth{0pt}
    \tablehead{
    \colhead{Name} &  \colhead{Hydro Solver}  & \colhead{$\Nres$} & \colhead{$\Delta x\, [{\rm pc}]$} & \colhead{$\mdotw/\Mstar\, [{\rm Myr}^{-1}]$}}
    \startdata
    \texttt{N128\_hydro} & HD  & 128 & $0.31$ & $2.965\times 10^{-3}$\\
    \texttt{N256\_hydro} & HD  & 256 & $ 0.16$ & $2.965\times 10^{-3}$\\
    \texttt{N512\_hydro} & HD  & 512 & $0.08$ & $2.965\times 10^{-3}$\\
    \texttt{N128\_mhd} & MHD  & 128 & $0.31$ & $2.965\times 10^{-3}$\\
    \texttt{N256\_mhd} & MHD  & 256 & $ 0.16$ & $2.965\times 10^{-3}$\\
    \texttt{N512\_mhd} & MHD  & 512 & $0.08$ & $2.965\times 10^{-3}$\\
    \texttt{N128\_hydro\_Mwmod} & HD  & 128 & $0.31$ & $ 10^{-2}$\\
    \texttt{N256\_hydro\_Mwmod} & HD  & 256 & $ 0.16$ & $ 10^{-2}$\\
    \enddata
    \tablecomments{Description of the parameters of our simulations. }
\end{deluxetable}

\subsection{Simulation Description}
\label{subsec:simulation_descibe}

In all of the simulations detailed here, we first set up a turbulent, multi-phase background before initializing a constant mechanical luminosity, $\Lwind$, and mass loss rate, $\mdotw$, wind at the center of the domain. Here we choose parameters of our background medium to be representative of a GMC with a total mass of $\Mcl = 10^5\, M_{\odot}$ and a radius of $\Rcl = 20\, {\rm pc}$. In particular, we initialize our simulations within a box of side length $\Lbox = 2\Rcl = 40\, {\rm pc}$ with a given number of resolving elements per grid-aligned direction $\Nres \equiv \Lbox/\delx$. We set the density everywhere to $n_{{\rm H},0} = 86.25\, \cc$ and initialize a Gaussian, random-phase turbulent velocity field with power spectrum $|v_k|^2 \propto k^{-4}$ for wavenumbers in the range $2\leq k \Lbox/(2\pi) \leq 64$. The normalization of the velocity field is chosen so that the total kinetic energy in the domain is twice the binding energy, $2W_{\rm cl}=(6/5)G\Mcl^2/\Rcl$, of a uniform-density spherical cloud of mass $\Mcl$ and radius $\Rcl$. The initial one-dimensional velocity dispersion is therefore $\sim 3\,\kms$. In our simulations with magnetic fields we additionally set a uniform magnetic field everywhere of $\mathbf{B}_0 = B_0 \mathbf{\hat{z}}$ with $B_0 = 13.5\, \mu{\rm G}$. This value was chosen based on setting the dimensionless mass-to-flux ratio equal to 2 as in \citet{JGK21}.

We then allow the simulation to evolve, without any further turbulent driving, until the total kinetic energy has become equal to $W_{\rm cl}$. The thermal pressure is initially set everywhere to $P/k_B = 6000\, {\rm K}\, \cc$. While this is above the equilibrium pressure at these densities, the simulation domain quickly relaxes to its equilibrium pressure during the initial turbulent evolution.

After the turbulent evolution, which lasts approximately $0.7\, {\rm Myr}$, the wind is turned on. For all simulations, we assume a stellar mass of $5000\, M_\odot$ with stellar wind luminosity per unit stellar mass of $\Lwind/\Mstar  = 9.75 \times 10^{33}\, {\rm erg}\, {\rm s}^{-1}\, M_{\odot}^{-1}$.  We consider two different values of the mass-loss rate per unit stellar mass: $\mdotw/\Mstar = 10^{-2}\,{\rm Myr}^{-1}$ and $ 2.965\times 10^{-3}\, {\rm Myr}^{-1}$. The first corresponds to values used in the simulations of \citet{Lancaster21b} while the latter is more representative of the value derived from Starburst99 for a solar metallicity population with a Kroupa IMF \citep{SB99,Leitherer92}. These choices of mass loss rate correspond to wind-velocities of $\Vwind = 1759$ and $3230 \, \kms$ and the momentum input rate of $\pdotw = 8.795\times 10^4$ and $4.788 \times 10^4 M_{\odot}\,{\rm km}\,{\rm s}^{-1}\,{\rm Myr}^{-1}$, respectively, for the given value of $\Lwind/\Mstar$. The higher wind velocity is our fiducial value. The simulations with $\mdotw/\Mstar = 10^{-2}\,{\rm Myr}^{-1}$ and $\Vwind = 1759 \, \kms$ will be referred to as our `mass loss rate modified' simulations, using the addendum \texttt{Mwmod}. The full parameter suite of the simulations is detailed in \autoref{tab:simulations}.

For all simulations we inject mass, momentum, and energy using the hybrid feedback method detailed in \citet{Lancaster21b}. Our effective injection radius, $r_{\rm fb}/\delx = 3.2,\, 3.2,$ and $ 6.4$ for the simulations with $\Nres = 128,\, 256,$ and $512$ respectively. \citet{Pittard21} demonstrated in one-dimensional simulations that it was necessary for $r_{\rm fb}$ to be less than a maximum value, $r_{\rm inj, max} = \sqrt{\pdotw/4\pi P_{\rm amb}}$, in order for a WBB to be successfully inflated, with $P_{\rm amb}$ the ambient pressure in the medium. For our default parameter values $r_{\rm inj, max} = 55\, {\rm pc}$, which is larger than the domain size of our simulations. Thus, all of our simulations lie in the regime $r_{\rm fb} \ll r_{\rm inj,max}/10$ recommended by \citet{Pittard21} in order to properly capture energy-driven bubbles.

\begin{deluxetable}{cc}
    \tablecaption{Definitions of gas phases.\label{tab:phase_defs}}
    \tablewidth{0pt}
    \tablehead{
    \colhead{Phase} &  \colhead{Temperature Condition} }
    \startdata
    WHIM & $ 10^4\, {\rm K}<T<10^6\, {\rm K}$ \\
    Warm Neutral Gas & $6085\, {\rm K}<T< 10^4\, {\rm K}$ \\
    Thermally Unstable Gas & $181\, {\rm K}<T<6085\, {\rm K}$ \\
    Cold Neutral Gas & $T<181\, {\rm K}$ \\
    \enddata
    \tablecomments{Temperature Definitions for colder phases. The wind phases are specified in the text. }
\end{deluxetable}

\subsection{Measurement Details}
\label{subsec:measurement}

Throughout the following section we will refer to various definitions of thermal phases of the gas in our simulations. These are generally similar to those outlined in \citet{Lancaster21b}, with the exception that we re-label `Ionized Gas' as `Warm-Hot Ionized Medium' (WHIM) and make a few small changes to the temperature ranges given our new cooling module; these are given in \autoref{tab:phase_defs}. The wind phases are as in \citet{Lancaster21b} where the shocked wind is defined as all gas hotter than $10^6\, {\rm K}$ and with a radial velocity less than half the wind velocity. The free wind is defined as all gas with a radial velocity that is greater than half $\Vwind$ as well as all gas contained within the feedback region.

In \autoref{fig:simplot} and \autoref{fig:simplot_mhd} we provide a rendering of our highest resolution fiducial simulation: \texttt{N512\_hydro} and \texttt{N512\_mhd} respectively. It is clear that the bubble's geometry is very complicated and highly `folded'. It is also clear from the top-right panel that the $T = 10^6 \, {\rm K}$ iso-temperature surface encloses nearly all of the WBB's volume.

We are particularly concerned in this work with properties of the gas near the bubble's surface. In order to precisely locate this surface within our simulations we follow \citet{FieldingFractal20} and \citet{Lancaster21b} by using the \texttt{marching\_cubes} algorithm from \texttt{scikit-image} \citep{van2014scikit}. Specifically, we measure iso-temperature surfaces at $\log_{10}\left( T/{\rm K}\right)= 3.75-6$ in logarithmic steps of $0.25$ (10 temperatures) and using step sizes of $\ell/\delx = 1,\, 2,\, 4,\, 8,\, 16,\, 32,\,\& \, 64$. We can only perform such measurements as post-processing on simulation snapshots, which we output from the simulations every $\sim 0.01\, {\rm Myr}$ (100 per simulation). A three-dimensional rendering of such an iso-temperature surface is given in the left-panel of \autoref{fig:simplot} for $T= 10^6\, {\rm K}$ and a spacing of $\ell/\delx = 1$ for the $\Nres = 512$ simulation.

After using the \texttt{marching\_cubes} algorithm to define the iso-temperature surfaces we calculate the total area of the surface $\Abub(T,\ell)$ and interpolate the fluid variables to each facet of the surface that is returned by the algorithm. We then calculate the component of the fluid velocity normal to each facet of the surface, $\vout$, and the dot product of the facet normal and the radial vector, $\mathbf{\hat{n}}\cdot \mathbf{\hat{r}}$. The quantities $\voutavg$ and $\foldedness$, presented below, correspond to the area-weighted average over all the faces of the iso-temperature surfaces. Unless it is stated otherwise, we use the iso-temperature surfaces defined by $T = 10^6\, {\rm K}$ and measured on the resolution scale $\ell = \delx$ for all analyses detailed below.

We will also be concerned with the distribution of gas in the simulation as a function of temperature. To investigate this in detail we take each snapshot of the simulations and compute the volume-weighted, pressure-weighted, and cooling-weighted histograms in temperature in 200 logarithmically spaced temperature bins between $10^{1.5}\, {\rm K} < T < 10^{8.5}\, {\rm K}$. For these measurement we restrict to gas that has a wind mass fraction $\fwind = \rho_{\rm wind}/\rho > 10^{-4}$ and a radial velocity $v_r < \Vwind/2$ to avoid contamination from the background and free wind respectively. However, the former constraint does somewhat bias measurement in our MHD simulations, where the momentum from the forward-shock of the wind bubble is carried away from wind-polluted material by MHD waves.

Further details on measurements of specific physical quantities are presented as needed in the remainder of the paper.

\section{Results}
\label{sec:results}

\begin{figure}
    \includegraphics[width=\columnwidth]{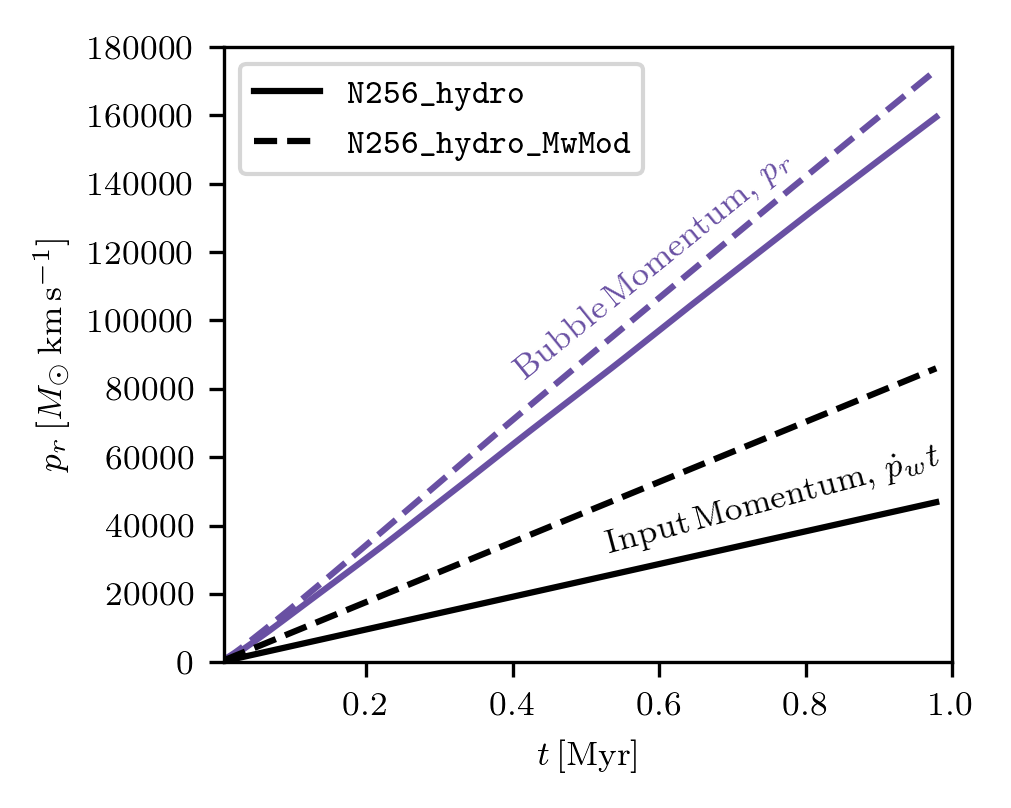}
    \caption{Comparison of the total radial momentum carried by WBBs in two of our simulations, \texttt{N256\_hydro} (solid) and \texttt{N256\_hydro\_Mwmod} (dashed). These simulations have the same energy input rate, $\Lwind$, but differ in their input wind mass-loss rates, $\mdotw$, and therefore have different wind velocities, $\Vwind$, and input momentum rates, $\pdotw$ (as indicated in the key). For each simulation blue lines indicate the total measured radial momentum as a function of time whereas black lines indicate the momentum that would result from $p_r = \pdotw t$, (i.e. $\alpha_p = 1$). Despite different input $\pdotw$, the resulting momentum carried by each bubble is quite similar.}
    \label{fig:pdot_comp}
\end{figure}

\begin{figure}
    \includegraphics[width=\columnwidth]{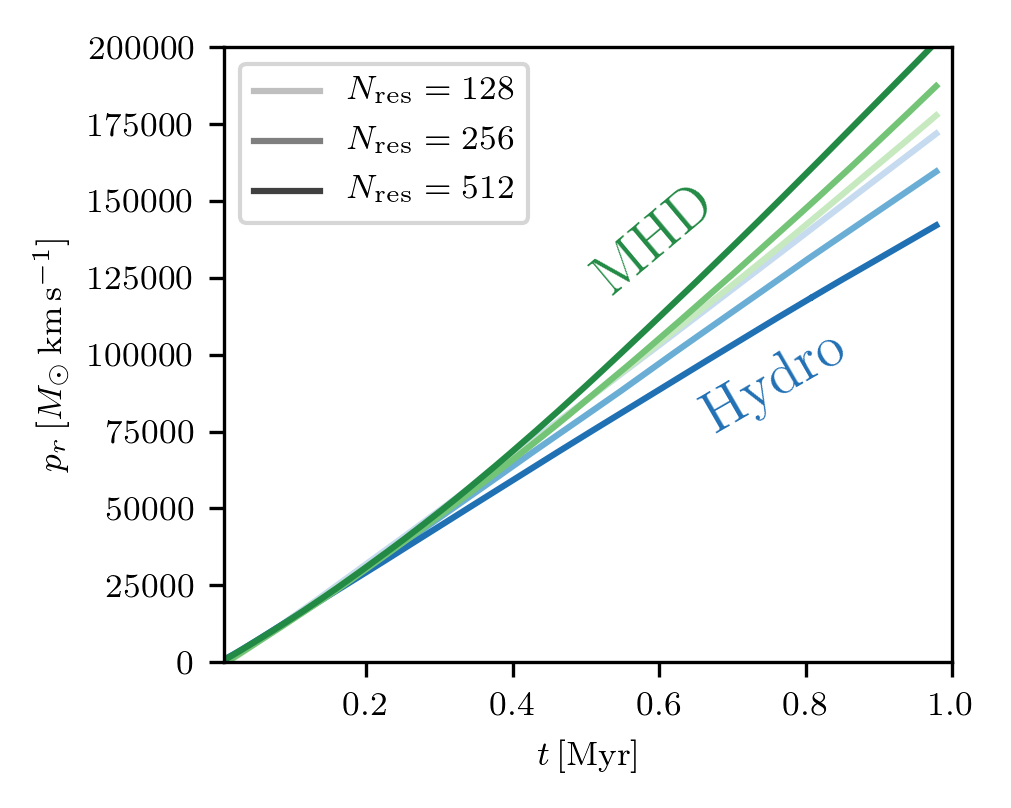}
    \caption{As in \autoref{fig:pdot_comp} but now comparing HD (blue) and MHD simulations (green). In each case, darker curves indicate higher numerical resolution. Notably, the HD and MHD simulations have the opposite behaviour as a function of resolution.}
    \label{fig:mhdVhydro}
\end{figure}

\subsection{Radial Momentum}
\label{subsec:momentum}

One of the most important aspects of WBBs is the dynamical effect they have on their surroundings, exemplified by the bulk momentum that they inject \citep{Keller22}, so it is natural to first investigate this aspect of our simulations. Here we measure the total radial momentum carried by the wind bubble by computing evolution of the total radial momentum within the simulation domain, and then subtracting out the evolution of the radial momentum from a simulation that is identical except it does not include wind feedback. This comparison to the no-feedback simulation is necessary since, due to our outflow boundary conditions and the turbulent conditions, the total radial momentum in the simulation is non-zero even when there is no feedback. We use this method instead of the method of \citet{Lancaster21b}, which measured total radial momentum in gas that included some fraction of wind material, as we found that in our MHD simulations momentum is able to be carried out of wind-polluted material by MHD waves, which has been observed previously \citep[e.g.][]{OffnerLiu18}.

In \citet{Lancaster21a} we posited that turbulent mixing in WBBs followed by radiative cooling was strong enough to reduce the bubble evolution to nearly the `momentum conserving' regime. That is, the momentum communicated to the surroundings by the bubble is very close to that put in by the wind, $p_r \sim \pdotw t$ or $\alpha_p \sim 1$. The simulations of \citet{Lancaster21b} showed measured values of $\alpha_p$ between 1 and 4 (see their Fig. 8). Additionally, these simulations showed that the measured energy retention fraction was consistent with the prediction for the measured values of $\alpha_p$. In \citet{Lancaster21a,Lancaster21b} we further argued that in the case of a radial flow in the bubble, the speed at the edge of the bubble, $\sim (3/16)\Vwind/\alpha_p$, should match the effective diffusion velocity that carries material into the shell. Here, from \autoref{eq:alphap_derive}, we have taken a further step to see that the value of $\alpha_p$ is actually determined by the interplay of the bubble's geometry and dissipation at its surface in time. The quasi-steady state assumption imposes the global constraint $\alpha_p \approx (\Rbub/\Rfree)^2$, which has a lower limit of unity, but there is no intrinsic reason why the ratio should be exactly 1 or why it should be constant in time. 

As we can see upon inspection of \autoref{fig:pdot_comp}, this is indeed not the case. Here we show the total radial momentum evolution in our default HD simulations with two different input wind momenta, $\pdotw$. As described in \autoref{sec:simulations}, this difference is mediated by a difference in $\mdotw$, while both simulations have the same mechanical wind luminosity, $\Lwind$. Though the input momenta, $\pdotw$, are different by about a factor of 2 between these simulations, the momentum they impart to background are the same to within 10\%. That is, $\alpha_p$ is a factor $\sim 2$ larger in the simulation with lower momentum injection rate (having lower $\mdotw$, higher $\Vwind$).  As we will discuss further in \autoref{subsec:interface}, the agreement in $\dot{p}_r$  is largely due to the fact that momentum input to the shell is governed by dissipation processes at the bubble's surface (which set $\Phot$ through through \autoref{eq:Phot_steady}), and these dissipation processes remain largely unchanged between these two simulations. For fixed $\voutavg \Abub/\Rbub^2$, \autoref{eq:pdot_main} says that $\dot{p}_r \propto \Lwind$, i.e. the same value for the two models shown, consistent with the simulation result of  \autoref{fig:pdot_comp}.  At the same time, \autoref{eq:alphap_derive} says that $\alpha_p\propto \Vwind$;  this implies a larger value for $\alpha_p$ in the fiducial simulation, which has higher $\Vwind$ and lower $\pdotw$. 

The second interesting issue we would like to introduce in \autoref{fig:mhdVhydro} is the \textit{opposite} resolution dependence of the total radial momentum in our HD and MHD simulations. While the HD simulations have decreasing momentum output with increased resolution, the MHD simulations have the opposite behaviour. As we will discuss below, and should be clear from \autoref{eq:alphap_derive}, this must be solely due to differences in the relative scaling of dissipation and geometry between the two types of simulation.

\begin{figure}
    \includegraphics[width=\columnwidth]{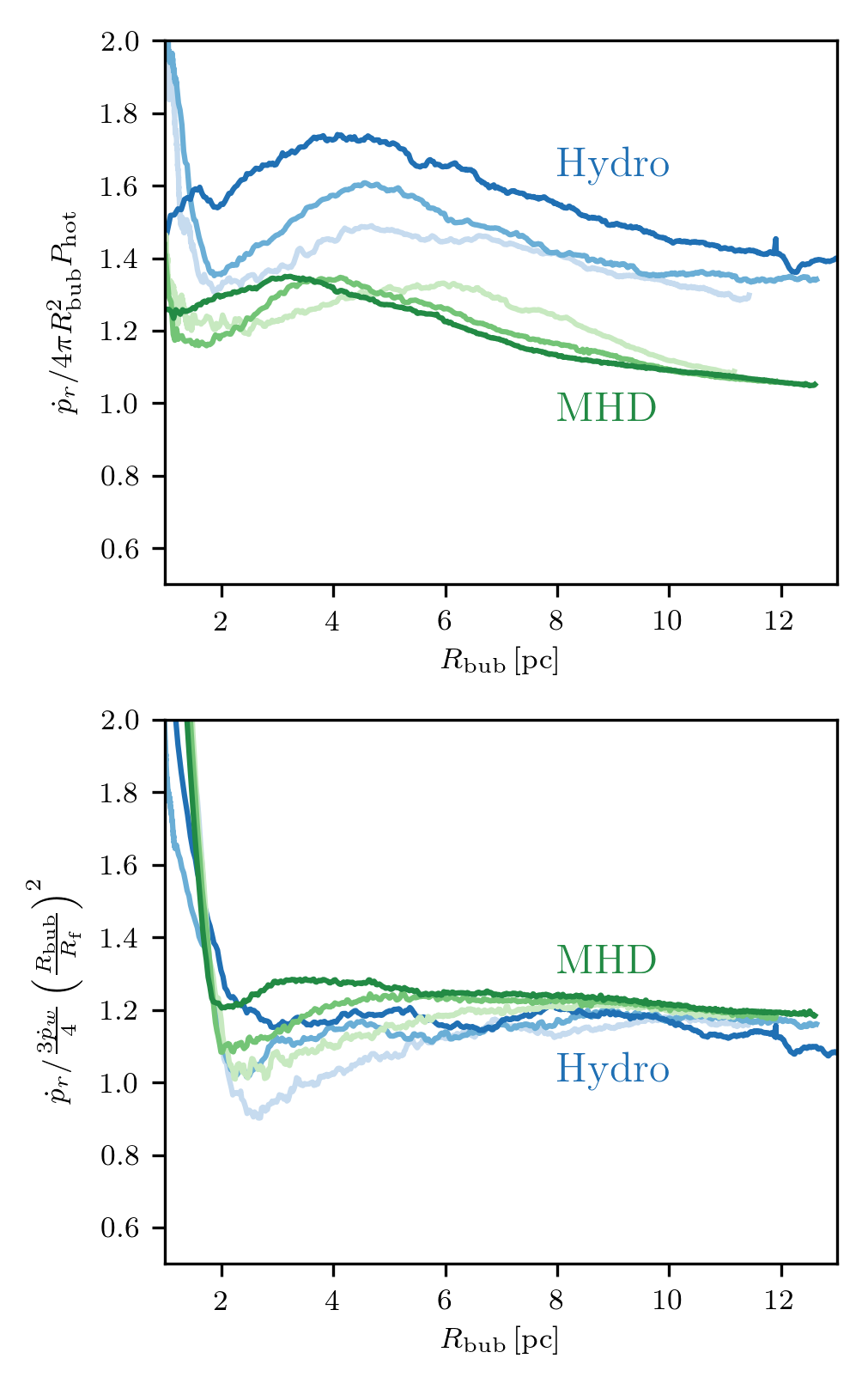}
    \caption{A validation of the driving dynamics of the bubble laid out in \autoref{subsec:structure_highlights}. Line colors in all panels are as in \autoref{fig:mhdVhydro} and all comparisons are shown as a function of $R_{\rm bub}$. \textit{Top Panel}: The rate of change of the total radial momentum carried by the WBBs divided by the radial force given in \autoref{eq:bubble_radial_force}. \textit{Bottom Panel}: Same as top panel, now dividing by \autoref{eq:bubble_force2}.}
    \label{fig:momentum_valid}
\end{figure}

\begin{figure*}
    \includegraphics[width=\textwidth]{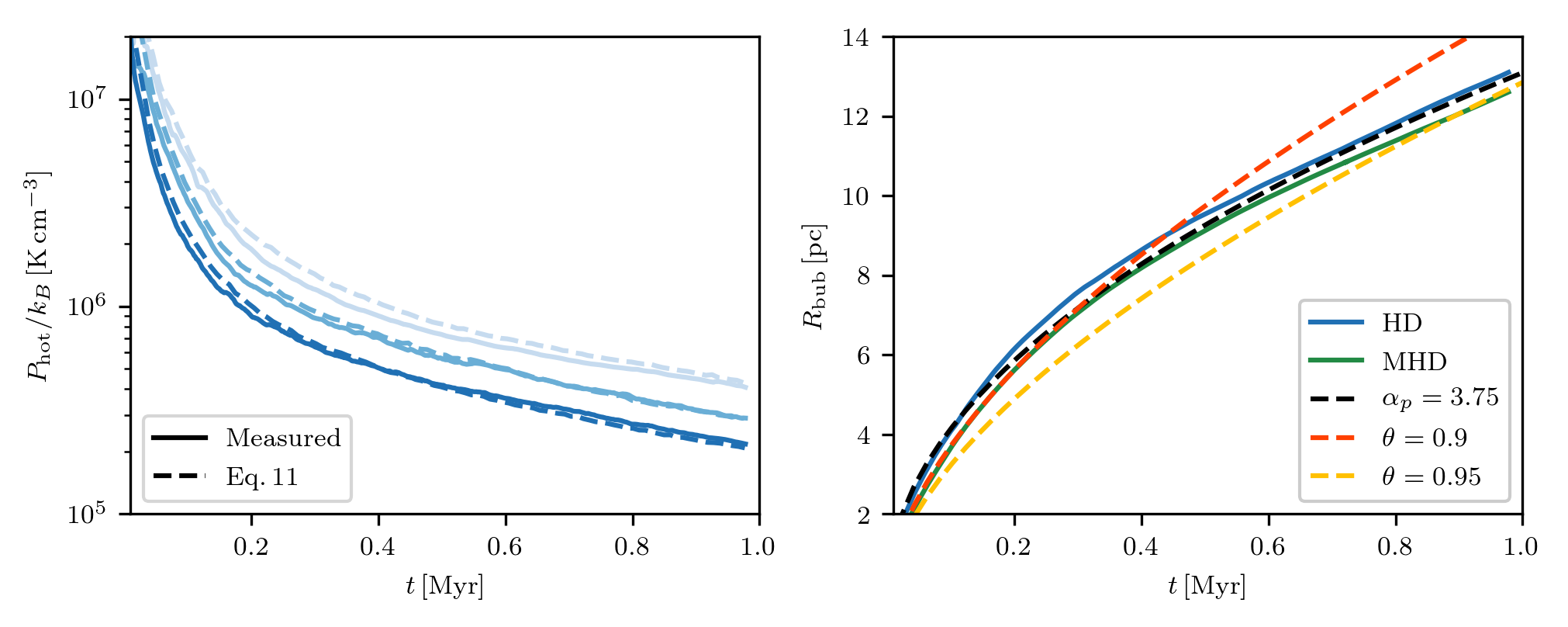}
    \caption{\textit{Left Panel}: The hot gas pressure as a function of time as measured in the simulations (solid lines), showing excellent agreement with the measurement derived from \autoref{eq:Phot_steady} (dashed lines). The MHD simulations are left out for clarity, but they also show excellent agreement with \autoref{eq:Phot_steady}. \textit{Right Panel:} We show the bubble's radius as a function of time in the highest resolution HD (blue) and MHD (green) runs. For comparison we also show, as dashed lines, the radial evolution derived for the constant-momentum input solution of \citet{Lancaster21a} with $\alpha_p = 3.75$ in black, as well as the constant cooling fraction solution of \citet{ElBadry19} with $\theta = 0.9,\, 0.95$ in red and yellow respectively.}
    \label{fig:pressure_valid}
\end{figure*}

\subsection{Validity of the Momentum Equation}
\label{subsec:momentum_valid}

In \autoref{subsec:structure_highlights} we argued that it is valid to describe the instantaneous force exerted on the bubble's surface using \autoref{eq:bubble_radial_force}. In \citet{Lancaster21b}, this was indeed shown to be a very good approximation in the HD simulations presented there. Before presenting the details of the bubble's structure we would like to validate this result in the current simulations.

In the top panel of \autoref{fig:momentum_valid} we show the rate of change of the total radial momentum carried by the bubbles (the time derivative of the curves in \autoref{fig:mhdVhydro}) divided by the force as computed from \autoref{eq:bubble_radial_force}. Specifically, the pressure is the volume-averaged value within the shocked wind region and $\Rbub$ is as computed from \autoref{eq:Rbub_def}. We can see that the momentum rate of change is always within an order unity factor from the prediction with smooth variations over time (represented here by $\Rbub$) and is an especially good prediction for the MHD simulations. In the bottom panel of \autoref{fig:momentum_valid} we show a comparison of the same force to the second prediction given in \autoref{eq:bubble_force2}, which shows even better agreement.

The fact that these predictions are slightly under-estimating the true force could be due, in part, to the fact that we are ignoring the contribution of the Reynolds stress term, which should be quite sub-dominant. Further under-estimates could be due to the slight inaccuracy of the `effective area' assumption behind \autoref{eq:bubble_radial_force}. This point is consistent with the fact that the equation is a better approximation in the MHD simulations (top panel of \autoref{fig:momentum_valid}), where the bubble's surface is less folded and one might expect $\Abub$ to be closer to $4\pi \Rbub^2$. Regardless, the agreement seen here is quite remarkable considering the extremely complex bubble geometry depicted in \autoref{fig:simplot}.

In the left panel of \autoref{fig:pressure_valid} we show $\Phot/k_B$, as a function of time along with the prediction of \autoref{eq:Phot_steady} with $\voutavg$ and $\Abub$ measured on iso-temperature surfaces with $T = 10^6 \, {\rm K}$. It is clear from the correspondence between the measured pressure and the predicted pressure that \autoref{eq:Phot_steady} is an excellent description of the bubble dynamics. This validates the picture discussed in \autoref{app:structure} and \autoref{subsec:structure_highlights}: that the bubble's interior is nearly steady and that its pressure is determined by the dissipation (leading to $\voutavg$) and the geometry ($\Abub$) of its interface with surrounding gas, as exemplified in \autoref{eq:Phot_steady} and \autoref{eq:alphap_derive}. Additionally, the fact that the prediction shown in \autoref{fig:momentum_valid} is given based on measurements from an iso-temperature surface at $T = 10^6\, {\rm K}$, which lies at the bubbles' furthest outer edges (see the iso-contour in the top right panel of \autoref{fig:simplot}), indicates that the steady-state bubble interior structure is a good approximation.

It is worth emphasizing that \autoref{fig:momentum_valid} and the left panel of \autoref{fig:pressure_valid} depict completely independent comparisons reflecting different aspects of bubble dynamics.  \autoref{fig:momentum_valid} shows that the measured radial force is equal to the bubble pressure multiplied by an {\it effective area} computed from the bubble volume (which is much smaller than the actual area of the bubble's surface). The left panel of \autoref{fig:pressure_valid} shows that the measured flow of thermal energy out of the bubble's (highly complex) bounding surface is consistent with the steady rate at which energy is being pumped in by the wind at its center.

It is also clear from the left panel of \autoref{fig:pressure_valid} that the pressure in the hot gas is not a converged quantity with resolution, even if we can consistently predict its value based on measurements at the bubble's interface. In particular, at a given time, higher resolution HD simulations have a lower bubble pressure. This resolution dependence then must be manifest in the resolution dependence of the quantities measured at the bubble's interface, which we explore below.

Finally, in the right panel of \autoref{fig:pressure_valid} we show the radial evolution of the bubbles in our highest resolution simulations in both HD and MHD as functions of time. For comparison we show the radial evolution expected for momentum-driven bubbles \citep{Steigman75,Lancaster21a}
\begin{equation}
    \label{eq:rbub_mc}
    \Rbub = \left(\frac{3\alpha_p\pdotw t^2}{2\pi \rhobar}\right)^{1/4} \, ,
\end{equation}
where, in \autoref{fig:pressure_valid}, we use $\alpha_p = 3.75$. We also show the radial evolution expected for a bubble which cools a set fraction, $\theta$, of its energy in its interface \citep{ElBadry19}
\begin{equation}
    \label{eq:rbubu_eb}
    \Rbub = \left( \frac{125\Lwind(1 - \theta)t^3}{154\pi \rhobar} \right)^{1/5} \, ,
\end{equation}
where in \autoref{fig:pressure_valid} we use $\theta = 0.9$ and $0.95$.

It is clear that the radial evolution of the bubbles in the right panel of \autoref{fig:pressure_valid} are better represented by the `momentum-driven' solution (which requires cooling to increase as time goes on \citep{Lancaster21a}) than by the constant $\theta$ solutions. We investigate the details of energy loss in the bubbles in the next section.

\begin{figure*}
    \includegraphics[width=\textwidth]{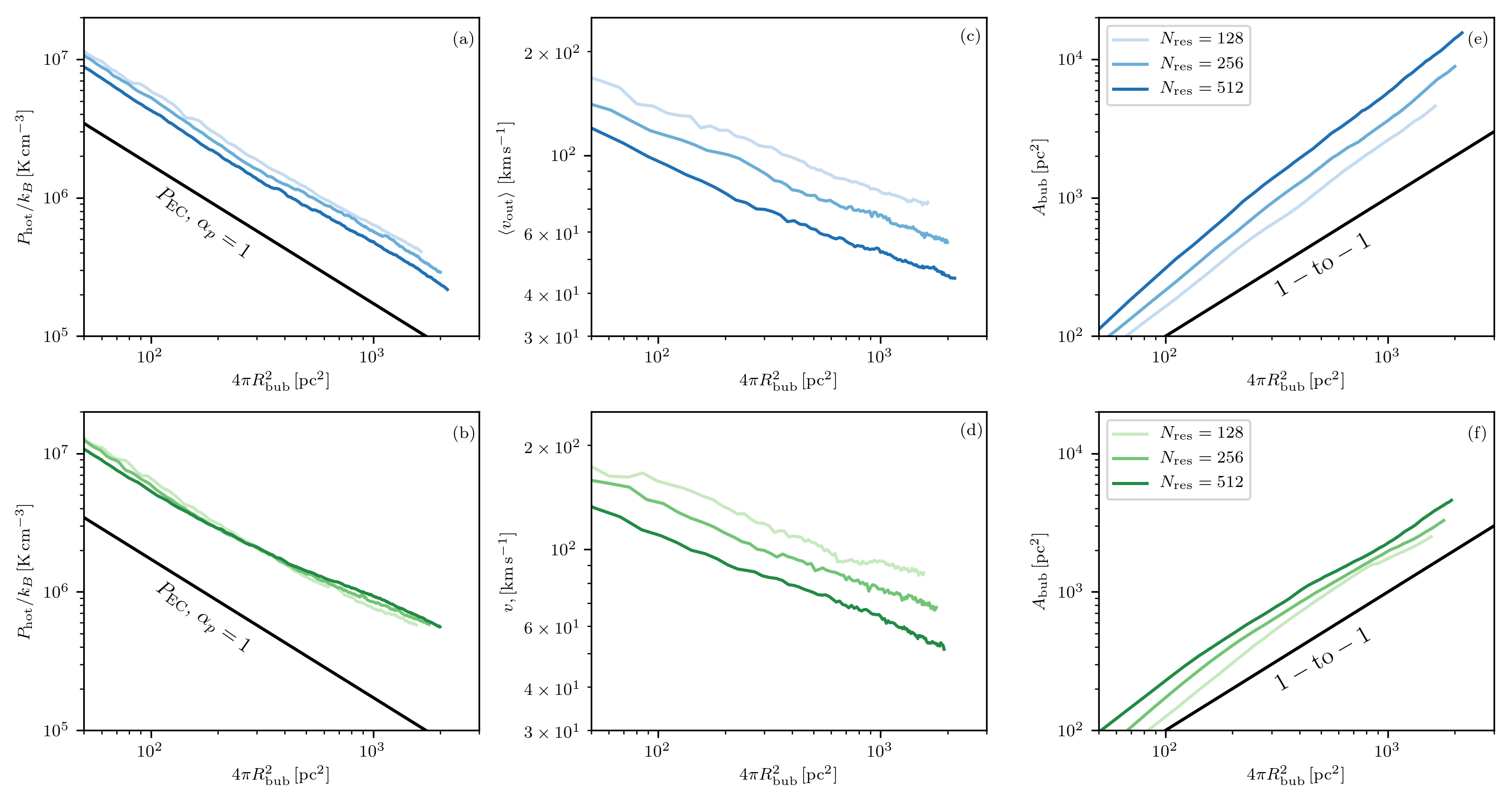}
    \caption{Summary characteristics of interior and interface properties of the bubble as a function of resolution and equivalent area $4\pi \Rbub^2$, both with (bottom panels) and without (top panels) magnetic fields. The resolution of each simulation is indicated by the legends in panels (e) and (f). \textit{Panels (a) \& (b)}: Hot gas pressure, which shows at later times the same opposing trends with resolution as in \autoref{fig:mhdVhydro}. For reference we show the pressure for an efficiently-cooled bubble (\autoref{eq:Pps} with $\Rfree \to \Rbub$) \textit{Panels (c) \& (d)}: The outward velocity averaged over the bubble's surface, $\voutavg$ (see \autoref{eq:voutavg_def}), decreasing as resolution increases. These values are all much smaller than the expected value for unimpeded radial flow $\Vwind\left(\Rfree/\Rbub \right)^2 /4$, which is nearly constant with time and in the range $200-300\, \kms$ in the simulations over the time range shown. \textit{Panels (e) \& (f)}: The area of the bubble surface, $\Abub$, increasing as resolution increases. We additionally include the $\Abub = 4\pi \Rbub^2$ 1-to-1 line in black to emphasize that the true bubble surface areas are much larger than their sphere-equivalent surface areas.}
    \label{fig:interface_resolution}
\end{figure*}

\subsection{The Interface}
\label{subsec:interface}

In \autoref{fig:interface_resolution} we show key properties of the bubble's interior and interface as a function of its sphere-equivalent area, $4\pi \Rbub^2$. In the left hand panels of \autoref{fig:interface_resolution} we can see the difference in HD  (top) vs. MHD (bottom) bubble dynamics that we saw in \autoref{fig:mhdVhydro}: the opposite trend with resolution of the bubble's pressure (and hence radial force through \autoref{eq:bubble_radial_force}; see \autoref{fig:momentum_valid}).

Since \autoref{fig:pressure_valid} shows that \autoref{eq:Phot_steady} is an accurate predictor of the bubble's interior pressure, we can attribute the difference in resolution dependence of $\Phot$ to a difference in resolution dependence of the quantities $\voutavg$ and $\Abub$ defined at the interface, which we show in the middle and right panels of \autoref{fig:interface_resolution}. In particular, since $\Phot \propto (\voutavg \Abub)^{-1}$ and consistently decreases at higher resolution in the HD simulations, the increase of $\Abub$ must be steeper than the decrease of $\voutavg$ at smaller $\Delta x$. In the HD simulations, we have found dependence on resolution varying roughly as $\voutavg \propto (\delx)^{ 1/4}$ and $\Abub \propto (\delx)^{ -1/2}$: thus, increased surface area wins over decreased outflow velocity normal to the surface, which in these simulations ultimately is due to numerical dissipation. 
In contrast, the relative scalings with resolution in the MHD simulations \textit{nearly} compensate for one another: $\voutavg \propto  (\delx)^{1/4}$ and $\Abub \propto (\delx)^{-1/4}$.  In the end, however, the decreased dissipation marginally wins over the increased bubble surface area, resulting in the increased pressure and therefore increased momentum input to the surroundings.

The question remains as to the origin of this dependence on numerical resolution, $\delx$. In addition, we would like to understand why $\voutavg$ decreases and $\Abub/(4 \pi \Rbub^2)$ increases for larger bubble radius (a proxy for later time in \autoref{fig:interface_resolution}). We investigate the scalings of $\Abub$ in \autoref{subsec:Abub_scaling} and $\voutavg$ in \autoref{subsec:voutavg_explained}, below.

\subsection{Scaling of $\Abub$}
\label{subsec:Abub_scaling}

As has been discussed extensively in the literature, dynamically unstable interfaces can lead to fractal, self-similar structures \citep{FieldingFractal20,TOG21,Lancaster21a,Lancaster21b}. We will show in this section that the scaling of $\Abub$ with time and resolution can be entirely explained through a fractal model for the geometry of these WBBs. We will leave discussion as to the origin of this fractal structure for \autoref{sec:discussion}.

There are several ways to think about fractal structure. Traditionally in mathematics \citep{Mandelbrot83,West97} and in the astrophysical literature \citep{MeneveauFractal87,ShibataKazunari01,Federrath09}, describing something as a fractal means parameterizing how some aspect of its structure, say $f$, varies as a function of scale, say $\ell$, as a power law in that scale, $f(\ell) \propto \ell^{a}$ for some $a$. For example, how the measurement of the size of the boundary of some region varies as a function of the physical scale on which you measure it: like measuring the coastline of England with a mile long ruler (not able to describe the structure of small bays and inlets) versus a foot long ruler (maybe not able to describe every rock). For our current purposes, and generally in the mixing-layer literature, we are interested in the area of the WBB interface and how this varies with scale. We can describe this scaling as \citep[see e.g.][]{Lancaster21b,FieldingFractal20}
\begin{equation}
    \label{eq:dell_def}
    d_{\ell} \equiv -\frac{d\log \Abub (\ell)}{d\log \ell} \, .
\end{equation}
We see that $d_{\ell} = 0$ corresponds to an interface with no finer-scale structure (area does not get larger when measured at smaller scales) and $d_{\ell} >0$ corresponds to the formation of smaller scale structure (area increases as scale on which it is measured decreases).

In mathematics, fractal structures can be infinitely self-similar, continuously forming structure on smaller and smaller scales. Physical structures, in contrast, have to be regulated and eventually become smooth on some scale. This implies that $d_{\ell}$ is not constant as a function of $\ell$ but must vary, as it must return to $0$ at the scale on which structure becomes smooth. In the context of real WBBs this is \textit{at least} the mean free path of individual particles, at which point the idea of a boundary between fluid elements becomes meaningless, and is more likely to have to do with dissipative physics. In our ideal (M)HD simulations, however, this scale is the resolution scale\footnote{It is interesting to note that, in the mathematical world of ideal fluid dynamics, there is no regulatory scale, and one \textit{can} form infinitely self-similar structures.}. 

\begin{figure}
    \includegraphics[width=\columnwidth]{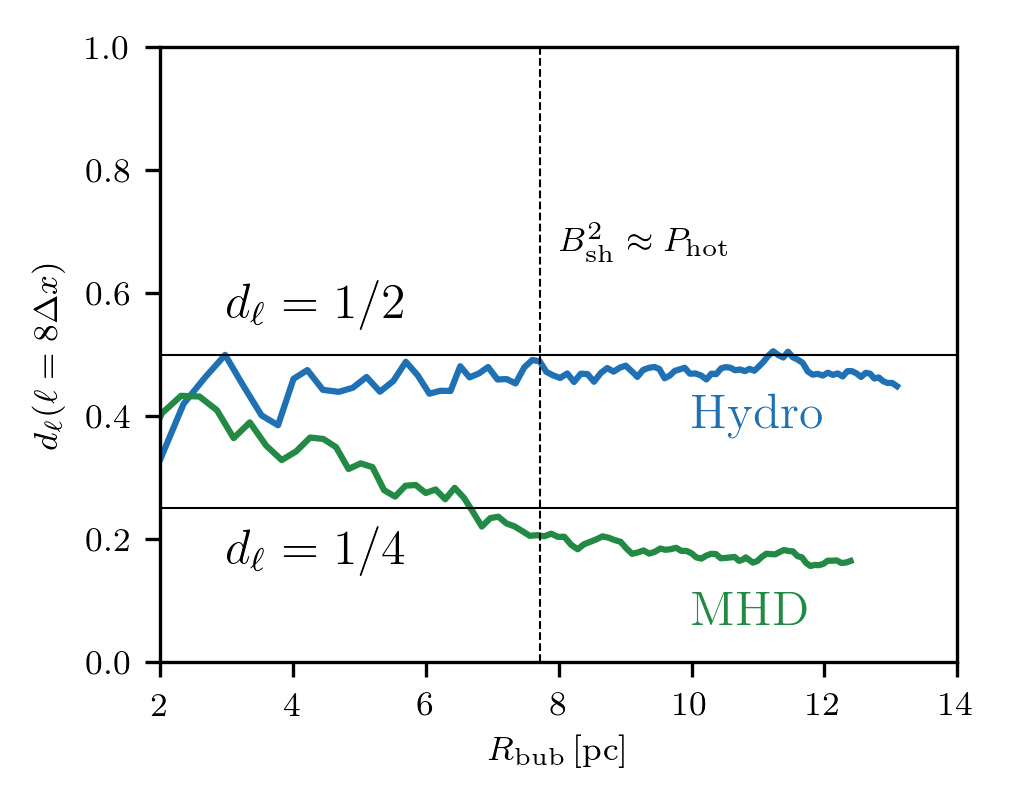}
    \caption{Fractal dimension of the WBB interfaces measured using \autoref{eq:dell_def} in the $\Nres = 512$ versions of both the HD (blue) and MHD (green) simulations as a function of $\Rbub$. Horizontal black lines indicate the fractal dimensions used for the scaling shown in \autoref{fig:Abub_frac}: $d = 1/2,\, 1/4$. The vertical dashed line indicates when the magnetic pressure of the bubble's shell comes into rough equilibrium with the bubble's interior thermal pressure in the MHD simulation.}
    \label{fig:frac_measure}
\end{figure}

In \autoref{fig:frac_measure} we show the fractal dimension as measured using \autoref{eq:dell_def} at $\ell = 8\delx$ in the $\Nres = 512$ versions of the HD and MHD simulations. Specifically, this is calculated as the logarithmic difference of $\Abub$ measured on $\ell = 4 \delx$ and $16\delx$. We choose this scale as it is the scale that seems to be least affected by the ends of the ``fractal dynamic range'' so that it is to the best extent probing the true fractal dimension. We see from \autoref{fig:frac_measure} that the fractal dimension is relatively constant as a function of radius/time in the HD simulations. In contrast, the MHD simulation has a clear decrease in its fractal dimension over time. If we attribute some part of the fractal structure of these bubbles to the growth of MHD instabilities at the interface, this decrease in the fractal structure can be understood as a suppression of these instabilities as the bubble expands, sweeping up the background magnetic field and causing the field strength to grow at the interface. This process is schematically shown in the bottom right panel of \autoref{fig:simplot_mhd}. Indeed, this period of decreasing $d$ coincides with the bubble's thermal pressure coming into equilibrium with the magnetic pressure of its shell, indicated by the vertical dashed line in \autoref{fig:frac_measure}. We indicate the time when the shell magnetic pressure first comes within 20\% of the bubble's thermal pressure with a vertical line in \autoref{fig:frac_measure}.

\begin{figure}
    \includegraphics[width=\columnwidth]{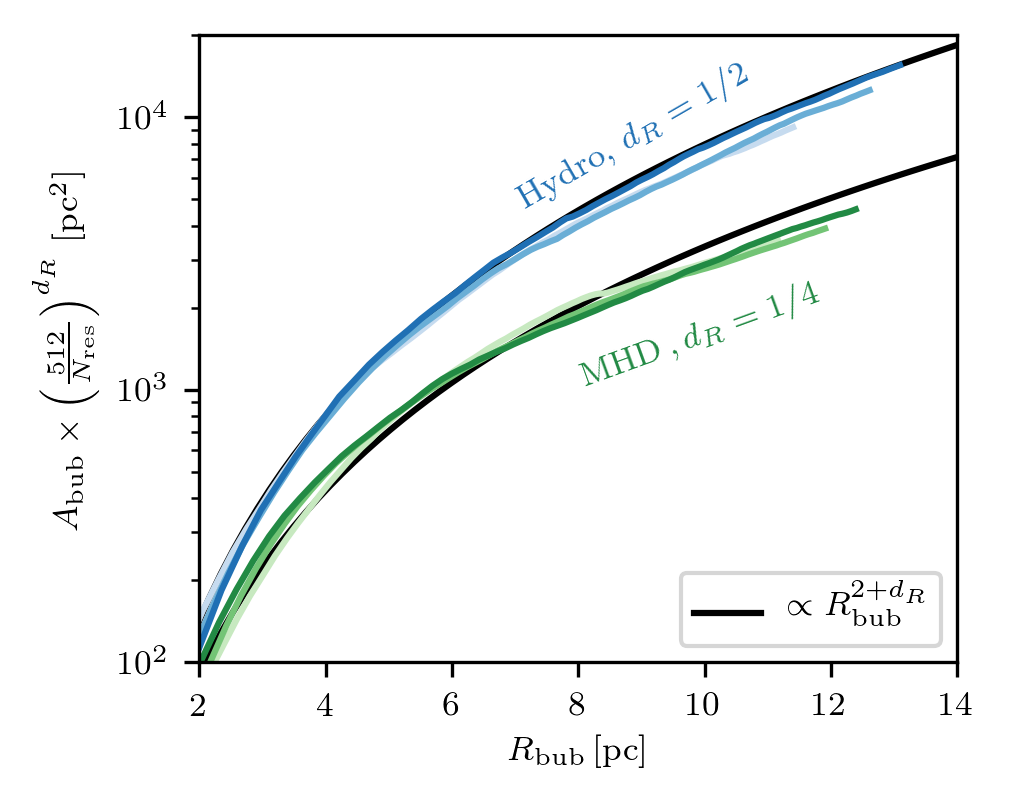}
    \caption{Demonstration of the fractal nature of the bubble's interfaces. The surface area is shown as a function of linear dimension, $\Rbub$, additionally scaling with resolution, in comparison to power law relations.  See text.}
    \label{fig:Abub_frac}
\end{figure}

Instead of considering the largest scales as fixed and investigating the fractal nature of the interface by examining the structure on smaller and smaller scales, we can also do the reverse. That is, we can consider the smallest scales as fixed (the resolution or dissipative scale) and investigate how the structure of the interface changes as it becomes larger. Investigating the fractal nature of the bubble in this way is a distinct feature of problems like the WBB (as opposed to cloud-crushing or TRML studies) as the outer scale actually varies in time. We can then alternatively define the excess fractal dimension as
\begin{equation}
    \label{eq:fracd_def}
    d_R \equiv \frac{d\log \Abub}{d\log \Rbub} - 2 \, .
\end{equation}
In the above we assume that $\Abub$ is measured on some fixed small scale. In practice we use the resolution scale of the simulations. This definition includes the $-2$ since we would expect that the surface area of a 3-dimensional structure scales with the square of its linear dimension; $d_R$ then quantifies the amount by which it grows \textit{more} than this.

In the context of \autoref{eq:fracd_def} we can show the areas of the bubble at fixed linear scale, $\Rbub$, as a pure geometrical measurement of the fractal qualities of the bubbles. However, at higher resolution there is a larger dynamic range on which to form structure. We can account for this using \autoref{eq:dell_def} assuming a given $d_R$ and backing out the factor needed to multiply the low-resolution simulations by in order to match the high resolution simulations: $\left(512/\Nres\right)^{d_R}$. 

This comparison is shown in \autoref{fig:Abub_frac} where we take $d_R=1/2$ for the HD simulations and $d_R=1/4$ for the MHD simulations for this resolution correction. These choices are generally consistent with the mean values shown in \autoref{fig:frac_measure}. The fact that the lines at differing resolution match very well with this correction points to the fact that it is simply the bubble's fractal structure that is causing this scaling.

We additionally show as black lines the scalings of $\Abub$ with $\Rbub$ for the same values of $d_R$ assumed in the resolution scaling. Specifically, we show 
\begin{equation}
    \Abub\left( \frac{N_{\rm res,hr}}{\Nres}\right)^{d_R} 
    = C 4\pi \Rbub^2 \left(\frac{\Rbub}{\Delta x_{\rm hr}} \right)^{d_R}
\end{equation}
where $N_{\rm res,hr}=512$ and $\Delta x_{\rm hr} \approx 0.08\, {\rm pc}$ is the resolution in the highest resolution simulations. We have set $C = 0.56, \, 0.80$ and $d_R = 1/2, \, 1/4$ for the HD and MHD simulations, respectively. The fact that these lines match very well with the simulations indicate the self-consistency of the fractal picture and how it explains the behaviour of $\Abub$ with resolution and scale. Furthermore, the fact that the black line is an over-estimate of the true $\Abub$ in the MHD simulations at late times is consistent with \autoref{fig:frac_measure} where we see a decrease in the fractal dimension, below $1/4$, at late times.

In conclusion, we find that the fractal structure of the bubble's surface, with excess dimension $d \approx 1/2$ in the HD case and $d \sim 0.25$ in the MHD case, is able to account for the scaling of $\Abub$ with numerical resolution seen in \autoref{fig:interface_resolution}e,f.

\begin{figure*}
    \includegraphics[width=\textwidth]{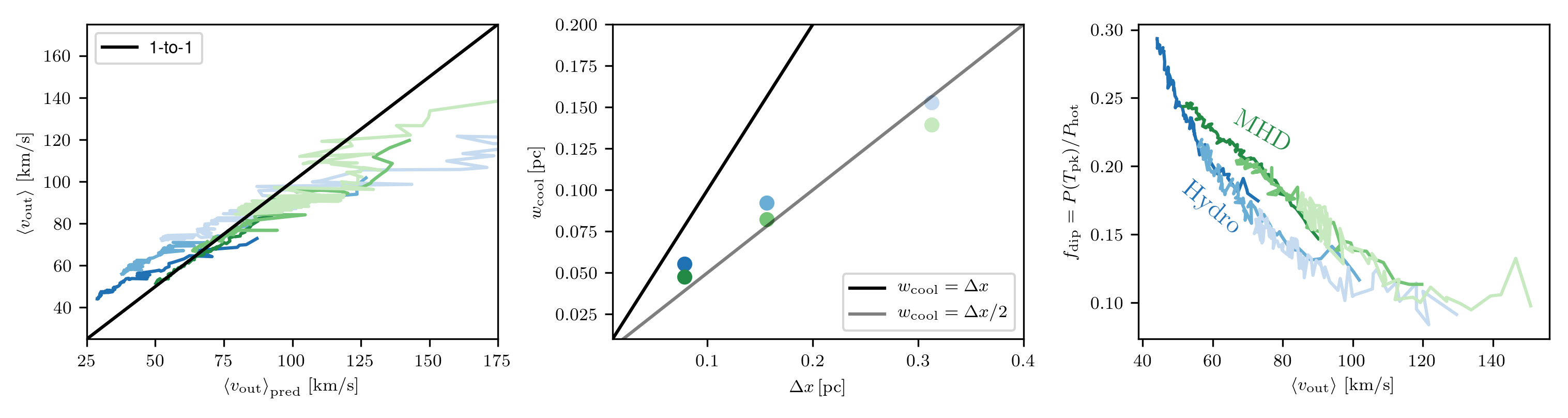}
    \caption{An empirical argument for the scaling of $\voutavg$. Line-styles are as specified in \autoref{fig:mhdVhydro}. \textit{Left Panel}:  The measured value of $\voutavg$ compared to that predicted by \autoref{eq:vout_cool_dip}. The one-to-one relation is shown for reference in black. \textit{Middle Panel}:  The width of the cooling layer, $\wcool$, as a function of the simulation's numerical resolution, $\delx$. As black lines we show scaling with resolution of $\wcool = \delx$ and $\wcool = \delx/2$. \textit{Right Panel}: The factor by which the pressure dips from the bubble interior to the pressure at peak cooling $\fdip = P(\Tpk)/\Phot$ as a function of $\voutavg$.}
    \label{fig:vout_scaling}
\end{figure*}

\subsection{Value and Scaling of $\voutavg$}
\label{subsec:voutavg_explained}

In \autoref{subsec:diss_scales} we discussed the dissipative scales that need to be resolved in order for $\voutavg$ to be set by physical processes. When these processes are not well resolved and cooling is strong enough so that the local value of $\vout$ (the velocity of the fluid into the surface, relative to the surface) is large compared to the velocity of the surface, $\Wbub$, then the local numerical dissipation is set by the local grid velocity, $\vout$. But, as we motivated in \autoref{subsec:diss_scales}, it is the dissipation which should be setting $\voutavg$ in the first place.

In \autoref{subsec:vout_cooling} we described how cooling provides a constraint on the value of $\voutavg$. In particular, through its role in setting $\Phot$ (through \autoref{eq:Phot_steady}), $\voutavg$ sets the pressure in the cooling layer (assuming the cooling layer is iso-baric with the bubble interior) and hence sets the cooling rate. Requiring a given cooling rate then requires a specific value of $\voutavg$.

For our numerical simulations, in fact the cooling layer is not perfectly iso-baric with the hot bubble interior, but instead has a pressure dip:
\begin{equation}
    \label{eq:fdip_def}
    \fdip \equiv \frac{\Ptpk}{\Phot} \, .
\end{equation}
Accounting for this in \autoref{eq:vout_cool} gives us
\begin{equation}
    \label{eq:vout_cool_dip}
    \voutavg_{\rm pred} = \frac{3\sqrt{\Ltpk \Vpk \Lwind}}{8 \Abub k_B \Tpk}\fdip \, ,
\end{equation}
where we have included the subscript for clarity of comparison in \autoref{fig:vout_scaling}. If we remove the factors that are explicitly dependent on the details of cooling we have
\begin{equation}
    \label{eq:vout_cool_scaling}
    \voutavg_{\rm pred} \propto \fdip \sqrt{\frac{\wcool}{\Abub}}
\end{equation}
where we have used $\Vpk = \wcool \Abub$ with $\wcool$ the thickness of the cooling layer.

In the left hand panel of \autoref{fig:vout_scaling} we show $\voutavg$ measured from the simulations against the prediction of \autoref{eq:vout_cool_dip}, where we take $\Vpk$ as the volume of gas with temperatures $10^4< T/{\rm K} < 4.8\times 10^4$. This is chosen to match the rough temperature range over which the highest rate of cooling occurs (see \autoref{app:cooling}). We measure $\Tpk$ and $\Ptpk$ using the cooling and pressure weighted temperature histograms as detailed in \autoref{subsec:measurement}, and $\Ltpk$ measured by interpolating the collisional ionization equilibrium (CIE) cooling curve of \citet{CGK_TIGRESS1}. 

Using this CIE $\Lambda(T)$ is not strictly faithful to the non-equilibrium cooling that we are employing in these simulations. While we could employ the equilibrium cooling curve from \citet{JGK_NCR23} for the non-photoionized case (e.g. top panel of Figure 3 there), this is negligibly different than \citet{CGK_TIGRESS1} for the temperature ranges of interest here. Additionally, the true cooling in our simulations will differ from the equilibrium curve of \citet{JGK_NCR23} due to the combination of turbulent and numerical mixing of the electron fraction, $x_e$, across the cooling layer. The CIE cooling curve is therefore a useful simplification. Using a single value of $\Tpk$ is of course a great simplification itself, since there is a wide range of temperatures where cooling occurs in these simulations. It is less obvious, but using a single value of $\Ptpk$ is also strictly inaccurate since, in these simulations, there are large pressure fluctuations in the cooling layer.

With these caveats in mind, the correspondence seen in the left panel of \autoref{fig:vout_scaling} acts as a test of \autoref{eq:vout_cool_scaling}, that is, the \textit{scaling} of $\voutavg$ and its dependence on the width of the cooling layer $\wcool$, the area of the bubble's surface, $\Abub$, and the pressure dip between the hot gas bubble and the cooling layer, $\fdip$. In particular, this shows that the resolution and time dependence of $\voutavg$ can be explained through the resolution and time scaling of these quantities.

We investigate the resolution scaling of $\wcool$ (which is independent of time in all of our simulations) in the middle panel of \autoref{fig:vout_scaling}. This width appears to scale roughly linearly with $\Delta x$. This is not unexpected; in the strong-cooling limit \citep{McCourt18,FieldingFractal20,TOG21} the cooling region will always occupy roughly a one-cell thickness volume over the interface unless the dissipative scales are resolved. The fact that $\wcool < \Delta x$ illustrates how quickly this transition occurs. Indeed, for a wider definition of the ``cooling region'' that includes hotter, not as rapidly cooling gas we would find $\wcool > \Delta x$ and scaling linearly with it. Together this means that the edge of the bubble always has approximately one cell at intermediate temperatures, but this is not always at the most rapidly cooling temperatures.

Finally, we'd like to investigate a relationship between the pressure at the cooling interface and $\voutavg$ or equivalently $\Phot$. In the right panel of \autoref{fig:vout_scaling} we show the fractional dip in pressure from the wind-bubble interior to the peak-cooling gas, $P(\Tpk)/\Phot$, as a function of $\voutavg$. While $\fdip$ is strongly dependent on resolution itself, it is clear from this panel that $\voutavg$ and $\fdip$ follow a relation to one another that is roughly independent of resolution but somewhat dependent on whether the medium is magnetized; thus, there may in principle be dependence on the numerical solver (MHD vs. HD).
As we discuss in \autoref{app:riemann}, this difference likely has to do with the details of the Riemann problem being solved at the interface and how these differ between the HD and MHD cases. With a fully resolved turbulent layer, this pressure dip should tend to unity \citep[i.e. disappear, see ][]{FieldingFractal20}. In any case, this relationship follows the regulatory pattern we set out at the beginning of this section: if $\voutavg$ increases not only does $\Phot$ decrease (through the bubble structure relationship laid out in \autoref{app:structure}) but the pressure dip also gets more extreme, leading to an \textit{even smaller} $P(\Tpk)$ and thus much decreased cooling. The opposite is  true for decreased values of $\voutavg$.

If we accept this final relation as empirical, then it closes the loop on our causality argument for the dynamics of the bubble: $\voutavg$ adjusts so that the pressure in the cooling gas produces a fixed $\edotpk$ given the volume of cooling gas $\Vpk = \wcool \Abub$. This adjustment of $\voutavg$ is not only dependent on how $\voutavg$ relates to $\Phot$ but also how it relates to $\fdip$. This means the detailed value of $\voutavg$ has a non-trivial dependence on the details of the Riemann solver and cooling, which determine $\fdip$. With $\voutavg$ determined, $\Phot$ then sets the dynamics of the bubble. In particular, the fact that the whole dynamics problem comes down to a question of energy regulation through cooling explains why the resulting momentum is independent of $\pdotw$ but has the same value as long as $\Lwind$ is the same, as we saw in \autoref{fig:pdot_comp}. That is, after $\Phot$ adjusts such that the cooling rate matches the input energy rate in \autoref{eq:edot_cool}, the same $\Phot$  sets the expansion rate through \autoref{eq:bubble_radial_force} (see \autoref{fig:momentum_valid}).

Given the complexity of the picture laid out above, and the fact that in the end it is dependent upon an empirical relationship between $\voutavg$ and $\fdip$, it is still quite possible that physical turbulent dissipation plays some role in the setting of $\voutavg$ as well. We will return to this in \autoref{subsec:vout_turb}.

Our detailed analysis has produced significant insight into bubble dynamics and the reasons for resolution dependence in numerical simulations, but a number of mysteries remain. In particular, the relationship between the exact fraction of energy lost in the cooling layer, the fractional dip in the pressure, $\fdip$, and $\voutavg$ remains uncertain. These issues are likely  explained by the detailed nature of numerical dissipation at the WBB interfaces and how this numerical dissipation is driven or modified by the presence of strong cooling and a strong mean velocity. As these details are not dependent upon the three-dimensional, dynamical nature of the simulations at hand, they are more likely to be made clear through the investigation of much simpler, one-dimensional simulations, where the parameters of the problem can be more easily tuned by hand. We leave this detailed investigation for future work.

\begin{figure}
    \includegraphics[width=\columnwidth]{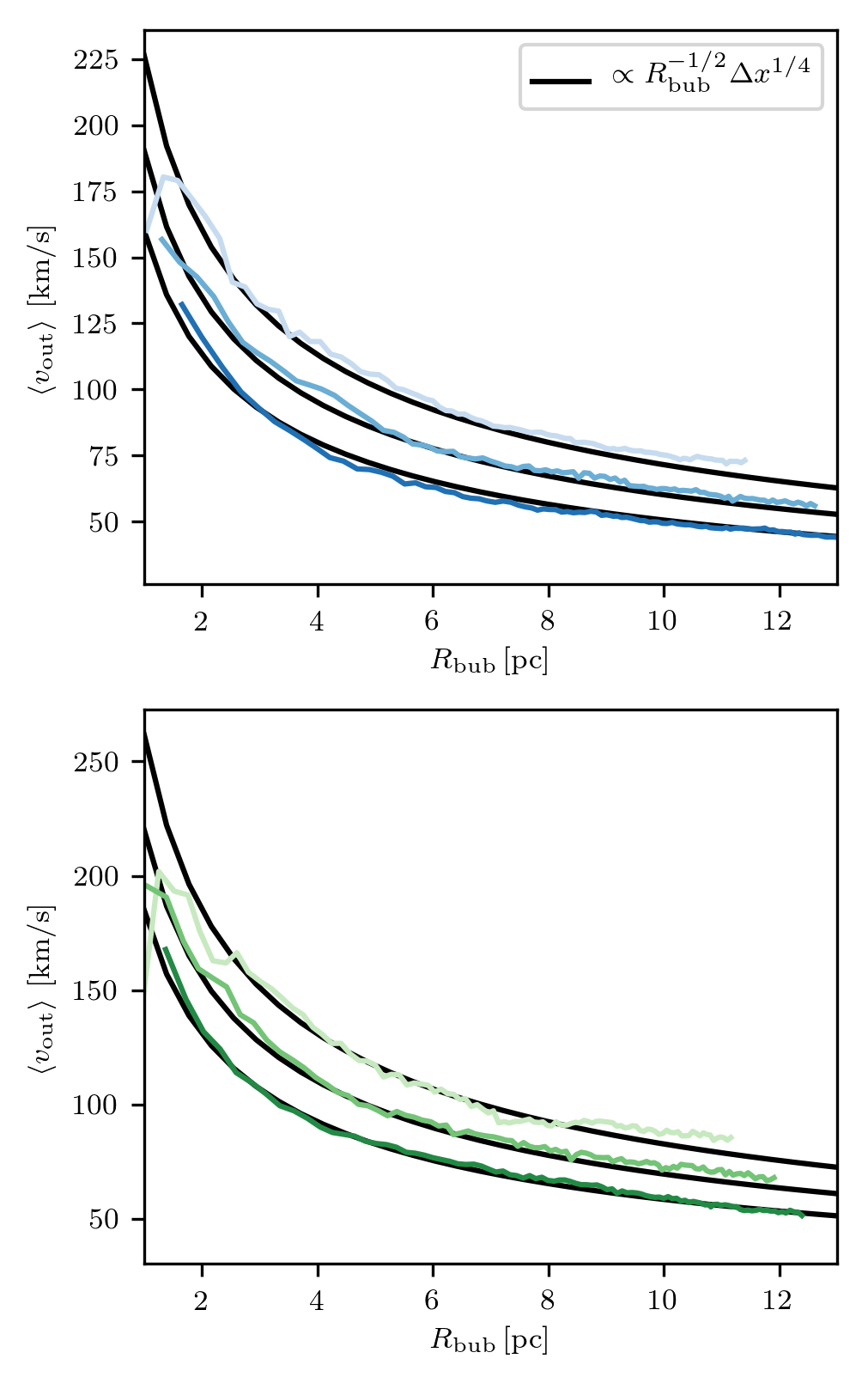}
    \caption{The mean outflow velocity $\voutavg$ as a function of $\Rbub$ for each of the simulations at different resolution. Both HD (top panel) and MHD (bottom panel) are shown. The empirical scaling relation of \autoref{eq:vout_scale_empricical} with a chosen norm (equal to $302,\, 350\, {\rm km/s}\, {\rm pc}^{1/4}$ in the HD and MHD simulations respectively) is shown as the black lines. Colored line styles for different resolution are as indicated in \autoref{fig:mhdVhydro}.}
    \label{fig:vout_geometry}
\end{figure}

Finally, we have found an empirical relation for $\voutavg$ based solely on $\Rbub$ and $\delx$ as
\begin{equation}
    \label{eq:vout_scale_empricical}
    \voutavg \propto \Rbub^{-1/2} \delx^{1/4} \, .
\end{equation}
The success of this empirical scaling is shown in \autoref{fig:vout_geometry}. We will additionally note that there is no hidden dependence in the above scaling relations on $\pdotw$, $\Vwind$, or $\mdotw$: the ``\texttt{Mwmod}" simulations outlined in \autoref{tab:simulations} have nearly identical values of $\voutavg$ to their unmodified counter-parts.

We emphasize that this empirical scaling of $\voutavg$ should be understood as describing simulations where the cooling length in the boundary layer is not resolved, so instead diffusion is numerical. Additionally, in the resolved case we would expect $\voutavg$ to depend on turbulence and/or the true thermal conductivity, as discussed in \autoref{subsec:momentum_resolved}.

\subsection{A Geometry-Driven Solution}
\label{subsec:geo_driven}

Combining the arguments from \autoref{subsec:Abub_scaling} and \autoref{subsec:voutavg_explained} with \autoref{eq:alphap_derive} we can derive a direct relationship between the momentum enhancement factor, $\alpha_p$, the geometric properties of the interface, and the numerical properties of diffusion across the interface. In particular, accepting a certain fractal dimension, $d$, that describes the bubble's surface
\begin{equation}
    \Abub \propto \Rbub^{2 + d} \delx^{-d} \, ,
\end{equation}
along with \autoref{eq:vout_scale_empricical} and \autoref{eq:alphap_derive}, we arrive at
\begin{equation}
    \label{eq:alphap_geometry}
    \alpha_p \propto \Vwind \Rbub^{1/2 - d} \delx^{d - 1/4} \, .
\end{equation}
If we then apply $d = 1/2$ for the HD simulations and $d= 1/4$ for the MHD simulations, as seems reasonable from \autoref{subsec:Abub_scaling}, we arrive at
\begin{equation}
    \label{eq:aphd}
    \alpha_{p,{\rm HD}} \propto \Vwind \delx^{1/4}
\end{equation}
\begin{equation}
    \label{eq:apmhd}
    \alpha_{p,{\rm MHD}} \propto  \Vwind \Rbub^{1/4} \, .
\end{equation}
If we instead apply $d= 1/5$, as seems appropriate at late times in the MHD simulations (see \autoref{fig:frac_measure}), then we have $\alpha_{p,{\rm MHD}} \propto \Vwind \Rbub^{3/10} \delx^{-1/20}$. We emphasize again that the scalings in \autoref{eq:aphd} and \autoref{eq:apmhd} are intended as an interpretation of our current simulations, in which numerical diffusion affects the value of $\voutavg$ as empirically characterized in \autoref{eq:vout_scale_empricical}.  For the fully resolved case, there could be no dependence of $\alpha_p$ on $\Delta x$.  

These relations elucidate several mysteries:
\begin{itemize}
    \item[1.] The ``momentum-conserving''-like behavior of the HD simulations, given the lack of $\Rbub$ dependence in \autoref{eq:aphd}, as well as the very weak dependence on resolution.

    \item[2.] As all of the above relations are linearly dependent on $\Vwind$, the resulting output momentum is $\dot{p}_r \propto \alpha_p \pdotw \propto \Lwind \Rbub^{1/2 - d} \delx^{d - 1/4}$, this explains the lack of $\pdotw$ (or $\Vwind$) dependence in \autoref{fig:pdot_comp}. In fact, this is explicitly \textit{not} a ``momentum-driven'' solution, but an energy driven solution with losses to cooling that are large and increase over time. We should note that there is likely some implicit additional dependence of $\voutavg$ on $\Lwind$ (and possibly $\Vwind$) given the causal structure of how $\voutavg$ is set, which we have not accounted for here. This dependence is apparent in the $\alpha_p$ dependence on $\Lwind$ in \citet{Lancaster21b} (see Figure 8).

    \item[3.] The different momentum scaling of the MHD simulations, which have explicit $\Rbub$ dependence, and their opposite resolution dependence behaviour in \autoref{fig:pdot_comp} from the HD models. This is apparent especially at late times where the fractal dimension is much reduced and the scaling of $\alpha_p$ with $\delx$ becomes a negative power.
\end{itemize}

\begin{figure*}
    \includegraphics[width=\textwidth]{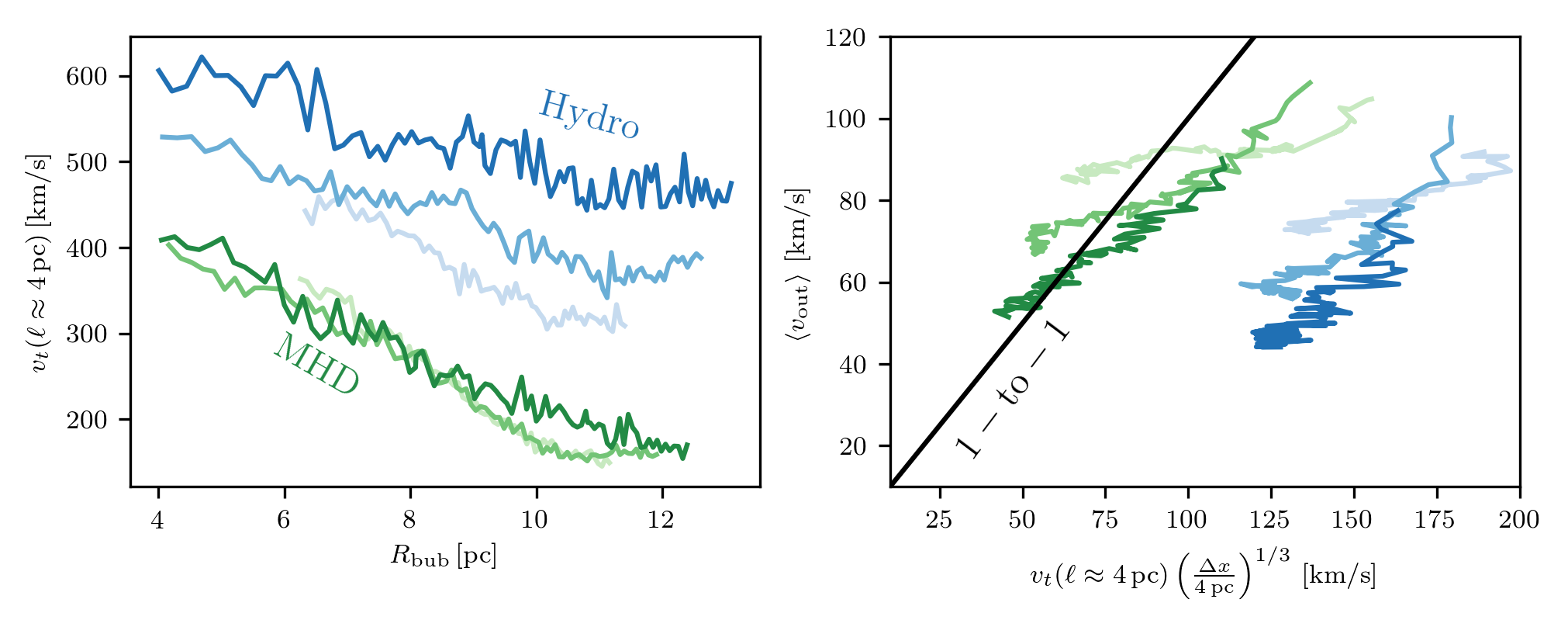}
    \caption{Turbulence in the hot, shocked wind. Line styles are as in \autoref{fig:mhdVhydro}. \textit{Left Panel}: We show the second order structure function of the turbulent velocity field, $v_t$ (details in text), on a scale $\ell = 4\, {\rm pc}$, which is roughly the energy containing scale of the turbulence across time in each of the simulations, as a function of the bubble's radius. We only show data for where the turbulence at this scale is well resolved: $\Rbub > 4\, {\rm pc}$ and $\Rbub > 20\, \delx$. \textit{Right Panel}: $\voutavg$ vs. the turbulent velocity, adjusted to match the predicted turbulent velocity on the resolution scale. We assume a power-law scaling of the turbulent velocity field of $p = 1/2$, which is consistent with the strucutre functions in our simulations, likely due to the fact that the turbulent dynamic range is not very well resolved, further details in text. We show the 1-to-1 relationship between $\voutavg$ and the measure of turbulence as a black line.}
    \label{fig:vt_vs_vout}
\end{figure*}

\subsection{Relation Between $\voutavg$ and Turbulence}
\label{subsec:vout_turb}

In the TRML literature it has often been shown that the amount of cooling in the interface layers is straightforwardly dependent upon the amount of turbulence in the mixing layer \citep{FieldingFractal20,TOG21,DasGronke23}. Indeed, in \citet{Lancaster21a} and \citet{Lancaster21b} the amount of turbulence in the hot, shocked wind gas is causally tied to the amount of mixing, and therefore dissipation, at the wind bubble's interface. While we will discuss nuances in this connection in \autoref{subsec:past_results}, it is worth reporting the relationship between turbulence and dissipation in the simulations at hand.

We quantify turbulence in our simulations by measuring the second-order structure function of the non-radial velocity field in the hot, shocked wind. We do this in exactly the same manner as described in Section 4.5 of \citet{Lancaster21b} except that we restrict solely to the $T > 10^6\, {\rm K}$ gas\footnote{In \citet{Lancaster21b} all gas with $v_r < \Vwind/2$ and $T>2\times 10^4\, {\rm K}$ was used.}. To give a brief description of the process here: we randomly choose 400 points in the shocked wind volume and calculate the average norm of the difference in the non-radial velocity field as a function of distance, $\ell$, away from the chosen point. We then use all 400 samplings of the square deviation as a function of distance to compute the mean over the samples and then take the square-root. This gives us $v_t(\ell)$, the second-order structure function as a function of scale.

In the left panel of \autoref{fig:vt_vs_vout} we show $v_t(4\, {\rm pc})$, which we found to be approximately the energy containing scale for the turbulence across all of the simulations, as a function of $\Rbub$. We see that, while the turbulent velocity is relatively well converged in the MHD simulations, it steadily increases with resolution in the HD simulations\footnote{We note that the turbulent velocity measured here is larger than in \citet{Lancaster21b} likely due to both (i) our measurement being restricted to higher temperature gas and (ii) $\Vwind$ being a factor of $\sim 1.8$ larger in these simulations.}. This is likely due, at least in part, to decreased numerical viscosity at higher resolution, allowing for a better resolved dynamic range in the HD simulations. In contrast, small-scale turbulence is likely suppressed in the MHD simulations due to the stabilization of instabilities from the magnetic field. If this small-scale suppression is well-resolved in even the lowest resolution simulations, then we would not expect a better resolved dynamic range, and therefore increased turbulence, at higher resolution.

There is an additional effect that can have a significant impact on the resolution dependence of turbulence in the HD simulations. As we can see in \autoref{fig:mhdVhydro}, in the HD simulations we have lower $\alpha_p$ at higher resolution. Through \autoref{eq:bubble_force2}, this implies a smaller ratio of $\Rbub/\Rfree$, which in turn implies more direct contact between the free-wind and dense clumps in the background cold gas. This process is directly observable in \autoref{fig:simplot}. This direct contact between the dense gas and the free wind creates oblique bow shocks which seed vorticity, and hence turbulence, in the shocked wind material. Indeed, such bow-shocks were posited as a main source of turbulence in the hot gas in \citet{Lancaster21b}. However, we still expect turbulent mixing at the interface, seeded by instabilities, even in the absence of this effect.

It is also interesting to compare
the turbulence and dissipation at the bubble's surface, quantified by $\voutavg$. To this end we take the energy-containing scale turbulence, which we believe to be well resolved, and extrapolate it to the expected turbulent velocity at the resolution scale using $v_t(\delx) = v_t(L) (\delx/L)^{1/3}$, the expected scaling for Kolmogorov turbulence. We can see from the right panel of \autoref{fig:vt_vs_vout}, where we compare $\voutavg$ with this scaled turbulent velocity, that they have a close to linear relation with one another, as one would expect from recent work in TRMLs \citep{FieldingFractal20,Lancaster21a,TOG21,DasGronke23}. It is encouraging that this relationship appears more linear in the higher resolution simulations. Thus, while our current simulations are not fully resolved, this result suggests that dissipation mediated by turbulence could be achieved with an enhancement in resolution.  The right panel of Figure 2 also indicates that at least for the pressures present at late time, we are close to being able to resolve $\lcool$.

\section{Discussion}
\label{sec:discussion}

\subsection{Relation to Past Work}
\label{subsec:past_results}


A key thesis of this work, laid out in \autoref{sec:theory} and \autoref{sec:results}, is that the dynamics of WBBs are intimately dependent upon the movement of energy across their bubble-shell interfaces. In particular, we have emphasized how this depends both on the multi-scale \textit{geometry} of these interfaces and the small-scale \textit{dissipation} processes that act across them. These two effects acting together make for a particularly difficult physical problem.

If we ignore geometry, considering one-dimensional problems, the role of dissipation across such interfaces has long been solved under the further assumptions of time-steady, isobaric flows \citep{ZDPN69,CowieMcKee77,Inoue06,KimKim13}. This is unfortunately not the picture that applies to WBBs. The full complications of the problem were circumvented in \citet{Weaver77} by ignoring the contribution of cooling at the interface, deriving a solution that required all conductive heat flux to be balanced by evaporation of mass from the shell into the hot, wind bubble.

\citet{ElBadry19} ameliorated this picture by running spherically symmetric, radiative, one-dimensional simulations of super-bubble evolution with both explicit Spitzer conductivity ($\propto T^{5/2}$) and a constant diffusivity at low temperatures, $\kappa_{\rm mix}$ (meant to account approximately for turbulent mixing) and showed that a predictable, constant fraction of the wind energy is lost to cooling within the interface layer, $\dot E_\mathrm{cool} \propto \kappa_{\rm mix}^{1/2}$ \citep[see Eq. 36 of][]{ElBadry19}. This essentially solves the problem from a `purely diffusive', one-dimensional picture, albeit by appealing, in an effective way, to a mechanism of diffusion that is inherently three-dimensional: turbulence. We note that the problem, of theoretical interest no doubt, that appeals to only Spitzer-like heat dissipation in the one-dimensional context but includes cooling in the dynamical WBB picture has not been fully resolved \citep[in the numerical sense, see Appendix A of][]{ElBadry19}.

Turbulence spans both sides of the divide we have drawn here between dissipation and geometry, and it is worth emphasizing which aspects of turbulence lie on either side in the context of these interfaces. One can treat turbulence purely as a micro-physical dissipation, as it is treated in \citet{ElBadry19} and countless other astrophysical applications \citep[e.g. ][]{Smagorinsky63,Ropke07}. However, the \textit{root} of this `effective' micro-physical dissipation is itself geometry: turbulence is \textit{stretching-enhanced diffusion}, it only acts to increase the surface area over which true micro-physical dissipation may act. As we discussed in \autoref{subsubsec:vout_turb_theory}, the scale on which turbulence effectively becomes micro-physical for these interfaces, is the cooling length, $\lcool$, at which point cooling effectively `stands still' as the turbulent cascade efficiently mixes material. Above this scale, larger scale turbulence acts to significantly enhance the surface area over which energy transport takes place, while being too slow at mixing to be of actual importance in mediating that energy transport. This is the geometric role that it serves. The distinction between these two regimes is essentially about the ratio of cooling time to the eddy-turnover time on the largest scales (quantified by $\xi$ in \citet{FieldingFractal20} and the Damk\"{o}hler number, ${\rm Da}$ in \citet{TOG21}).

While \citet{ElBadry19} simply treated $\kappa_{\rm mix}$ as a parameter, \citet{Lancaster21a} linked its expected value to predictions from the theory of \citet{FieldingFractal20} discussed in \autoref{subsubsec:vout_turb_theory}. Folded into this theory, and the model for diffusivity given in \citet{Lancaster21a}, is the geometric role that turbulence plays, accounted for using an additional factor for the increase in fractal area as a bubble expands. While, in this case, the energy lost to cooling would no longer necessarily be a constant fraction of the energy input rate, modifying the WBB expansion solution, the evaporative flow and temperature of the hot medium could still be computed following \citet{ElBadry19}.

The true complexity of this problem is realized when we allow resolved dissipation to act in concert with complicated geometry. Here the geometry changes the temperature and velocity structure of the flow in three dimensions, changing the nature of the conductive heat transport. In particular, while either condensation or evaporation is allowed exclusively in one-dimensional problems, in this complicated three-dimensional picture, both could occur simultaneously.

The largest further complication to this picture is the presence of magnetic fields, which alter the turbulent cascade \citep{MHDTurbulence22,Grete23}, reduce the strength of physical conduction processes \citep{SpitzerHarm53,Spitzer62,Meinecke22}, and alter the geometric enhancement of surfaces, as we have demonstrated here. While the MHD picture of TRMLs has begun to be investigated in recent numerical works \citep{DasGronke23,ZhaoBai23}, the picture of the roles of physical versus numerical dissipation in these simulations still appears unclear.

In the present work, we have emphasized that the resolution of the relevant micro-physical diffusive scales (outlined in \autoref{subsec:diss_scales}) is important to accurately capture the dynamics of WBBs. We have presented a detailed study of the origin of the numerical diffusivity in these simulations, as well as why it has a weak scaling with resolution and leads to momentum-driven-like behavior (summarized in \autoref{subsec:geo_driven}). In particular, the decrease in dissipation at higher resolution is accomplished by a near equal increase in surface area, making the net result the same in our MHD simulations. At the same time, from the results of \autoref{subsec:vout_turb}, it is likely that turbulent dissipation may play some role in the setting of $\voutavg$ in our simulations.

These results may also provide an explanation for the lack of resolution dependence in simulations of Kelvin-Helmholz mixing with cooling, even when $\lcool$ is not resolved. In particular, for the Kelvin-Helmholz problem in which there is not a strong mean velocity field flowing into the interface (unlike the present situation), we would expect the numerical dissipation velocity to scale as $\delx^{1/2}$, as was demonstrated for one-dimensional problems in \citet{TOG21}. Similarly, as is demonstrated in \citet{FieldingFractal20} and here, we would expect the area of the fractal interface to scale as $\delx^{-1/2}$. These effects perfectly balance one another in this case. We emphasize that the Kelvin-Helmholz problem is quite different from the present case, where there \emph{is} a strong flow into the mixing layer, with a comparable effective velocity of dissipation.

Finally, what does the numerical dependence demonstrated here tell us about our ability to use numerical simulations to predict the evolution of WBBs? In \citet{Lancaster21a,Lancaster21b} we described the root of the ``momentum-driven''-like behaviour present in both \citet{Lancaster21b} and the simulations presented here as due to extreme cooling, with $\alpha_p$ approaching its lower bound of unity. We argued that the strong cooling was enabled by turbulence-driven mixing of hot gas into denser material at the interface with the shell.

What we have demonstrated in this work is that the turbulence, while likely playing a role in the enhancement of the fractal structures in our simulations, is not the only, or necessarily dominant, source of diffusivity. This is clearly evidenced by the resolution dependence of the total momentum carried by the bubble, as laid out in \autoref{fig:pdot_comp}, as well as the resolution dependence of $\voutavg$ (laid out in \autoref{subsec:voutavg_explained}), which plays a major role in setting the bubble's interior pressure.

While numerical dissipation is playing a role in setting $\dot{p}_r$ in the current simulations, given the available resolution, this does not change the overall physical picture for WBB evolution outlined in \citet{Lancaster21a}. As discussed in \autoref{subsec:diss_scales} and above, in the real world turbulence should play a major role in both enhancing the surface area of the interface and in the transport that leads to intermingling of hot and cool gas at small scales. The main difference between \citet{Lancaster21a} and the picture suggested by this work is that this evolution need not be ``momentum-driven" in the $\alpha_p = 1$ or even $\alpha_p=const.$ sense, but rather could have $\alpha_p$ that varies with time, as suggested by \autoref{eq:alphap_derive}.

However, as we showed in \autoref{subsec:momentum_resolved}, we can use \autoref{eq:alphap_derive} to derive the momentum scaling that we would expect in WBBs where dissipation at the interface is mediated through turbulent mixing (described in \autoref{eq:vout_turb}), and the overall bubble geometry is given by a fractal geometry. If we apply a Kolmogorov scaling for the turbulence and an excess fractal dimension of $d=1/2$ (as seems relevant for our hydro simulations) this results in an $\alpha_p$ which is constant in time (``momentum-driven'') with an order unity value for the parameters relevant to the simulations described here. This validates and lends further meaning to the physical picture laid out in \citet{Lancaster21a}.

\subsection{Prospects for Future Work}
\label{subsec:future_work}

While the argument laid out in \autoref{subsec:momentum_resolved} resolves the issue of the evolution of WBBs whose interiors can be thought of as quasi-steady (details in \autoref{app:structure}) and whose turbulent properties and fractal geometry are known, several open questions remain. First among these is when these conditions may apply. Since the relevant speeds involved in setting the interior and boundary conditions -- the sound speed and flow speeds within the bubble and the turbulent velocities in the mixing layer -- are large compared to the bubble expansion velocity, a quasi-steady state should be easy to establish.

The question of the nature of energy transport and geometry at turbulent interfaces has some solid theoretical basis \citep{FieldingFractal20,TOG21}. However, some open questions remain as to the details of the resolution (in)dependence of the results obtained from these simulations and the interpretation of the scaling of the total cooling with the relevant physical parameters ($\tcool$, $v_t(L)$, etc.). Understanding the nature of the resolution dependence of these simulations is, in our view, essential to a confident theoretical model for dissipation in these systems. To settle this issue it is essential that future numerical work spatially resolves the dynamics of these interfaces while including explicit, physically motivated, diffusive properties \textit{and} demonstrating that physical diffusion is dominant over numerical diffusion.

This problem is made even harder by the ubiquitous presence of magnetic fields in astrophysical fluids, which these interfaces are likely not exempt from. While this problem has begun to be explored numerically \citep{DasGronke23,ZhaoBai23}, questions of numerical dependence still remain. The magnetic field not only changes the geometry of these interfaces, as we have demonstrated here, but also reduces the strength of thermal conduction \citep{RobergClark16,Komarov18,Drake21,Meinecke22}. If magnetic fields act to significantly dampen the turbulent diffusivity, through altering the cascade at these interfaces by stabilizing small-scale instabilities, as seems likely from our study as well as \citet{ZhaoBai23} and \citet{DasGronke23}, this makes the resolution requirements even stricter. Numerical studies that resolve the dynamics and dissipative processes, especially in the presence of magnetic fields, will likely be quite costly but are absolutely necessary.

Even if we have an understanding of the small-scale diffusive processes mentioned above, we still need a way of accurately accounting for them in larger-scale simulations where we have little hope of resolving the appropriate scales. While one way forward is to potentially include a modified diffusivity that accounts for this unresolved physics \citep{ElBadry19,Lancaster21a}, one still needs to ensure that the numerical diffusivity is less than this new, modified diffusivity. One promising avenue for ensuring this is to explicitly separate the high and low entropy components of the fluid in a subgrid manner \citep{arkenstone1,WeinbergerHernquist23,Butsky24}. The validity of this approach is subject to the details of the coupling between these subgrid components, which must be tuned to match the understanding gained from resolved interface simulations.

\section{Conclusions}
\label{sec:conclusions}

We end with a summary of our main conclusions:
\begin{itemize}
    \item[1.] We demonstrate that the evolution of a WBB is governed by boundary conditions at its surface (see \autoref{subsec:structure_highlights} and \autoref{app:structure}). These boundary conditions are a combination of \textit{geometry}, embodied by the area of the bubble's surface, $\Abub$, and \textit{dissipation}, embodied by the interface's ability to absorb high-entropy gas, $\voutavg$, as seen in \autoref{eq:alphap_derive}. Increased $\Abub$ gives more surface area over which energy can be dissipated, reducing the degree to which momentum is enhanced above that input by the wind, $\alpha_p$. At the same time decreased $\voutavg$ limits losses at each point on the surface, increasing $\alpha_p$.
 
    \item[2.] We demonstrate that solutions with constant $\alpha_p$ are not necessarily ``momentum-dirven'' (see \autoref{fig:pdot_comp}). Our past work suggested, with evidence from numerical simulations, that turbulent mixing at the bubble's interface leads to efficient enough cooling that these WBBs always enter a $\alpha_p\approx 1$ ``momentum-driven" limit, \autoref{fig:pdot_comp} demonstrates that this is not actually a ``momentum-driven" evolution, as is also apparent from a fair interpretation of \autoref{eq:alphap_derive}. In \autoref{sec:results} we showed how this apparent ``momentum-driven''-like behaviour is simply determined by the relative scaling of $\Abub$ and $\voutavg$ both in time and with resolution, $\delx$. In particular, differences in these scaling relations between the HD and MHD simulations, due mostly to differences in $\Abub$ from differing fractal structures, can explain the opposite convergence behaviour apparent in these simulations, as seen in \autoref{fig:mhdVhydro}. This is summarized in \autoref{subsec:geo_driven}.

    \item[3.] In analyzing the scaling of $\Abub$, \autoref{subsec:Abub_scaling} demonstrated, using three different measurement techniques, how the surfaces of the simulated WBBs are well described as fractals with an excess fractal dimension of $d=1/2,\, 1/4$ for the HD and MHD simulations respectively. The fractal dimension of the MHD WBBs is also clearly time-dependent and this is likely related to the growth of magnetic pressure in the shell of the WBB as the bubble evolves. This strong magnetic pressure both halts dynamical instabilities such as Rayleigh-Taylor and Kelvin-Helmholtz but also creates a `thick' magnetized shell which also suppresses `thin-shell' Vishniac instabilities \citep{Vishniac83}.

    \item[4.] In analyzing the scaling of $\voutavg$,  \autoref{subsec:voutavg_explained} demonstrated how the resolution and time dependence of $\voutavg$ is likely set by a consistency condition (discussed in \autoref{subsec:vout_cooling}) on the cooling occurring in the interface for `steady' wind bubbles. The value of $\voutavg$ is then determined by the volume of gas that is cooling rapidly, $\Vpk = \Abub \wcool$, and the pressure of the rapidly cooling gas, $\Ptpk$, which, for isobaric layers where $\Ptpk = \Phot$, is also dependent on the value of $\voutavg$ through how $\voutavg$ acts as a boundary condition on the bubble's interior structure. The straightforward relationship outlined in \autoref{eq:vout_cool} is somewhat complicated in the simulations due to a dip in the pressure at the peak cooling gas, leading to a prediction following \autoref{eq:vout_cool_scaling}. If the turbulence were fully resolved, $\voutavg$ would be independent of numerical details. However, at current resolutions, $\voutavg$ and the dip in pressure, $\fdip$, seem to be related through the details of how the Riemann problem is solved at the interface (see \autoref{fig:vout_scaling} and \autoref{app:riemann}) this does not lead to a straightforward prediction for the scaling of $\voutavg$. To circumvent this we give an empirical relation for the dependence of $\voutavg$ on the bubble's size, $\Rbub$, and the physical resolution, $\delx$ presented in \autoref{eq:vout_scale_empricical} and \autoref{fig:vout_geometry}.

    \item[5.] The explicit dependence on numerical resolution in our simulations, detailed above, is only due to the fact that the relevant dissipative scales (\autoref{subsec:diss_scales}) are not fully resolved in our simulations (see \autoref{fig:scales}). While running a simulation that that resolves the length scales of interest here is prohibitively expensive, in \autoref{subsec:momentum_resolved} we used our understanding of the momentum scaling of bubbles derived in \autoref{app:structure} along with our understanding of the turbulent dissipative physics discussed in \autoref{subsubsec:vout_turb_theory} to derive the momentum evolution we would expect if our simulations did resolve the relevant dissipative scales. The result, given in \autoref{eq:alphap_specific}, suggests that resolved simulations are expected to exhibit a ``momentum-like'' scaling with $\alpha_p$ of order unity for the parameters relevant to the problem explored here. This further validates the physical picture drawn in \citet{Lancaster21a}.
\end{itemize}

In \autoref{sec:discussion} we provide an in-depth discussion of the consequences of these results in the context of past work. In particular, in \autoref{subsec:past_results} we emphasize one key point: numerical dissipation is not physical dissipation and this can lead to important differences when the relevant dissipative scales are not resolved. A corollary of this discussion is our speculation that the lack of resolution dependence seen in simulations of the Kelvin-Helmholz mixing layers with cooling \citep{FieldingFractal20,TOG21}  -- even when dissipative scales are not resolved -- is due to a conspiracy between the relative scaling of the fractal structure of the layer ($\Abub \propto \delx^{-1/2}$ for $d=1/2$) and the numerical dissipation ($\propto \delx^{1/2}$ see \citet{TOG21} Section 4). Confirmation of this suggestion remains to be shown clearly in the mixing layer context.

We further emphasize that future work should investigate the effects of temperature dependent, Spitzer-like conductivity in 3D numerical simulations, which can potentially lead to mass-loading of the bubble's interior (evaporative flows) combined with net cooling in the interface layer, as is suggested by the 1D simulations of \citet{ElBadry19}. While the bubble expansion dynamics may be unchanged, the thermal state has important observational consequences for these systems. In particular, X-ray emission from these bubbles \citep{Townsley03,Townsley11,Lopez11,Lopez14,Rosen14,Sasaki23,Webb24} is likely dominated by the intermediate temperature gas that exists as a result of such evaporative flows \citep{ChuMacLow90,ParkinPittard10,ToalaArthur18}.

\acknowledgments

The authors would like to thank Drummond Fielding, Romain Teyssier, Max Gronke, Ulrich Steinwandel, Peng Oh, Brent Tan, Hitesh Das, Jim Stone, Mordecai-Mark Mac Low, Laura Lopez, Anna Rosen, and Michael Grudi\'{c} for insightful discussions. L.L. gratefully acknowledges the support of the Simons Foundation under grant 965367. This work was also partially supported by grant No. 510940 from the Simons Foundation to E. C. Ostriker. C.-G.K. acknowledges support from a NASA ATP award No. 80NSSC22K0717. J.-G.K. acknowledges support from the EACOA Fellowship awarded by the East Asia Core Observatories Association. G.L.B. acknowledges support from the NSF (AST-2108470 and AST-2307419, ACCESS), a NASA TCAN award, and the Simons Foundation through the Learning the Universe Collaboration. The numerical work of this paper would not have been possible without the computational resources provided by the Princeton Institute for Computational Science and Engineering (PICSciE) and the Office of Information Technology’s High Performance Computing Center at Princeton University.

\software{
{\tt scipy} \citep{scipy},
{\tt numpy} \citep{harrisNumpy2020}, 
{\tt matplotlib} \citep{matplotlib_hunter07},
{\tt adstex} (\url{https://github.com/yymao/adstex})
}

\appendix

\section{Fluid Structure of The Bubble Interior}
\label{app:structure}

In this appendix we focus on the structure of the fluid in the wind bubble interior -- i.e. the free and shocked winds. This generalizes the analysis given in Appendix A of \citet{Lancaster21a} and provides more justification and explanation of the assumptions.

The free-wind structure can be thought of as the limit of the \citet{ChevalierClegg85} wind solution far outside the ``injection region" of mass and energy. This region is steady and super-sonic at constant velocity $\Vwind$, with negligible thermal energy. The only  relevant parameter of the fluid that varies is its density, which is determined by mass conservation as
\begin{equation}
    \rho(r) = \frac{\mdotw}{4\pi r^2 \Vwind} \, .
\end{equation}
The free-wind ends at the shock surface where the Rankine-Hugoniot jump conditions tell us the post-shock fluid variables as
\begin{equation}
    \label{eq:rhpos}
    \rhops = \rho_{\rm hot} = 4\rho_{\rm free}(\Rfree) = \frac{\mdotw}{\pi \Rfree^2 \Vwind}  \, ,
\end{equation}
\begin{equation}
    \label{eq:vps}
    \vps = \frac{\Vwind}{4}
    \, ,
\end{equation}
and
\begin{equation}
    \label{eq:Pps}
    \Pps = P_{\rm hot} = 3 \rhops \vps^2 = \frac{3\pdotw}{16\pi \Rfree^2}
    \, .
\end{equation}
In \autoref{eq:rhpos}--\autoref{eq:Pps} we have treated the free wind as a spherical volume of radius $\Rfree$ bounded by a perpendicular shock. More generally, one may allow for oblique wind shocks with $\mu_\mathrm{f} \equiv \foldedness_{\rm free}$ the mean obliquity of the free wind's surface, in which case \autoref{eq:rhpos}--\autoref{eq:Pps} respectively have additional factors $\left(4 - 3\mu_\mathrm{f}^2\right)^{-1}$, $\left( 4- 3\mu_\mathrm{f}^2\right)$, and $\mu_\mathrm{f}^2$ on the right-hand side of each equation. With weighting by solid angle, it is straightforward to show that $1/4 \le \mu_\mathrm{f} \le 1$.

With the above strong-shock jump conditions the shocked wind will be subsonic with Mach number $\mathcal{M} \approx 1/\sqrt{5}$, with $\mathcal{M}$ decreasing as we go outward in to the shocked wind (as we will see below). Having restricted to the case of adiabatic shocked winds, we can make progress by assuming that this region is isobaric.
To justify the isobaric assumption we need to show that the sound-crossing time for the shocked wind, $\tschot \approx \Rbub/\cshot$ is much smaller than the expansion time of the bubble $\texp \equiv \Rbub/\dot{R}_{\rm bub}$, or equivalently $\cshot \gg \dot{R}_{\rm bub}$. Since $\cshot\approx 0.6\Vwind$, this is equivalent to assuming that $\dot{R}_{\rm bub} \ll \Vwind$, which is equivalent to assuming that we've left the `free-expansion' stage, which we assume in the main text.

Given the assumed lack of spatial symmetry of the post-shock flow in the bubble interior, to make progress we must adopt a further simplifying assumption. We shall assume that the shocked wind flow is steady, with all partial time derivatives vanishing from Equations \ref{eq:mass_cons}, \ref{eq:momentum_cons}, and \ref{eq:energy_cons}. This is of course not strictly true: since the bubble is expanding, the structure has to evolve with time. Our steady assumption is then that the time-scale on which the fluid structure changes (or is non steady) $\tns\equiv \rho/|\frac{\partial \rho}{\partial t}|$ is long compared to the time-scale for structure in the flow to be established, which is again the sound-crossing time. We will justify \textit{a posteriori} that $\tns \approx \texp$ and thus, $\tns \gg \tschot$ holds.

The assumption of a (quasi-)steady flow is more restrictive than is readily apparent. In particular, for the time rate of change of the fluid variables to be fractionally small throughout the shocked wind, nearly all mass, energy, and momentum that enters through the source terms must be carried out of the bubble. Otherwise, material would build up within the bubble interior, changing the local fluid variables. This means that, under these assumptions, we cannot think of the edge of the bubble as a contact discontinuity, across which mass/density cannot pass. Instead, the steady assumption necessitates the presence of strong diffusive transport across the bubble's surface.

In the general case there would be small-scale turbulence within the shocked wind, which would break the steady assumption as the eddy-turnover time is small compared to the evolution time of the bubble. However, this turbulence will likely only be dominant at the surface of the bubble. Indeed, this turbulence should be one way in which the diffusion mentioned above is mediated.  Also, while individual turbulent eddies would evolve more rapidly than the bubble expands, the statistical properties of the turbulence may be assumed to evolve more slowly, consistent with the quasi-steady assumption. Thus, our {\it ansatz} is that there will be a turbulent boundary layer with quasi-steady properties that effectively impose boundary conditions on the interior of the WBB.

The structure of the shocked wind will then be determined by the evolution of an adiabatic, steady, subsonic flow \citep[e.g. Appendix A of][]{Lancaster21a}. The characteristics of such a flow along with the matching of boundary conditions at its inner and outer boundary surfaces will determine the extent of the volume. These boundary conditions will be the Rankine-Hugoniot relations at the interior shock and the matching of the velocity and pressure at the bubble's outer edges.

\begin{figure*}
    \centering
    \includegraphics[width=0.5\textwidth]{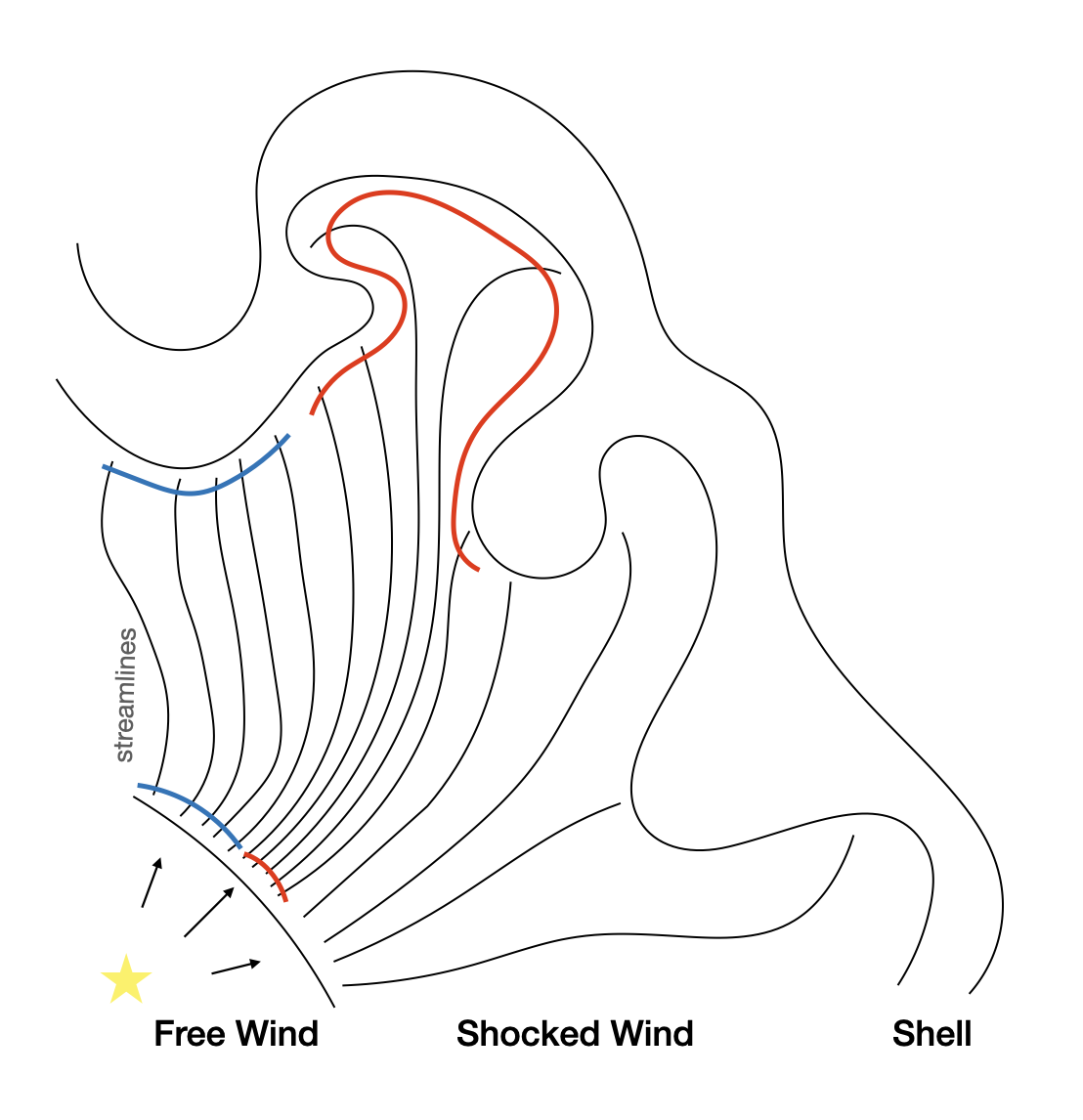}
    \caption{An illustration of  gas streamlines in the shocked wind in a general, non-spherical geometry. Consider the blue and red material surfaces that enter the shocked wind at its boundary with the free wind, each covering roughly similar areas and corresponding to five streamlines as shown. Following the streamlines through the shocked wind, we can see how each material surface is deformed as it approaches the interface between the shocked wind and the shell. In particular the red material surface is clearly much more deformed and has reached a much larger surface area than the blue surface. The dilation factor, $\varphi$, accounts for this deformation along with the `normal' increase in area of a material surface that one would get in a spherical geometry so that $\varphi$ is much larger for the `red' streamlines than it is for the `blue' streamlines.}
    \label{fig:streamline_divergence}
\end{figure*}

We now generalize the arguments from Appendix A of \citet{Lancaster21a} to remove the assumption of a radial outflow. Given this lack of symmetry, we consider properties of the gas on streamlines as a parcel of wind material moves from the shock front towards the contact discontinuity. Our assumptions allow us to give an adiabatic equation of state 
\begin{equation}
    \label{eq:adiabatic_eos}
    \frac{P}{\rho} = 
    \frac{\Pps}{\rhops}
    \left( \frac{\rho}{\rhops} \right)^{2/3}
\end{equation}
and a constant Bernoulli parameter along the flow line (steady, assuming no diffuisve transport across streamlines)
\begin{equation}
    \label{eq:bernoulli}
    \frac{1}{2}v^2 + \frac{5}{2}\frac{P}{\rho}  = \frac{1}{2}\vps^2 + \frac{5}{2}\frac{\Pps}{\rhops}  \, .
\end{equation}
The final relevant equation comes from steady flow along with conservation of mass (\autoref{eq:mass_cons}), which we will write here for any individual streamline as
\begin{equation}
    \label{eq:steady_mass}
    \frac{\rho}{\rhops} = \frac{\vps}{v \varphi}
\end{equation}
where $\varphi$ is the `area dilation factor' proportional to the area normal to a given flux tube, and $v$ is the magnitude of the flow velocity. For a radial flow we would have $\varphi = (r/\Rfree)^2$ as in \citet{Lancaster21a}, but for the more general case this accounts for how the area over which mass flux is spread can be much larger than the $\propto r^{2}$ dependence if we allow for large deformations, as illustrated in \autoref{fig:streamline_divergence}. 

As in \citet{Lancaster21a} we can put together \autoref{eq:adiabatic_eos}, \autoref{eq:bernoulli}, and \autoref{eq:steady_mass} along with the jump conditions to get
\begin{equation}
    \label{eq:vsteady}
    \frac{v}{\vps} = \varphi^{-1}
    \left(\frac{15}{16 - (v/\vps)^2} \right)^{3/2} \, .
\end{equation}
This technically has an analytic solution, but if we consider flow far into the shocked wind region, $v\ll \vps$, we can simplify the above to
\begin{equation}
    \label{eq:vprof}
    \frac{v}{\vps} = \left(\frac{15}{16}\right)^{3/2} \varphi^{-1} \, .
\end{equation}
Our assumption that we are `far' into the shocked wind $v \ll \vps$ is then equivalent to assuming $\varphi \gg 1$, which is even more true in the aspherical case.

It should be noted that a similar conclusion would have been drawn from assuming that the shocked wind is isobaric.  In this case, the Bernoulli equation is not used, but  the adiabatic assumption implies constant density and therefore (with \autoref{eq:steady_mass})
\begin{equation}
    \label{eq:steady_mass2}
    \frac{v}{\vps} = \varphi^{-1} \, .
\end{equation}
As we justified above, it is very reasonable to assume the shocked wind is isobaric, so we will use this even simpler version of the velocity profile. 

If $\shat = \mathbf{v}/v$ is the direction of a given streamline and $\nhat$ is the normal to the bubble surface with local area $dA_\mathrm{bub}$, the area element perpendicular to the streamline is $\shat \cdot \nhat\, dA_\mathrm{bub}=dA_\perp =\varphi d\Omega_\mathrm{f} \Rfree^2$, for $d\Omega_\mathrm{f}$ the corresponding solid angle element of the streamline at the free wind surface. We then have $\mathbf{v}\cdot \nhat\,   dA_\mathrm{bub} = v\, dA_\perp =\vps\, d\Omega_\mathrm{f} \Rfree^2$, where we have used \autoref{eq:steady_mass2}. Integrating over area for all streamlines in the fluid at the edge of the bubble, we arrive at
\begin{equation}
    \label{eq:boundary_condition}
    4\pi \vps\Rfree^2 = \left\langle  \mathbf{v}\cdot \nhat\right\rangle \approx \voutavg \Abub
     \, ,
\end{equation}
where
\begin{equation}
    \label{eq:voutavg_def_app}
    \voutavg \equiv\Abub^{-1} \int_{\Abub} \left(\mathbf{v} - \mathbf{W}_\mathrm{bub}\right)\cdot \nhat dA
\end{equation}
is the velocity normal to the bubble's surface, the frame co-moving with the surface ($\mathbf{W}_\mathrm{bub}$ is the local velocity of surface in the lab frame) averaged over that surface. 
Here, in defining the area $\Abub$ of the bubble surface (and the corresponding $\voutavg$), we must choose a `ruler' scale $\ell$ in the case that the surface is fractal. We will discuss this further in \autoref{app:frac}. If $d$ is the `excess dimension' of the fractal, we will have $\Abub = 4 \pi \Rbub^2 (\Rbub/\ell)^d$.

If we take $\voutavg$ and $\Abub$ as given at any time, we can see that the free wind radius, $\Rfree$, is then determined through \autoref{eq:boundary_condition}. We can then use the jump conditions (Eqs. \ref{eq:rhpos}, \ref{eq:vps}, and \ref{eq:Pps}) to infer the post-shock pressure based on $\voutavg$ and $\Abub$. Since the shocked wind is assumed isobaric, this provides the pressure of all of the hot, shocked wind as
\begin{equation}
    \label{eq:Phot_steady_app}
    \Phot = \frac{3}{8} \frac{\Lwind}{\voutavg \Abub} \, .
\end{equation}
We note that, if we had simply assumed that the enthalpy flux into the shell were all of $\Lwind$ the prefactor on the right-hand side of the above equation would be $2/5$ instead of $3/8$, which are the same within $\sim 6\%$. This difference arises from our matching of the outward enthalpy flux from the shocked wind ($5\Phot \voutavg \Abub/2$) to the enthalpy flux through the interior wind shock ($15\Lwind/16$), instead of $\Lwind$ itself. The former is more consistent with the assumption of steady mass flow combined with assumptions that the flow is and isobaric, so we use that instead here.

It is worth emphasizing what \autoref{eq:Phot_steady_app} is telling us and how exactly this condition is enforced within the dynamical picture of the bubble. Suppose, at a given point in the bubble's evolution with a fixed surface area, $\Abub$, that the velocity $\voutavg$ (as set by the boundary condition in the mixing layer) is high enough that the flux through the surface, $\Phot \voutavg$, is larger than the steady-state value as given in \autoref{eq:Phot_steady_app}, such that there is a greater rate of energy leaving the shocked wind than is being provided through the interior shock. This would lower the energy density and pressure within the bubble, and this de-pressurization would cause the interior wind shock to move outward. With larger $\Rfree$, a new equilibrium at lower pressure as set by \autoref{eq:Pps} could be established, which for the given $\Abub$ and $\voutavg$ would be compatible with a balance between energy flow into and out of the bubble. Conversely, if $\voutavg$ (as limited by turbulent mixing in the physical case) is low enough that the energy outflow from the bubble to the boundary layer is smaller than the energy inflow at the free wind surface, energy buildup and the resulting pressurization of the bubble interior will move $\Rfree$ inward.  Once a new steady state in the bubble interior is established, with pressure consistent with \autoref{eq:Pps} everywhere, the specific energy of the flow into the boundary layer will be high enough that the total losses can match the total gains.  In this way, under the assumption of a steady bubble, $\voutavg$ and $\Phot$ are necessarily inversely proportional to one another.

Using the projection of the radial component of the momentum flux on the bounding surface of the bubble, we may compute the rate of radial momentum leaving the bubble as
\begin{equation}
    \label{eq:app_pr1}
    \dot{p}_r = \Abub \left(\Phot\foldedness + 
    \rhohot \left\langle v_r (\mathbf{v}\cdot \nhat - W_\mathrm{bub}) \right\rangle \right) \, ,
\end{equation}
where $\mathbf{W}_\mathrm{bub} = \Wbub\nhat$ is the velocity of the bubble's surface, which, by the steady assumption, is much less than the gas velocity, so that the second term in the above is essentially $\left\langle v_r \vout\right\rangle$. Additionally, following \citet{Lancaster21b}, we define $\foldedness$ as the ``foldedness'' of the bubble: the average value of the dot product of the radial unit vector $\rhat$ with the unit vector normal to the bubble's surface $\nhat$.

Continuing from the above assuming $\Phot = \Pps$, $\rhohot = \rhops$ and using the shock jump conditions we have
\begin{equation}
    \label{eq:app_pr2}
    \dot{p}_r = 
    \frac{\pdotw}{4} \frac{\Abub}{4 \pi \Rfree^2}
    \left(3\foldedness + \frac{\left\langle v_r \vout\right\rangle}{\vps^2} \right)
\end{equation}
where the factor 
\begin{equation}
    \label{eq:arearatio}
    \frac{\Abub}{4 \pi \Rfree^2} = \frac{\dsteady}{\langle \shat \cdot \nhat\rangle}    
\end{equation}
represents the ratio of the surface area of the bubble to the surface area of the interior shock surface (assumed spherical), and $\dsteady$ is the `dilation factor' of the flow averaged over all streamlines at the outer limit of the WBB. 
From \autoref{eq:steady_mass2} we can write $\langle v_r \vout \rangle/\vps^2 =\langle \rhat \cdot \shat\, \shat \cdot \nhat /\varphi^2\rangle $ for the Reynolds stress term in \autoref{eq:app_pr2}.

As quantified in \citet{Lancaster21b} (see Figure 16), it seems generally true for the case of WBBs that $\Abub \foldedness \approx 4 \pi \Rbub^2$; that is, the product of the total area and its average foldedness is approximately equal to is sphere-equivalent surface area. This implies that
\begin{equation}
    \label{eq:fold_scaling}
    \foldedness \approx \frac{4\pi \Rbub^2}{\Abub}
    \, .
\end{equation}
Using \autoref{eq:arearatio} and \autoref{eq:fold_scaling} in \autoref{eq:app_pr2}, we obtain 
\begin{equation}
    \label{eq:app_pr4}
    \dot{p}_r = \frac{\pdotw}{4} \left[3 \left( \frac{\Rbub}{\Rfree}\right)^2 + \dsteady
    \frac{\langle \rhat \cdot \shat\, \shat \cdot \nhat /\varphi^2\rangle}{\langle \shat \cdot \nhat\rangle} \right]
\end{equation}
We note that in the case of a purely radial post-shock flow,  $ \langle \rhat \cdot \shat\, \shat \cdot \nhat \rangle /\langle \shat \cdot \nhat \rangle = 1$ and $\varphi=\dsteady=(\Rbub/\Rfree)^2$ for the second term in \autoref{eq:app_pr4}. In this case, the square bracket is equal to $3(\Rbub/\Rfree)^2 + (\Rfree/\Rbub)^2$, which recovers Equation A11 of \citet{Lancaster21a}. More generally, the term representing Reynolds stresses depends on the orientation of streamlines and the bubble surface.  If $\nhat$ and $\rhat$ are aligned (i.e.~the surface is spherical), while $\shat$ is randomly oriented (but still outward) with $\varphi$ independent of $\shat$,  $ \langle \rhat \cdot \shat\, \shat \cdot \nhat \rangle /\langle \shat \cdot \nhat \rangle = \int_0^1 \mu^2 d\mu/\int_0^1 \mu d\mu = 2/3$.  If the bubble surface is highly folded so that $\nhat$ and $\rhat$ are independent, and  $\varphi$ is independent of $\shat$, then  $ \langle \rhat \cdot \shat\, \shat \cdot \nhat \rangle /\langle \shat \cdot \nhat \rangle = \langle \rhat \cdot \shat\rangle$, which has an upper limit of unity.  Thus, for a range of circumstances, the second term in square brackets is $\dsteady^{-1}$ times a factor no larger than 1.  If, furthermore, the surface is so folded that $\nhat$ has no preferred orientation, then $\langle \shat \cdot \nhat\rangle \sim \foldedness$ and from  \autoref{eq:arearatio} and \autoref{eq:fold_scaling} we have $\dsteady\approx (\Rbub/\Rfree)^2$.

In practice, we are frequently in the $\dsteady \gg 1$ limit (and hence, implicitly, $\Rbub^2 \gg \Rfree^2$), so that the first term in parentheses in \autoref{eq:app_pr4} (associated with the pressure force) is much greater than the second term (associated with Reynolds stresses). In this case, we can disregard the Reynolds stress term to arrive at
\begin{equation}
    \label{eq:app_pr3}
    \dot{p}_r =\frac{3}{4}\pdotw\left( \frac{\Rbub}{\Rfree}\right)^2 = \frac{3}{4}\pdotw\dsteady\frac{\foldedness}{\langle \shat \cdot \nhat \rangle } \, .
\end{equation}
We note, however,that \autoref{eq:app_pr3} does not apply in the extreme cooling limit.  For this case, the post-shock flow is radial and it immediately enters the shell.  Thus, $\Rbub/\Rfree =1$, the square bracket in \autoref{eq:app_pr4} is equal to 4, and $\dot{p}_r \to \pdotw$. This corresponds to the ``Efficiently Cooled" limit where $\alpha_p = 1$ in \citet{Lancaster21a}. 

In the case where the thermal pressure term dominates, we can recast the radial stress communicated out of the bubble (\autoref{eq:app_pr3}) as
\begin{equation}
    \label{eq:Fr_eff}
    \dot{p}_r  \approx
    4\pi \Rbub^2 \Phot
    = \frac{3}{4} \pdotw\frac{\Vwind/4}{\voutavg} \frac{4\pi \Rbub^2}{\Abub} \, .
\end{equation}
The first equality is from \autoref{eq:app_pr3} with \autoref {eq:Pps},  while in the last equality we have applied \autoref{eq:Phot_steady_app} again. This tells us that the `normal' force term, $4\pi \Rbub^2 \Phot$ still applies in the completely aspherical, messy picture, as long as \autoref{eq:fold_scaling} applies.

It is worthwhile to remark on the cause and effect of how this structure is set up within the shocked wind gas. Considering $\Abub$ to be set at a fixed time $t$ by the past evolution of the bubble, and that the flow in the shocked wind is subsonic, we see from \autoref{eq:boundary_condition} that $\Rfree$ is set by the velocity at which gas is moving out of the `steady' wind region. As is explicit in all of the derivations of this section, this condition is maintained on timescales that are short compared with the expansion time of the bubble (over which $\Abub$ may vary). It is then $\Rfree$, through \autoref{eq:Pps}, that determines $\Phot$, which determines the instantaneous radial momentum evolution of the bubble in \autoref{eq:Fr_eff}.

Causality requires that the dynamics are determined by the properties of the flow at the bubble's surface. This can be seen by dividing \autoref{eq:Fr_eff} by $\pdotw$ to obtain an explicit prediction for the momentum enhancement factor as 
\begin{equation}
    \alpha_p = \frac{3}{4} \frac{\Vwind/4}{\voutavg} \frac{4\pi \Rbub^2}{\Abub} \, .
\end{equation}
We then see that the momentum enhancement is determined solely by the average outward velocity at the surface, $\voutavg$ which we will show is governed by dissipation, and the geometry of its surface, given by $\Abub$. 

\begin{figure*}
    \includegraphics[width=0.5\textwidth]{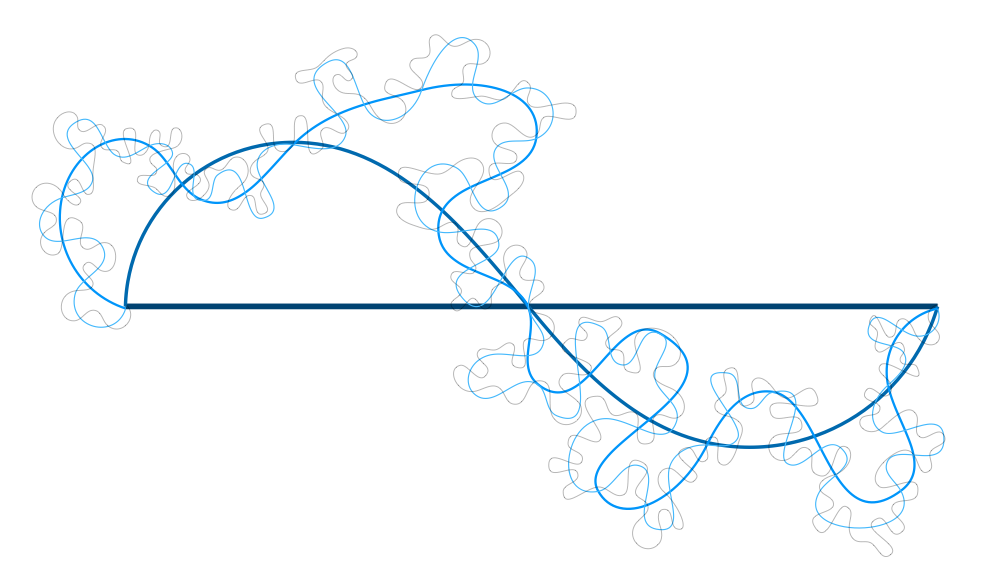}
    \includegraphics[width=0.5\textwidth]{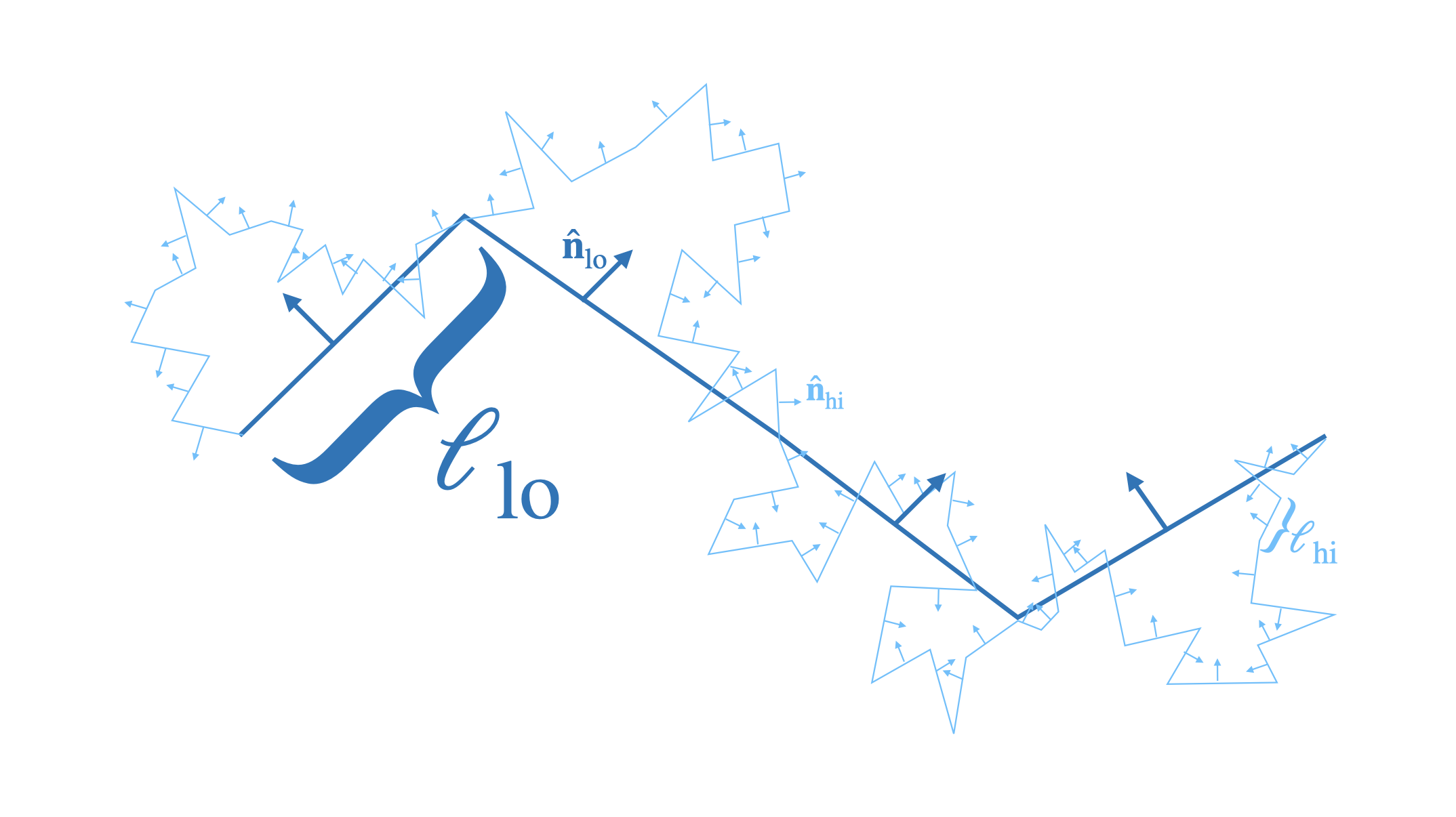}
    \caption{Illustration of the multi-scale structure of fractal bubble surfaces. In the left panel we show a schematic of the fractal bubble/ambient interface  measured on multiple scales with thinner line styles indicating measurements on smaller scales. In the right panel we show  how the velocity through this fractal surface can also be thought of as having a fractal structure, as described in the text.}
    \label{fig:folded}
\end{figure*}

\section{Fractal Description of the Surface}
\label{app:frac}

The quantities $\Abub$ and $\voutavg$ respectively quantify the area and mean outflow velocity at the bubble's surface. These are used in computing mass and thermal energy flows through the surface, respectively proportional to $\rhohot$ and  $\Phot $, and the momentum flow given in \autoref{eq:Fr_eff}.  Special care is required when the surface has a `fractal' structure. In reality this fractal structure will only be present over some range of scales, and in particular it will be regulated or smoothed out by dissipative physics on the smallest scales. Despite that, we can think of measuring both $\Abub$ and $\voutavg$ as a function of scale. For $\Abub$, this is the usual measurement for fractals where we only allow for surface elements of a given linear dimension, $\ell$. We can also think of this as insisting that the surface describing the interface has some minimum radius of curvature, $\ell$. This is shown schematically in the left panel of \autoref{fig:folded}.

The outflow velocity, $\vout$, also has a fractal structure, and therefore $\voutavg$ is dependent on the measurement scale, $\ell$. We depict this schematically for two given scales, $\ell_{\rm hi} < \ell_{\rm lo}$ in the right panel of \autoref{fig:folded}, where $\ell_{\rm hi}$ represents the high resolution picture, at small scale, and $\ell_{\rm lo}$ represents the low resolution picture, at large scale. In particular, imagine the light blue surface in \autoref{fig:folded}, measured on scale $\ell_{\rm hi}$, is the `true' interface between hot and cold gas and we can define a normal component of the outflow velocity $v_{\rm out, hi} = \mathbf{v}\cdot \hat{\mathbf{n}}_{\rm hi}$ at each facet of this interface. We would then define $\langle \vout (\ell_{\rm hi})\rangle$ as the average of this quantity over the facets. We can then ascribe a value $v_{\rm out, lo}$ to a given facet defined on the larger scale $\ell_{\rm lo}$ by associating each smaller scale facet with a larger scale one and setting $v_{\rm out, lo} = N^{-1} \sum_i^N v_{{\rm hi},i}\left(\hat{\mathbf{n}}_{{\rm hi},i} \cdot \hat{\mathbf{n}}_{\rm lo}\right)$, where the sum is over all smaller facets associated with the larger facet (and we have implicitly assumed that all facets at a given scale have the same area).  We note that provided $\Phot$ and $\rhohot$ are effectively spatially uniform, there is no need to apply additional weighting in the average.

With the above definition of the fractal nature of $\voutavg$ we can then note that there is a natural relationship between the fractal scaling of $\Abub$ and $\voutavg$ at fixed time provided by the mass continuity condition. Through \autoref{eq:boundary_condition}, the product $\Abub\voutavg$ must be independent of  $\ell$ at a fixed $\Rbub$ or time. In particular, if we measure both $\Abub$ and $\voutavg$ at the same $\ell$, and we assume $\Abub = 4\pi \Rbub^2 (\Rbub/\ell)^d$ then we have
\begin{equation}
    \label{eq:outflux_scaling}
    \langle \vout \rangle \frac{\Abub}{4 \pi \Rbub^2} = \langle \vout \rangle \left(\frac{\Rbub}{\ell}  \right)^d \, . 
\end{equation}
Thus, in order for \autoref{eq:Fr_eff} to be well-defined for $\ell \rightarrow 0$, we must have $\langle \vout \rangle \propto \ell^d$. That is, the mean outflow velocity through the bubble surface must decrease as the `ruler' scale shrinks.

We can see how foldedness leads to a decrease of $\voutavg$ at smaller $\ell$ by considering the case of a purely radial flow with constant speed (relative to bubble's surface) $v_r$ in the bubble interior, for which (using \autoref{eq:fold_scaling}) 
\begin{equation}
    \label{eq:vout_rad}
    \langle \vout \rangle =v_r\Abub^{-1}\int_{\Abub} \rhat\cdot \nhat \, dA \approx v_r\frac{4\pi \Rbub^2}{\Abub}= v_r \left(\frac{\ell}{\Rbub}\right)^d\, .
\end{equation}
Furthermore, if the fractal structure has no preferred orientation, the above scaling would be true if the velocity in the hot gas has any direction that is uncorrelated with the interface structure, not just a purely-radial direction, since $\langle \shat \cdot \nhat\rangle \sim \foldedness$. We caution, however, that while this limiting case illustrates the effect of the bubble's fractal geometry on the scaling of $\voutavg$ with the length of the ruler $\ell$, it does not provide information about the overall normalization of $\voutavg$. In actuality, the velocity in the bubble interior will not be purely radial, and the magnitude of the outflow velocity through the surface must depend not just on properties in the bubble interior, but also on dissipative processes and cooling within the interaction region of the shell near the interface (see \autoref{subsec:diss_scales} and \autoref{subsec:vout_cooling}).    

\begin{figure*}
    \includegraphics[width=\textwidth]{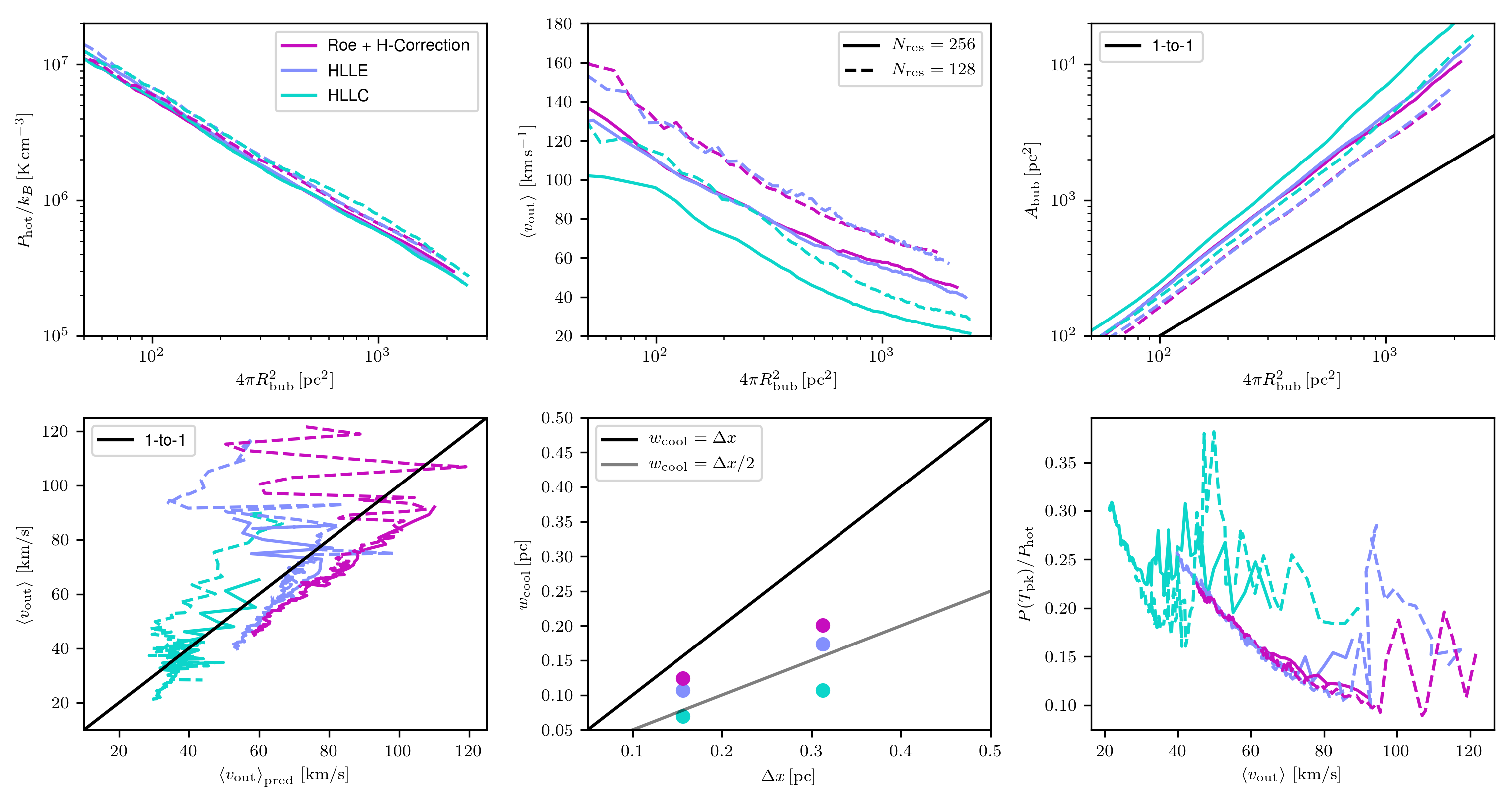}
    \caption{A summary of the key WBB characteristics presented in \autoref{fig:interface_resolution} and \autoref{fig:vout_scaling}, now for our series of simulations run with different choices for the Riemann Solver. \textit{Top Left:} The hot gas pressure as a function of the WBB's ``effective surface area'' $4\pi \Rbub^2$. \textit{Top Middle:} The average velocity at the bubble's surface with $4\pi \Rbub^2$. \textit{Top Right:} The area of the bubble's surface. \textit{Bottom Left:} $\voutavg$ versus the predicted value from \autoref{eq:vout_cool_dip} with the same linear relationship as that shown in the left panel of \autoref{fig:vout_scaling}. \textit{Bottom Middle:} $\wcool$ as a function of resolution. \textit{Bottom Right:} The pressure dip from the bubble interior to the peak cooling temperature $P(\Tpk)/\Phot$ as a function of $\voutavg$.}
    \label{fig:rieman_summary}
\end{figure*}

\section{Effect of the Riemann Solver}
\label{app:riemann}

The results of the main text indicate that the evolution of simulated WBBs, at the resolutions probed here, is dependent upon the effects of numerical dissipation. It is therefore informative to investigate simulations that change the nature of that dissipation. To that end, we run a suite of 6 hydrodynamical simulations corresponding to two choices of resolution $\Nres = 128,\, 256$ and three choices of Riemann solver: the Roe Solver with H-Correction (as in the main text) \citep{Roe81,SandersHcorr98}, the Harten-Lax-van Leer (HLL) method \citep{HLL83} with the Einfeldt estimate for wave-speeds \citep{Einfeldt88} (HLLE), and finally the HLLE method with the contact-wave restored (HLLC) \citep{HLLC94}. These methods are reviewed pedagogically in \citet{ToroRiemann} and their implementation in the \textit{Athena} code is explained in \citet{Stone08_Athena}. We run these simulations without the non-equilibrium cooling and chemistry module of \citet{JGK_NCR23} and instead use the collisional-ionization equilibrium cooling used in \citet{Lancaster21b} and originally described in \citet{CGK_TIGRESS1}. As we will see, these different choices in solvers lead to small quantitative differences but do not make a difference to the overall dynamics of these bubbles.

In \autoref{fig:rieman_summary} we present the results of these simulations. This figure portrays the key quantities of the WBB's evolution that are highlighted in \autoref{fig:interface_resolution} and \autoref{fig:vout_scaling}. Other aspects of the WBB's that are shown in the main text remain broadly consistent here: (i) \autoref{eq:bubble_radial_force} and \autoref{eq:bubble_force2} are good descriptions of the force exerted by the bubbles, (ii) \autoref{eq:Phot_steady} provides a good prediction for the bubble's interior pressure, illustrating that the interiors are nearly steady, and (iii) the convergence properties of the total momenta carried by these bubbles are similar to the hydrodynamical simulations shown in \autoref{fig:mhdVhydro}.

Two factors indicate that the HLLC Riemann solver is much less diffusive for the present application than the other two Riemann solvers:
\begin{itemize}
    \item[1.] As shown in the top middle panel of \autoref{fig:rieman_summary}, $\voutavg$ is much smaller with the HLLC solver at fixed bubble properties, indicating a decreased aptitude to allow diffusion across the bubble's interface.
    
    \item[2.] The thickness of the cooling layer, $\wcool$, is much smaller when using the HLLC solver, by a factor $\approx 2$ compared to the HLLE solver, and slightly more compared to Roe. This indicates that the interface is less `smeared-out' by numerical diffusion.
\end{itemize}
This is not surprising as the HLLC solver was explicitly developed to help better resolve contact discontinuities like the bubble's surface that we are investigating here. Notably, the more expensive Roe solver with the H-correction seems to be as diffusive as the simple HLLE solver for these types of problems. This may not be as surprising when one considers that the H-correction, which effectively adds diffusion, is only applied when differences in wave-speeds are large in different grid-aligned directions, as they almost always are at the interface. While it is still discouraging that this makes the method essentially as diffusive as the HLLE solver, this likely only applies at the WBB interface; other regions of the solution are still less diffusive (at least from visual inspection) with the Roe + H-correction solver as with the HLLE solver.

We also see from the top right panel of \autoref{fig:rieman_summary} that the surface area of the bubble is noticeably larger with the HLLC solver, especially at late times. If one considers that the HLLC solver is less diffusive and therefore less likely to smooth out small scale structure at the bubble's surface, this development is not particularly surprising. What is very intriguing is that the competition between geometry (increased $\Abub$) and diffusion (decreased $\voutavg$) here seems to be nearly completely balanced as we see in the top left panel: the resulting interior pressure in the HLLC solution is nearly exactly the same at given $\Rbub$, which means that $\dot{p}_r$ is nearly independent of the solver. It remains to be seen whether this is a coincidence or points to a more fundamental dynamical relationship that is as yet undiscovered here.

Turning our attention to the bottom left panel of \autoref{fig:rieman_summary}, we see that the relationship between the measured $\voutavg$ and that predicted by \autoref{eq:vout_cool_dip} remains consistent between the various Riemann solvers, as in the results in the main section of the paper. As we explained in \autoref{eq:vout_cool}, this consistency condition should always roughly hold, even if it varies slightly due to the assumption that all gas is cooling at a single temperature, which cannot be the case.

One clear indication of strong differences between the Riemann solvers is given in the final, bottom-right panel of \autoref{fig:rieman_summary}. Here we see that the relationship between the pressure dip and the average velocity at the interface is different amongst all the Riemann solvers: a clear indication that this is a numerical artifact, as has been reported elsewhere \citep{Ji19,FieldingFractal20}. The higher value of $P(\Tpk)/\Phot$ in the HLLC solver is yet another indication of its lower diffusivity. It is interesting that these relationships do seem to remain resolution independent even with the different Riemann solvers.

The exact origin of the relationships shown here for different Riemann solvers may be more fully explicated with targeted, simplified, one-dimensional simulations.

\begin{figure*}
    \includegraphics[width=\textwidth]{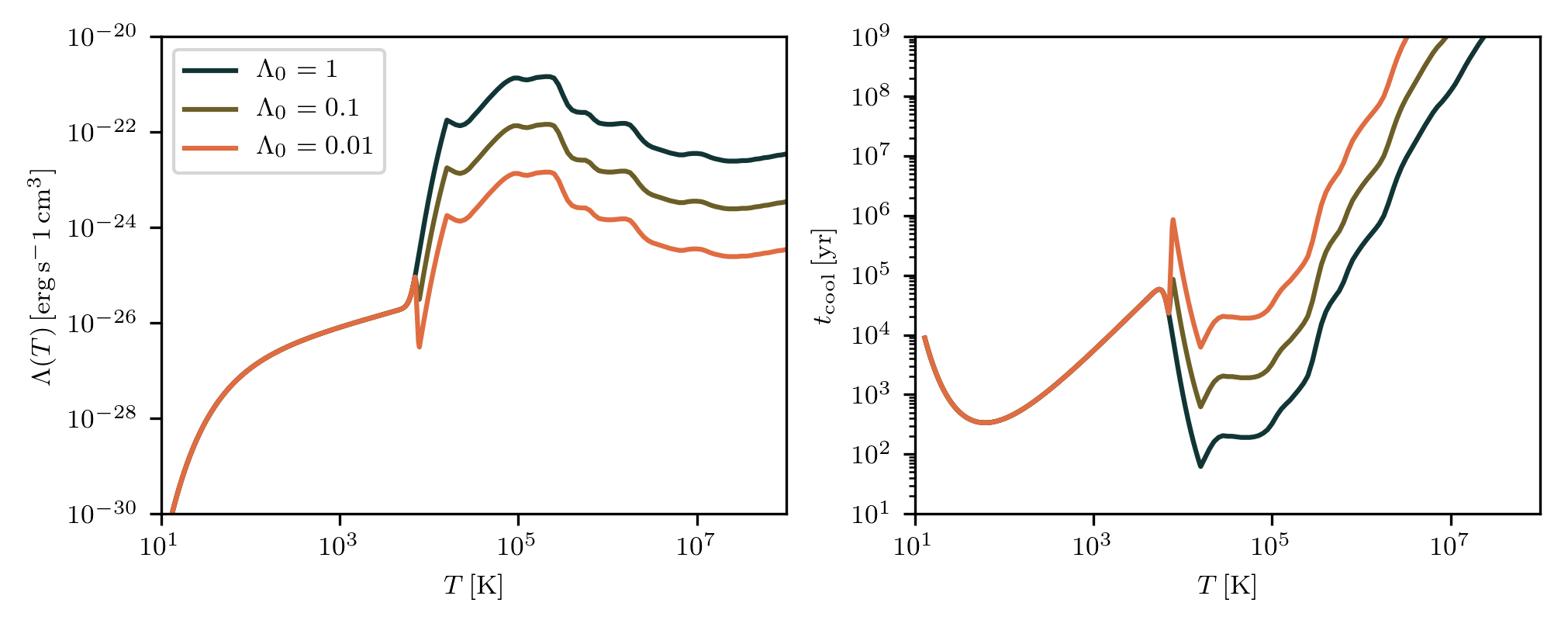}
    \caption{Our modifications to the cooling function that are used in the simulations presented in \autoref{app:cooling}. \textit{Left}: The cooling function, $\Lambda(T)$ for the three cases investigated here, details in text. \textit{Right}: The corresponding changes in the cooling time of the gas, $\tcool$, as given by \autoref{eq:tcooldef} with an assumed pressure of $P/k_B = 10^5\, {\rm K}\, {\rm cm}^-3$.}
    \label{fig:cool_modify}
\end{figure*}

\begin{figure*}
    \includegraphics[width=\textwidth]{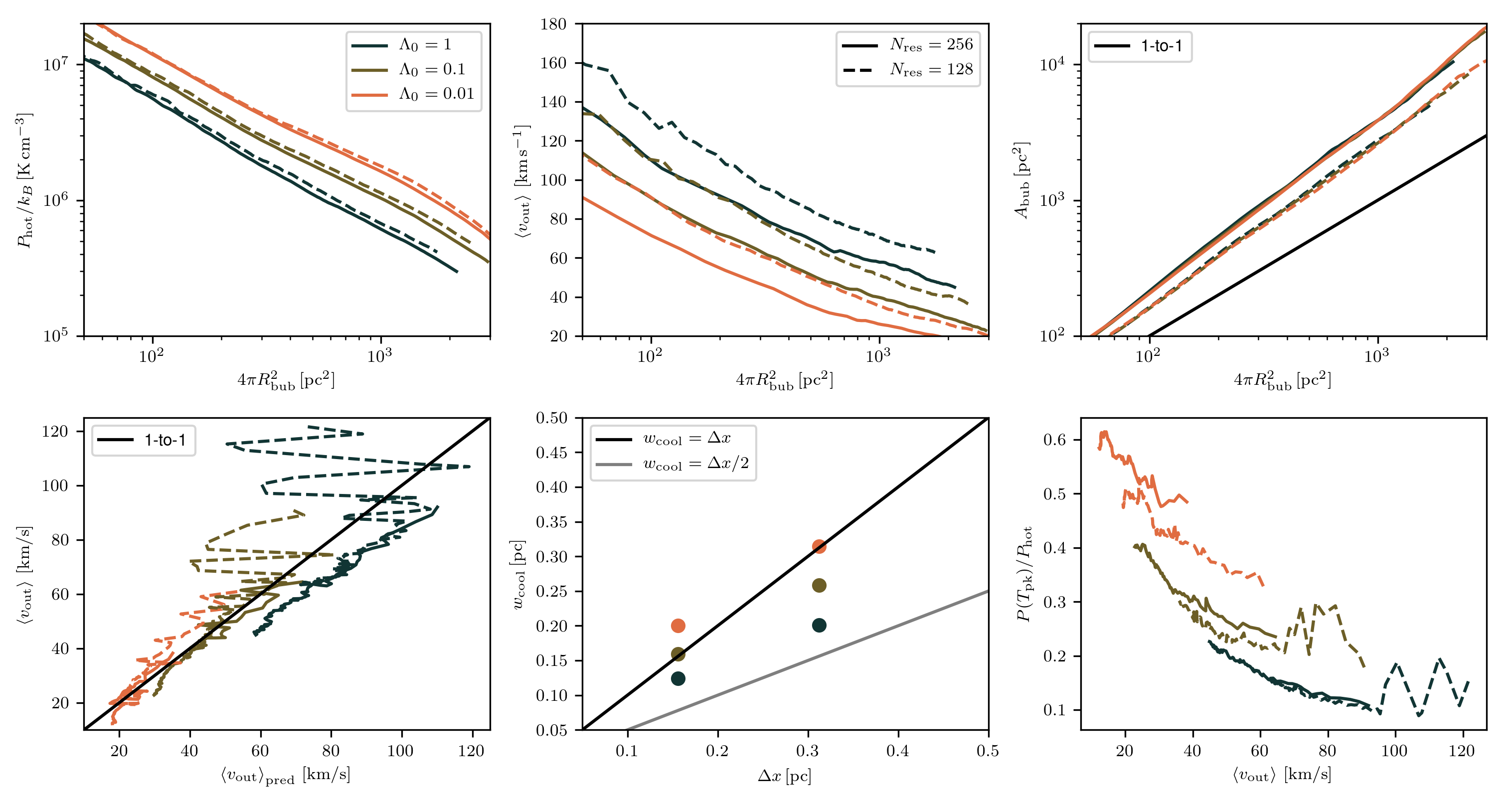}
    \caption{Same as \autoref{fig:rieman_summary} but now for simulations that vary the strength of cooling in the interface layer. Details on the modification of the cooling function are given in the text.}
    \label{fig:cooling_summary}
\end{figure*}

\section{Cooling Function and the Effects of its Variation}
\label{app:cooling}

Finally, we present an additional series of simulations where we have modified the cooling function by hand. All other parameters of the simulations are as in the main text. We modify the cooling function by beginning with the CIE cooling function of \citet{CGK_TIGRESS1} (as in \autoref{app:riemann}), which we will denote $\Lambda_T(T)$ here. We then modify it by adopting
\begin{equation}
    \label{eq:cool_modify}
    \Lambda(T) = 
    \begin{cases} 
      \Lambda_T(T) & T <  T_{\rm red} \\
      \Lambda_T(T)10^{-\frac{T-T_{\rm red}}{\Delta T}} & T_{\rm red} < T < 1.1 T_{\rm red}\\
      \Lambda_0 \Lambda_T(T)  & T > 1.1T_{\rm red}
   \end{cases}
\end{equation}
where $T_{\rm red} = 7000\, {\rm K}$ is the temperature above which we reduce the cooling function, $\Lambda_0$ is the factor by which we reduce it and
\begin{eqnarray}
    \Delta T = - \frac{T_{\rm red}}{10 \log_{10}\left( \Lambda_0\right)}
\end{eqnarray}
is a factor to account for a smooth transition between the two regimes (the middle term in \autoref{eq:cool_modify}). The modified cooling functions and the changes they impose to the cooling time of the gas are presented in \autoref{fig:cool_modify}. We note that, for $\Lambda_0 = 1$ we simply use $\Lambda_T(T)$ (these results are the same simulations as the ``Roe + H-Corection" simulations of \autoref{app:riemann}).

The main summary results of these simulations are presented in \autoref{fig:cooling_summary} in an exactly analogous way to that presented in \autoref{fig:rieman_summary}. The main takeaway from the results presented here is that the formalism presented in the main body of the text still applies in this cooling-modified picture. In particular, the pressure in the hot gas is still straightforwardly predicted from \autoref{eq:Phot_steady} with the area of the bubble, $\Abub$, being described by a fractal and the small scale dissipative velocity, $\voutavg$, being determined by numerical dissipation as given by \autoref{eq:vout_cool_dip}. In particular, we see here that $\Abub$ is independent of the strength of the cooling in the interface layer (top right of \autoref{fig:cooling_summary}) indicating that this surface is simply determined by the dynamical evolution of the bubble, surface hydrodynamical instabilities, and the structure of the background gas into which the bubble expands.

In the bottom row of \autoref{fig:cooling_summary} we can see that, while \autoref{eq:vout_cool_dip} still accurately predicts the velocity into the surface at all levels of cooling, the surface itself is becoming better resolved as cooling becomes weaker. This is indicated by an increase in the effective width of the cooling layer, $\wcool$ (middle bottom panel), and an increase in the pressure as measured by $\fdip$ closer to 1 (right bottom panel), as cooling becomes weaker. It seems that cooling does not have a large effect on $\voutavg$, as shown in the top middle panel. In particular, reduction of $\Lambda$ by a factor of 10 does not decrease $\voutavg$ by a factor $1/\sqrt{10}$, as \autoref{eq:vout_cool} would imply if the cooling volume $\Vpk$ were fixed.  This is because, at a lower cooling rate, the correct cooling length is better resolved, leading to $\Vpk$ larger and $\fdip$ closer to 1, such that \autoref{eq:vout_cool_dip} predicts only a modest decrease in $\voutavg$.

%

\bibliography{bibliography}{}
\bibliographystyle{aasjournal}

\end{document}